\documentclass[12pt]{article}
\usepackage{epsf}

\def\bm#1{\mbox{\boldmath$#1$\unboldmath}}
\def\pslash{\rlap{\hspace{0.02cm}/}{p}}
\def\kslash{\rlap{\hspace{0.02cm}/}{k}}
\def\lslash{\rlap{\hspace{0.011cm}/}{\ell}}
\def\Pslash{\rlap{\hspace{0.065cm}/}{P}}

\def\spose#1{\hbox to 0pt{#1\hss}}
\def\lsim{\mathrel{\spose{\lower 3pt\hbox{$\mathchar"218$}}
 \raise 2.0pt\hbox{$\mathchar"13C$}}}
\def\gsim{\mathrel{\spose{\lower 3pt\hbox{$\mathchar"218$}}
 \raise 2.0pt\hbox{$\mathchar"13E$}}}

\newcommand\epjc[3]{Eur.\ Phys.\ J.\ C {\bf #1}, #3 (#2)}

\newcommand\jhep[3]{J.\ High Ener.\ Phys.\ {\bf #1} (#2) #3}
\newcommand\npb[3]{Nucl.\ Phys.\ B {\bf #1} (#2) #3}
\newcommand\npps[3]{Nucl.\ Phys.\ B (Proc.\ Suppl.) {\bf #1} (#2) #3}
\newcommand\plb[3]{Phys.\ Lett.\ B {\bf #1} (#2) #3}
\newcommand\prd[3]{Phys.\ Rev.\ D {\bf #1} (#2) #3}

\newcommand\prl[3]{Phys.\ Rev.\ Lett.\ {\bf #1} (#2) #3}
\newcommand\rmp[3]{Rev.\ Mod.\ Phys.\ {\bf #1} (#2) #3}
\newcommand\sjnp[3]{{Sov.\ J.\ Nucl.\ Phys.\ }{\bf#1} (#2) #3}
\newcommand\yf[3]{{Yad.\ Fiz.\ }{\bf#1} (#2) #3}
\newcommand\zpc[3]{Z.\ Phys.\ C {\bf #1} (#2) #3}
\newcommand{\hepph}[1]{{\tt hep-ph/#1}}
\newcommand{\heplat}[1]{{\tt hep-lat/#1}}
\newcommand{\hepex}[1]{{\tt hep-ex/#1}}

\begin{document}

\begin{titlepage}

\begin{flushright}
{\small
CERN-TH/2001-107\\
CLNS~01/1728\\
PITHA~01/01\\
SHEP~01/11\\
hep-ph/0104110\\
April 11, 2001}
\end{flushright}

\vspace{0.4cm}
\begin{center}
\Large\bf\boldmath
QCD Factorization in $B\to\pi K,\,\pi\pi$ Decays and\\
Extraction of Wolfenstein Parameters
\unboldmath
\end{center}

\vspace{0.5cm}
\begin{center}
{\sc M. Beneke$^a$, G. Buchalla$^b$, M. Neubert$^c$ and C.T. 
Sachrajda$^d$}\\
\vspace{0.7cm}
{\sl ${}^a$Institut f\"ur Theoretische Physik E, RWTH Aachen\\
D--52056 Aachen, Germany\\
\vspace{0.3cm}
${}^b$Theory Division, CERN, CH--1211 Geneva 23, Switzerland\\
\vspace{0.3cm}
${}^c$Newman Laboratory of Nuclear Studies, Cornell University\\
Ithaca, NY 14853, USA\\
\vspace{0.3cm}
${}^d$Department of Physics and Astronomy, University of Southampton\\
Southampton SO17 1BJ, UK}
\end{center}

\vspace{0.3cm}
\begin{abstract}
\vspace{0.2cm}\noindent
In the heavy-quark limit, the hadronic matrix elements entering 
nonleptonic $B$-meson decays into two light mesons can be calculated
from first principles including ``nonfactorizable'' strong-interaction
corrections. The $B\to\pi K,\pi\pi$ decay amplitudes are computed 
including electroweak penguin contributions, SU(3) violation in the 
light-cone distribution amplitudes, and an estimate of power corrections 
from chirally-enhanced terms and annihilation graphs. The results are 
then used to reduce the theoretical uncertainties in determinations of 
the weak phases $\gamma$ and $\alpha$. In that way, new  
constraints in the $(\bar\rho,\bar\eta)$ plane are derived. Predictions 
for the $B\to\pi K,\,\pi\pi$ branching ratios and CP asymmetries are 
also presented. A good global fit to the (in part preliminary) 
experimental data on the branching fractions is obtained without taking 
recourse to phenomenological models. 
\end{abstract}

\vfil
\end{titlepage}

\section{Introduction}

The study of nonleptonic two-body decays of $B$ mesons is of primary
importance for the exploration of CP violation and the determination of 
the flavour parameters of the Standard Model. Because of the 
interference of several competing amplitudes, these processes allow for 
the presence of different weak and strong-interaction phases, which 
play a crucial role for CP violation. In the Standard Model, all 
CP-violating observables are related to the complex phase of the quark 
mixing matrix, which in turn implies nontrivial angles in the 
``unitarity triangle'' $V_{ud} V_{ub}^* + V_{cd} V_{cb}^* 
+ V_{td} V_{tb}^*=0$. With the standard choice of phase conventions, 
one defines the weak phases $\beta=-\mbox{arg}(V_{td})$ and $\gamma
=\mbox{arg}(V_{ub}^*)$, as well as $\alpha=180^\circ-\beta-\gamma$. In 
the Standard Model, $\sin2\beta$ can be extracted in a theoretically 
clean way by measuring the time-dependent rates for the decays 
$B^0,\bar B^0\to J/\psi\,K_S$. The measurement of $\gamma$ (or $\alpha$) 
is more difficult, since it requires controlling the hadronic dynamics 
in nonleptonic $B$ decays. 

A promising strategy for the determination of $\gamma$ is based on rate 
measurements for the charged modes $B^\pm\to(\pi K)^\pm$ and 
$B^\pm\to\pi^\pm\pi^0$ \cite{NR2}. Hadronic uncertainties in this method 
can be reduced to a minimum by exploiting the structure of the effective 
weak Hamiltonian and using isospin and SU(3) flavour symmetries 
\cite{Fl96,NR1,Mat98,Fl98}. If only measurements of CP-averaged 
branching ratios are available, it is still possible to derive bounds on 
$\gamma$ \cite{NR1,FM97}, or to determine $\gamma$ under the assumption 
of only a moderate strong-interaction phase $\phi$ between penguin and 
tree amplitudes. A different strategy for extracting $\gamma$ uses 
SU(3)-symmetry relations between the various contributions to the 
time-dependent $B_d,\bar B_d\to\pi^+\pi^-$ and $B_s,\bar B_s\to K^+ K^-$ 
decay amplitudes \cite{Fl99}. The main theoretical limitation of these 
methods is in the accuracy with which the effects of SU(3) breaking can 
be estimated. 

The angle $\alpha$ can be determined from the time-dependent CP 
asymmetry in the decays $B^0,\bar B^0\to\pi^+\pi^-$, if the subdominant 
penguin contribution to the decay amplitudes can be subtracted in some 
way. This can be achieved by an isospin analysis \cite{GL90}; however, 
in practice that route is extremely challenging due to the difficulty in 
measuring the very small $B^0,\bar B^0\to\pi^0\pi^0$ branching ratios. 
One must therefore rely on dynamical input for the penguin-to-tree ratio 
\cite{SiWo94,GQ97,Ch98}. Alternatively, $\alpha$ can be measured in 
related decays such as $B\to\rho\,\pi$, for which the penguin 
contribution can be eliminated using a time-dependent analysis of the 
$B\to\pi^+\pi^-\pi^0$ Dalitz plot \cite{QS93}.

Most of the above-mentioned determinations of $\gamma$ or $\alpha$ 
have theoretical limitations, which would be reduced if some degree of 
theoretical control over two-body nonleptonic $B$ decays could be 
attained. However, in the past this has proven to be a very difficult 
problem. Even advanced methods such as lattice gauge theory, QCD sum 
rules, or the large-$N_c$ expansion have little to say about the QCD 
dynamics relevant to hadronic $B$ decays. In recent work, we have 
developed a systematic approach to this problem. It is based on the 
observation that, in the heavy-quark limit $m_b\gg\Lambda_{\rm QCD}$, a 
rigorous QCD factorization formula holds for the two-body decays 
$B\to M_1 M_2$, if the ``emission particle'' $M_2$ (the meson not 
obtaining the spectator quark from the $B$ meson) is a light meson. (It 
has been argued that perhaps the large-$N_c$ limit may be more relevant 
to factorization than the heavy-quark limit \cite{Zoltan}. However, the 
dramatic decrease of ``nonfactorizable'' effects seen when comparing 
$K\to\pi\pi$, $D\to\pi K$ and $B\to\pi K$ decays shows that the 
heavy-quark limit is of crucial importance.) We have previously applied 
the factorization formula to $B\to\pi\pi$ decays and obtained results 
for the decay amplitudes at next-to-leading order in $\alpha_s$ and to 
leading power in $\Lambda_{\rm QCD}/m_b$ \cite{BBNS1}. The conceptual 
foundations of our approach have been discussed in detail in 
\cite{BBNS2}, which focused on the simpler case of decays into 
heavy--light final states (where $M_1$ is a charm meson). In the present 
work, the QCD factorization formula is applied to the general case of 
$B$ decays into a pair of light, flavour-nonsinglet pseudoscalar mesons. 
(Preliminary results of this analysis have been presented in 
\cite{BBNS3}.) The present analysis contains three new theoretical 
ingredients in addition to a much more detailed phenomenological 
analysis:

\vspace*{-0.3cm}
\paragraph{\rm 1.} 
Matrix elements of electroweak penguin operators are included, which 
play an important role in charmless decays based on $b\to s\bar q q$ 
transitions. This is a straightforward extension of our previous 
analysis. However, a sensible implementation of QCD corrections to 
electroweak penguin matrix elements implies that one departs from the 
usual renormalization-group counting, in which the initial conditions 
for the electroweak penguin operators at the scale $\mu=M_W$ are 
treated as a next-to-leading order effect. 

\vspace*{-0.3cm}
\paragraph{\rm 2.} 
Hard-scattering kernels are derived for general, asymmetric meson 
light-cone distribution amplitudes. This is important for addressing 
the question of nonfactorizable SU(3)-breaking corrections, since the 
distribution amplitudes of strange mesons are, in general, not symmetric 
with respect to the quark and antiquark momenta. 

\vspace*{-0.3cm}
\paragraph{\rm 3.} 
The leading power corrections to the heavy-quark limit are estimated by
analyzing ``chirally-enhanced'' power corrections related to certain 
twist-3 distribution amplitudes for pseudoscalar mesons. We also 
discuss potentially large power corrections arising from annihilation 
topologies, noted first in \cite{KLS00}. This is essential for 
controlling the theoretical uncertainties of our approach.

\vspace*{0.2cm}
\noindent
The second and third items have not been considered in previous 
generalizations of the results of \cite{BBNS1} to the case of 
$B\to\pi K$ decays \cite{MSYY,YY1,YY2,HMW}. Chirally-enhanced power 
corrections were discussed in \cite{DYZ,CY2}, but we disagree with some 
of the results obtained by these authors.

The QCD factorization approach provides us with model-independent 
predictions for the decays amplitudes including, in particular, their 
strong-interaction phases. The same method can also be applied to other 
charmless decays, such as vector--pseudoscalar \cite{YY1} or 
vector--vector modes. Our main focus here is on the development of the 
new conceptual aspects of the approach that are important for a 
comprehensive phenomenological analysis. This includes a detailed 
discussion of various sources of potentially large power corrections to 
the heavy-quark limit. We then perform a comprehensive study of 
CP-averaged branching fractions and direct CP asymmetries in decays 
to $\pi K$ and $\pi \pi$ final states, including a detailed discussion
of the theoretical uncertainties from all inputs to the QCD 
factorization approach. In many of the phenomenological applications 
discussed in this work the dynamical information obtained using the QCD 
factorization formalism is used in a ``minimal way'', to reduce the 
hadronic uncertainties in methods that are theoretically clean up to 
``nonfactorizable'' SU(3)-breaking effects. Most importantly, these 
strategies do not suffer from uncertainties related to weak annihilation 
contributions. In this way, it is possible to reduce the hadronic 
uncertainties in the strategies for determining $\gamma$ from 
$B^\pm\to\pi K,\pi\pi$ decays proposed in \cite{NR2,Mat98} to the 
level of ``nonfactorizable'' corrections that simultaneously violate 
SU(3) symmetry and are power suppressed in the heavy-quark limit. These 
corrections are parametrically suppressed by the product of three small 
quantities: $1/N_c$, $(m_s-m_d)/\Lambda_{\rm QCD}$, and 
$\Lambda_{\rm QCD}/m_b$. As a consequence, we argue it will eventually
be possible to determine $\gamma$ with a theoretical accuracy of 
about $10^\circ$ (unless $\gamma$ is much different from its expected
value in the Standard Model). More accurately, the strategies discussed 
here can constrain the Wolfenstein parameters $\bar\rho$ and $\bar\eta$ 
with accuracies similar those obtained from the standard global fit to 
$|V_{ub}|$, $\epsilon_K$, and $B$--$\bar B$ mixing. Given the
theoretical input discussed in this paper even the present, 
preliminary data on the rare hadronic decays exclude at 
95\% confidence level half of the parameter space obtained from the 
standard fit.

The QCD factorization approach provides a complete theoretical 
description of all $B\to PP$ decay amplitudes. This allows for a large 
variety of predictions, which go far beyond those explored in the 
present work. In the future, this will offer the possibility of several 
nontrivial experimental tests of the factorization formula. A more 
detailed discussion of these predictions will be presented elsewhere.

The remainder of this paper is organized as follows: In 
Section~\ref{sec:2}, we collect some basic formulae and express the
$B\to\pi K,\pi\pi$ decay amplitudes in terms of parameters $a_i$ and 
$b_i$ appearing in the effective, factorized transition operators for 
these decays (including weak annihilation contributions). 
Section~\ref{sec:3} contains the technical details of the calculations 
based on the factorization formula, a discussion of annihilation 
effects, and a compilation of the relevant formulae for the numerical 
evaluation of our results. Readers not interested in the technical 
aspects of our work can proceed directly to Sections~\ref{sec:numerics} 
and \ref{sec:4}, where we present numerical values for the amplitude 
parameters $a_i$ and $b_i$ (Section~\ref{sec:numerics}) and discuss 
phenomenological applications of our results (Section~\ref{sec:4}).
Specifically, we consider strategies to bound and determine the weak 
phase $\gamma$ and to extract $\sin 2\alpha$ from mixing-induced 
CP violation in $B\to\pi^+\pi^-$ decay. We also present predictions for
CP-averaged branching fractions and CP asymmetries, and perform a 
global fit in the $(\bar\rho,\bar\eta)$ plane to all measured 
$B\to\pi K,\pi\pi$ branching fractions. A critical comparison of our 
formalism with other theoretical approaches to hadronic $B$ decays is 
performed in Section~\ref{sec:comp}.

\section{Parameterizations of the decay amplitudes}
\label{sec:2}

The effective weak Hamiltonian for charmless hadronic $B$ decays 
consists of a sum of local operators $Q_i$ multiplied by short-distance 
coefficients $C_i$ and products of elements of the quark mixing matrix, 
$\lambda_p=V_{pb} V_{ps}^*$ or $\lambda_p'=V_{pb} V_{pd}^*$. Below we 
will focus on $B\to\pi K$ decays to be specific; however, with 
obvious substitutions a similar discussion holds for all other $B$ decays
into two light, flavour-nonsinglet pseudoscalar mesons. Using the 
unitarity relation $-\lambda_t=\lambda_u+\lambda_c$, we write
\begin{equation}\label{Heff}
   {\cal H}_{\rm eff} = \frac{G_F}{\sqrt2} \sum_{p=u,c} \!
   \lambda_p \bigg( C_1\,Q_1^p + C_2\,Q_2^p
   + \!\sum_{i=3,\dots, 10}\! C_i\,Q_i + C_{7\gamma}\,Q_{7\gamma}
   + C_{8g}\,Q_{8g} \bigg) + \mbox{h.c.} \,,
\end{equation}
where $Q_{1,2}^p$ are the left-handed current--current operators arising 
from $W$-boson exchange, $Q_{3,\dots, 6}$ and $Q_{7,\dots, 10}$ are 
QCD and electroweak penguin operators, and $Q_{7\gamma}$ and $Q_{8g}$ 
are the electromagnetic and chromomagnetic dipole operators. They are 
given by
\begin{eqnarray}
   Q_1^p &=& (\bar p b)_{V-A} (\bar s p)_{V-A} \,,
    \hspace{2.5cm}
    Q^p_2 = (\bar p_i b_j)_{V-A} (\bar s_j p_i)_{V-A} \,, \nonumber\\
   Q_3 &=& (\bar s b)_{V-A} \sum{}_{\!q}\,(\bar q q)_{V-A} \,,
    \hspace{1.7cm}
    Q_4 = (\bar s_i b_j)_{V-A} \sum{}_{\!q}\,(\bar q_j q_i)_{V-A} \,,
    \nonumber\\
   Q_5 &=& (\bar s b)_{V-A} \sum{}_{\!q}\,(\bar q q)_{V+A} \,, 
    \hspace{1.7cm}
    Q_6 = (\bar s_i b_j)_{V-A} \sum{}_{\!q}\,(\bar q_j q_i)_{V+A} \,,
    \nonumber\\
   Q_7 &=& (\bar s b)_{V-A} \sum{}_{\!q}\,{\textstyle\frac32} e_q 
    (\bar q q)_{V+A} \,, \hspace{1.11cm}
    Q_8 = (\bar s_i b_j)_{V-A} \sum{}_{\!q}\,{\textstyle\frac32} e_q
    (\bar q_j q_i)_{V+A} \,, \nonumber \\
   Q_9 &=& (\bar s b)_{V-A} \sum{}_{\!q}\,{\textstyle\frac32} e_q 
    (\bar q q)_{V-A} \,, \hspace{0.98cm}
    Q_{10} = (\bar s_i b_j)_{V-A} \sum{}_{\!q}\,{\textstyle\frac32} e_q
    (\bar q_j q_i)_{V-A} \,, \nonumber\\
   Q_{7\gamma} &=& \frac{-e}{8\pi^2}\,m_b\, 
    \bar s\sigma_{\mu\nu}(1+\gamma_5) F^{\mu\nu} b \,,
    \hspace{0.81cm}
   Q_{8g} = \frac{-g_s}{8\pi^2}\,m_b\, 
    \bar s\sigma_{\mu\nu}(1+\gamma_5) G^{\mu\nu} b \,,
\end{eqnarray}
where $(\bar q_1 q_2)_{V\pm A}=\bar q_1\gamma_\mu(1\pm\gamma_5)q_2$, 
$i,j$ are colour indices, $e_q$ are the electric charges of the quarks 
in units of $|e|$, and a summation over $q=u,d,s,c,b$ is implied. (The
definition of the dipole operators $Q_{7\gamma}$ and $Q_{8g}$ corresponds 
to the sign convention $iD^\mu=i\partial^\mu+g_s A_a^\mu t_a$ for the 
gauge-covariant derivative.) The Wilson coefficients are calculated at 
a high scale $\mu\sim M_W$ and evolved down to a characteristic scale 
$\mu\sim m_b$ using next-to-leading order renormalization-group 
equations. The essential problem obstructing the calculation of 
nonleptonic decay amplitudes resides in the evaluation of the hadronic 
matrix elements of the local operators contained in the effective 
Hamiltonian.

Applying the QCD factorization formula and neglecting power-suppressed 
effects, the matrix elements of the effective weak Hamiltonian can be 
written in the form \cite{BBNS1,BBNS2}
\begin{equation}\label{Top}
   \langle\pi K|{\cal H}_{\rm eff}|\bar B\rangle
   = \frac{G_F}{\sqrt2} \sum_{p=u,c} \lambda_p\,
   \langle\pi K|{\cal T}_p+{\cal T}_p^{\rm ann}|\bar B\rangle \,,
\end{equation}
where
\begin{eqnarray}\label{Toper}
   {\cal T}_p &=& a_1(\pi K)\,\delta_{pu}\,
    (\bar u b)_{V-A} \otimes (\bar s u)_{V-A} \nonumber\\
   &+& a_2(\pi K)\,\delta_{pu}\,
    (\bar s b)_{V-A} \otimes (\bar u u)_{V-A} \nonumber\\
   &+& a_3(\pi K) \sum{}_{\!q}\, (\bar s b)_{V-A} \otimes
    (\bar q q)_{V-A} \nonumber\\
   &+& a_4^p(\pi K) \sum{}_{\!q}\, (\bar q b)_{V-A} \otimes
    (\bar s q)_{V-A} \nonumber\\
   &+& a_5(\pi K) \sum{}_{\!q}\, (\bar s b)_{V-A} \otimes
    (\bar q q)_{V+A} \nonumber\\
   &+& a_6^p(\pi K) \sum{}_{\!q}\, (-2)(\bar q b)_{S-P} \otimes
    (\bar s q)_{S+P} \nonumber\\
   &+& a_7(\pi K) \sum{}_{\!q}\, (\bar s b)_{V-A} \otimes
    {\textstyle\frac32} e_q(\bar q q)_{V+A} \nonumber\\
   &+& a_8^p(\pi K)\sum{}_{\!q}\, (-2)(\bar q b)_{S-P} \otimes
    {\textstyle\frac32} e_q(\bar s q)_{S+P} \nonumber\\
   &+& a_9(\pi K)\sum{}_{\!q}\, (\bar s b)_{V-A} \otimes
    {\textstyle\frac32} e_q(\bar q q)_{V-A} \nonumber\\
   &+& a_{10}^p(\pi K) \sum{}_{\!q}\, (\bar q b)_{V-A} \otimes
    {\textstyle\frac32} e_q(\bar s q)_{V-A} \,.
\end{eqnarray}
Here $(\bar q_1 q_2)_{S\pm P}=\bar q_1(1\pm\gamma_5)q_2$, and a 
summation over $q=u,d$ is implied. The symbol $\otimes$ indicates that 
the matrix elements of the operators in ${\cal T}_p$ are to be evaluated 
in the factorized form $\langle\pi K|j_1\otimes j_2|\bar B\rangle\equiv
\langle\pi|j_1|\bar B\rangle\,\langle K|j_2|0\rangle$ or 
$\langle K|j_1|\bar B\rangle\,\langle\pi|j_2|0\rangle$, as appropriate.
``Nonfactorizable'' corrections are, by definition, included in the
coefficients $a_i$. The matrix elements for $B$ mesons (i.e., mesons 
containing a $\bar b$-antiquark) are obtained from (\ref{Top}) by 
CP conjugation.  A corresponding result, with obvious substitutions, 
holds for other decays such as $B\to\pi\pi$. 

The term ${\cal T}_p^{\rm ann}$ in (\ref{Top}) arises from weak 
annihilation contributions and introduces a set of coefficients 
$b_i(\pi K)$, which we shall  define and discuss in detail in 
Section~\ref{subsec:annihilation}. Annihilation contributions are 
suppressed by a power of $\Lambda_{\rm QCD}/m_b$ and not calculable 
within the QCD factorization approach. Nevertheless, we will include 
the coefficients $b_i$ in the amplitude parameterizations in this 
section.

The coefficients $a_i$ multiplying products of vector or axial-vector 
currents are renor\-ma\-li\-zation-scheme invariant, as are the 
hadronic matrix elements of these currents. For the coefficients $a_6$ 
and $a_8$ a scheme dependence remains, which exactly compensates the 
scheme dependence of the hadronic matrix elements of the scalar or 
pseudoscalar densities associated with these coefficients. These matrix 
elements are power suppressed by the ratio
\begin{equation}\label{rKdef}
   r_\chi^K(\mu)
   = \frac{2 m_K^2}{\overline{m}_b(\mu)\,
      (\overline{m}_q(\mu)+\overline{m}_s(\mu))} \,,
\end{equation}
which is formally of order $\Lambda_{\rm QCD}/m_b$ but numerically
close to unity.  In the following we shall use the same notation for 
charged ($q=u$) and neutral kaons ($q=d$), since the difference is 
tiny. A corresponding ratio
\begin{equation}\label{rpidef}
   r_\chi^\pi(\mu)
   = \frac{2 m_\pi^2}{\overline{m}_b(\mu)\,
      (\overline{m}_u(\mu)+\overline{m}_d(\mu))}
\end{equation}
appears in the discussion of the $B\to\pi\pi$ decay amplitudes. For a 
phenomenological analysis of nonleptonic $B$ decays it is necessary to 
estimate these ``chirally-enhanced'' corrections despite the fact that 
they are formally power suppressed. A detailed discussion of these 
corrections will be presented in Section~\ref{subsec:power}, and their
numerical importance will be investigated in Section~\ref{sec:numerics}. 

In terms of the parameters $a_i$, the $B\to\pi K$ decay amplitudes 
(without annihilation contributions) are expressed as
\begin{eqnarray}
\label{BKpi}
   {\cal A}(B^-\to\pi^-\bar K^0)
   &=& \lambda_p \left[ \left( a_4^p - \frac12\,a_{10}^p \right) 
    + r_\chi^K \left( a_6^p - \frac12\,a_8^p \right) \right]
    A_{\pi K} \,, \nonumber\\
   - \sqrt2\,{\cal A}(B^-\to\pi^0 K^-)
   &=& \left[ \lambda_u\,a_1 + \lambda_p\,(a_4^p + a_{10}^p) 
    + \lambda_p\,r_\chi^K\,(a_6^p + a_8^p) \right]
    A_{\pi K} \nonumber\\
   &&\mbox{}+ \left[ \lambda_u\,a_2
    + \lambda_p \,\frac32\,(- a_7 + a_9) \right] A_{K\pi}
    \,, \nonumber\\
   - {\cal A}(\bar B^0\to\pi^+ K^-)
   &=& \left[ \lambda_u\,a_1 + \lambda_p\,(a_4^p + a_{10}^p) 
    + \lambda_p\,r_\chi^K\,(a_6^p + a_8^p) \right] A_{\pi K}
    \,, \nonumber\\ 
   \sqrt2\,{\cal A}(\bar{B}^0\to\pi^0 \bar{K}^0)
   &=& {\cal A}(B^-\to\pi^-\bar K^0)
    + \sqrt2\,{\cal A}(B^-\to\pi^0 K^-) \nonumber\\
   &&\mbox{}- {\cal A}(\bar B^0\to\pi^+ K^-) \,.
\end{eqnarray}
Here $\lambda_p=V_{pb} V_{ps}^*$, $a_i\equiv a_i(\pi K)$, and a 
summation over $p=u,c$ is implicitly understood in expressions like 
$\lambda_p\,a_i^p$. The last relation follows from isospin symmetry. 
The CP-conjugate decay amplitudes are obtained from the above by 
replacing $\lambda_p\to\lambda_p^*$. We have defined the factorized 
matrix elements
\begin{eqnarray}\label{As}
   A_{\pi K} &=& i\frac{G_F}{\sqrt2}\,(m_B^2-m_\pi^2)\,
    F_0^{B\to\pi}(m_K^2)\,f_K \,, \nonumber\\
   A_{K\pi} &=& i\frac{G_F}{\sqrt2}\,(m_B^2-m_K^2)\,
    F_0^{B\to K}(m_\pi^2)\,f_\pi \,,
\end{eqnarray}
where $F_0^{B\to M}(q^2)$ are semileptonic form factors. Weak 
annihilation effects contribute further terms to the decay amplitudes, 
which can be parameterized as
\begin{eqnarray}\label{BKPiann}
   {\cal A}_{\rm ann}(B^-\to\pi^-\bar K^0)
   &=& \Big[ \lambda_u\, b_2+(\lambda_u+\lambda_c)(b_3+ b_3^{\rm EW})
    \Big] B_{\pi K} \,, \nonumber\\
   - \sqrt2\,{\cal A}_{\rm ann}(B^-\to\pi^0 K^-)
   &=& {\cal A}_{\rm ann}(B^-\to\pi^-\bar K^0) \,, \nonumber\\
   - {\cal A}_{\rm ann}(\bar B^0\to\pi^+ K^-)
   &=& (\lambda_u+\lambda_c)\left(b_3-\frac{1}{2}\, b_3^{\rm EW}
       \right) B_{\pi K} \,, \nonumber\\ 
   \sqrt2\,{\cal A}_{\rm ann}(\bar{B}^0\to\pi^0 \bar{K}^0)
   &=& \mbox{}- {\cal A}_{\rm ann}(\bar B^0\to\pi^+ K^-) \,,
\end{eqnarray}
where 
\begin{equation}
   B_{\pi K} = i\frac{G_F}{\sqrt{2}}\,f_B f_\pi f_K \,.
\end{equation}
The coefficients $b_i\equiv b_i(\pi K)$ will be defined in 
Section~\ref{subsec:annihilation}, but we may note here that $b_{1,2}$ 
are related to the current--current operators $Q_1^p$ and $Q_2^p$ in 
the effective Hamiltonian (\ref{Heff}), and $b_{3,4}$ 
($b_{3,4}^{\rm EW}$) are related to QCD (electroweak) penguin operators.

The $B\to\pi\pi$ decay amplitudes are given by
\begin{eqnarray}\label{Bpipi}
   - {\cal A}(\bar B^0\to\pi^+\pi^-)
   &=& \left[ \lambda_u'\,a_1 + \lambda_p'\,(a_4^p + a_{10}^p) 
    + \lambda_p'\,r_\chi^\pi\,(a_6^p + a_8^p) \right] A_{\pi\pi}
    \,, \nonumber\\
   - \sqrt2\,{\cal A}(B^-\to\pi^-\pi^0)
   &=& \left[ \lambda_u' (a_1 + a_2) + \frac32\,\lambda_p'
    (-a_7 + r_\chi^\pi\,a_8^p + a_9 + a_{10}^p) \right] A_{\pi\pi}
    \,, \nonumber\\[0.15cm]
   {\cal A}(\bar{B}^0\to\pi^0\pi^0)
   &=& \sqrt2\,{\cal A}(B^-\to\pi^-\pi^0)
    - {\cal A}(\bar B^0\to\pi^+\pi^-) \,,
\end{eqnarray}
where now $\lambda_p'=V_{pb} V_{pd}^*$, $a_i\equiv a_i(\pi\pi)$, and
\begin{equation}
   A_{\pi\pi} = i\frac{G_F}{\sqrt2}\,(m_B^2-m_\pi^2)\,
   F_0^{B\to\pi}(m_\pi^2)\,f_\pi \,.
\end{equation}
The additional annihilation contributions are
\begin{eqnarray}\label{Bpipiann}
   - {\cal A}_{\rm ann}(\bar B^0\to\pi^+\pi^-)
   &=& \left[ \lambda_u'\,b_1 + (\lambda_u'+\lambda_c') \left(
    b_3 + 2 b_4 - \frac{1}{2}\,b_3^{\rm EW} + \frac{1}{2}\,b_4^{\rm EW}
    \right) \right] B_{\pi \pi} \,, \nonumber\\
   - \sqrt2\,{\cal A}_{\rm ann}(B^-\to\pi^-\pi^0) &=& 0  \,,
    \nonumber\\[0.2cm]
   {\cal A}_{\rm ann}(\bar{B}^0\to\pi^0\pi^0)
   &=& - {\cal A}_{\rm ann}(\bar B^0\to\pi^+\pi^-) \,,
\end{eqnarray}
where $b_i\equiv b_i(\pi\pi)$, and 
\begin{equation}
   B_{\pi\pi} = i\frac{G_F}{\sqrt{2}}\,f_B f_\pi^2 \,.
\end{equation}
Neglecting tiny mass corrections of order $(m_{\pi,K}/m_B)^2$,
\begin{equation}\label{ratios}
   R_{\pi K}\equiv \frac{A_{K\pi}}{A_{\pi K}}
   \simeq \frac{F_0^{B\to K}(0)\,f_\pi}{F_0^{B\to\pi}(0)\,f_K} \,,
   \qquad
   \frac{A_{\pi K}}{A_{\pi\pi}}\simeq \frac{f_K}{f_\pi} \,.
\end{equation} 
These ratios will play an important role in the discussion of SU(3)
violations in Section~\ref{sec:4}.

The expressions collected above provide a complete description of the 
decay amplitudes in terms of the parameters $a_i$ and $b_i$ for the 
various processes. They are the basis for most of the phenomenological
applications discussed in this work. However, in the literature several 
alternative parameterizations of the $B\to\pi K$ decay amplitudes have 
been introduced, which are sometimes useful when considering CP 
asymmetries or ratios of branching fractions. We briefly elaborate on 
one such parameterization here, adopting the notations of \cite{Mat98}. 
The dominant contributions to the $B\to\pi K$ decay amplitudes come 
from QCD penguin operators. Because the corresponding operators in the 
effective weak Hamiltonian preserve isospin, this contribution is the 
same (up to trivial Clebsch--Gordon coefficients) for all decay modes. 
Isospin-violating contributions to the decay amplitudes are subdominant 
and arise from the current--current operators $Q_1^u$ and $Q_2^u$ 
(so-called ``tree'' contributions), and from electroweak penguins. The 
latter are suppressed by a power of $\alpha/\alpha_s$, whereas the 
former are suppressed by the ratio
\begin{equation}
   \epsilon_{\rm KM}\,e^{-i\gamma}
   \equiv \frac{\lambda_u}{\lambda_c} 
   = \tan^2\!\theta_C\,R_b\,e^{-i\gamma} \,,
\end{equation}
where $\theta_C$ is the Cabibbo angle, 
\begin{equation}
   R_b = \cot\theta_C\,\frac{|V_{ub}|}{|V_{cb}|} 
   = \sqrt{\bar\rho^2 + \bar\eta^2}
\end{equation}
is one of the sides of the unitarity triangle, and $\bar\rho$ and 
$\bar\eta$ are the Wolfenstein parameters. A general parameterization 
of the decay amplitudes is 
\begin{eqnarray}\label{para}
   {\cal A}(B^-\to\pi^-\bar K^0) &=& P \left( 1 + \varepsilon_a\,
    e^{i\phi_a} e^{-i\gamma} \right) , \nonumber\\
   - \sqrt2\,{\cal A}(B^-\to\pi^0 K^-) &=& P\,\Big[ 1
    + \varepsilon_a\,e^{i\phi_a} e^{-i\gamma}
    - \varepsilon_{3/2}\,e^{i\phi} (e^{-i\gamma} - q\,e^{i\omega})
    \Big] \,, \nonumber\\
   - {\cal A}(\bar B^0\to\pi^+ K^-) &=& P\,\Big[ 1
    + \varepsilon_a\,e^{i\phi_a} e^{-i\gamma}
    - \varepsilon_T\,e^{i\phi_T} (e^{-i\gamma} - q_C\,e^{i\omega_C})
    \Big] \,.
\end{eqnarray}
(The phase $e^{i\phi_a}$ was denoted $-e^{i\eta}$ in \cite{Mat98}.) 
The amplitude ${\cal A}(\bar B^0\to\pi^0\bar K^0)$ is then determined 
by the isospin relation shown in the last line of (\ref{BKpi}). The 
dominant penguin amplitude $P$ is defined as the sum of all 
contributions to the $B^-\to\pi^-\bar K^0$ amplitude that are not
proportional to $e^{-i\gamma}$. This quantity cancels whenever one
takes ratios of decay amplitudes, such as CP asymmetries or ratios of
branching fractions. The parameters $\varepsilon_{3/2}$ and 
$\varepsilon_T$ measure the relative strength of tree and QCD penguin 
contributions, $q$ and $q_C$ measure the relative strength of 
electroweak penguin and tree contributions, and $\varepsilon_a$ 
parameterizes a rescattering contribution to the $B^-\to\pi^-\bar K^0$ 
amplitude arising from up-quark penguin topologies. Moreover, $\phi$, 
$\phi_T$, $\omega$, $\omega_C$, and $\phi_a$ are strong rescattering 
phases. (The strong-interaction phase of $P$ is not observable and can 
be set to zero.) All parameters except $q\,e^{i\omega}$ receive weak 
annihilation contributions. 

It is straightforward to express the various amplitude parameters in 
terms of the parameters $a_i\equiv a_i(\pi K)$ and 
$b_i\equiv b_i(\pi K)$ defined earlier. We obtain
\begin{equation}\label{Pampl}
   P = \lambda_c \left\{ \left[ \left( a_4^c - \frac12\,a_{10}^c \right)
   + r_\chi^K \left( a_6^c - \frac12\,a_8^c \right) \right] A_{\pi K}
   + \left( b_3 + b_3^{\rm EW} \right) B_{\pi K} \right\}
\end{equation}
for the leading penguin amplitude, and for the remaining parameters
\begin{eqnarray}\label{params}
   \varepsilon_{3/2}\,e^{i\phi} &=& - \epsilon_{\rm KM}\,
    \frac{(a_1+R_{\pi K} a_2) + \frac32[a_{10}^u+r_\chi^K a_8^u
          +R_{\pi K}(a_9-a_7)]}
         {(a_4^c+r_\chi^K a_6^c) - \frac{1}{2}(a_{10}^c+r_\chi^K a_8^c)
          + r_A (b_3+b_3^{\rm EW})} \,, \nonumber\\ 
   \varepsilon_T\,e^{i\phi_T} &=& - \epsilon_{\rm KM}\,
    \frac{a_1 + \frac32(a_{10}^u+r_\chi^K a_8^u)
          - r_A (b_2+\frac32\,b_3^{\rm EW})}
         {(a_4^c+r_\chi^K a_6^c) - \frac{1}{2}(a_{10}^c+r_\chi^K a_8^c)
          + r_A (b_3+b_3^{\rm EW})} \,, \nonumber\\
   \varepsilon_a\,e^{i\phi_a} &=& \epsilon_{\rm KM}\,
    \frac{(a_4^u+r_\chi^K a_6^u) - \frac{1}{2}(a_{10}^u+r_\chi^K a_8^u)
          + r_A (b_2+b_3+b_3^{\rm EW})}
         {(a_4^c+r_\chi^K a_6^c) - \frac{1}{2}(a_{10}^c+r_\chi^K a_8^c)
          + r_A (b_3+b_3^{\rm EW})} \,, \nonumber\\
   q\,e^{i\omega} &=& - \frac{3}{2\epsilon_{\rm KM}}\,
    \frac{a_{10}^c+r_\chi^K a_8^c + R_{\pi K}(a_9-a_7)}
         {(a_1+R_{\pi K} a_2) + \frac32[a_{10}^u+r_\chi^K a_8^u
          +R_{\pi K}(a_9-a_7)]} \,, \nonumber\\
   q_C\,e^{i\omega_C} &=& - \frac{3}{2\epsilon_{\rm KM}}\,
    \frac{a_{10}^c + r_\chi^K a_8^c -r_A b_3^{\rm EW}}
         {a_1 + \frac32(a_{10}^u+r_\chi^K a_8^u)
          - r_A (b_2+\frac32\,b_3^{\rm EW})} \,,
\end{eqnarray}
where $R_{\pi K}=A_{K\pi}/A_{\pi K}$ is the ratio of the two factorized 
amplitudes given in (\ref{ratios}), and 
\begin{equation}\label{rAdef}
   r_A = \frac{B_{\pi K}}{A_{\pi K}} 
   \simeq \frac{f_B f_\pi}{m_B^2\,F_0^{B\to\pi}(0)} \,.
\end{equation} 
Our notation for amplitude ratios is such that the ratio $R_{\pi K}$ 
(denoted by a capital $R$) deviates from 1 only by SU(3)-breaking 
corrections, whereas the ratios $r_A$ and $r_\chi^K$ (denoted by a 
lower-case $r$) are formally of order $\Lambda_{\rm QCD}/m_b$ in the 
heavy-quark limit. However, whereas $r_A\approx 0.003$ is indeed very 
small, $r_\chi^K\approx 0.7$ (at a scale $\mu\approx 1.45$\,GeV) is 
numerically large for real $B$ mesons. Finally, note that the 
electroweak penguin coefficients $a_{7,\dots, 10}$ could be safely 
neglected in all quantities other than $q\,e^{i\omega}$ and 
$q_C\,e^{i\omega_C}$, because they are tiny compared with the other 
coefficients $a_{1,\dots,6}$. In (\ref{params}), they are included only 
for completeness. (The systematics of including electroweak penguin 
contributions will be discussed in more detail later.) 

An important quantity affecting the determination of $\sin 2\alpha$ 
from the time-dependent CP asymmetry in the decays 
$B^0,\bar B^0\to\pi^+\pi^-$ is the ratio of penguin to tree amplitudes. 
In this case the tree contribution is no longer CKM suppressed, since
\begin{equation}
   \frac{\lambda_u'}{\lambda_c'} = - R_b\,e^{-i\gamma}
\end{equation}
is of order unity. The $\bar B^0\to\pi^+\pi^-$ decay amplitude in 
(\ref{Bpipi}) can be written as
\begin{equation}
   - {\cal A}(\bar B^0\to\pi^+\pi^-)
   \propto e^{-i\gamma} + \frac{P_{\pi\pi}}{T_{\pi\pi}} \,,
\end{equation}
where 
\begin{equation}\label{TPpipi}
   \frac{P_{\pi\pi}}{T_{\pi\pi}}
   = - \frac{1}{R_b}\,
   \frac{(a_4^c+r_\chi^\pi a_6^c) + (a_{10}^c+r_\chi^\pi a_8^c)
         + r_A [b_3+2b_4-\frac12(b_3^{\rm EW}-b_4^{\rm EW})]}
        {(a_1+a_4^u+r_\chi^\pi a_6^u) + (a_{10}^u+r_\chi^\pi a_8^u)
         + r_A [b_1+b_3+2b_4-\frac12(b_3^{\rm EW}-b_4^{\rm EW})]} \,,
\end{equation}
and $r_A=B_{\pi\pi}/A_{\pi\pi}$. In this case the electroweak penguin 
terms are very small, because they are not CKM enhanced with respect to 
the tree contribution. 

This concludes the discussion of parameterizations of the decay 
amplitudes. The following section is devoted to a detailed description 
of the QCD factorization formalism, the calculation of the parameters 
$a_i$ for the various nonleptonic decay amplitudes, and an estimation of 
the annihilation parameters $b_i$. The reader mainly interested in the 
phenomenological applications of our results can proceed directly to 
Section~\ref{sec:numerics}, where we present numerical results for these 
parameters, which will be used in the subsequent analysis in 
Section~\ref{sec:4}.

\section{\boldmath QCD factorization in $B\to\pi K$ decays\unboldmath}
\label{sec:3}

\begin{figure}[t]
\epsfxsize=11cm
\centerline{\epsffile{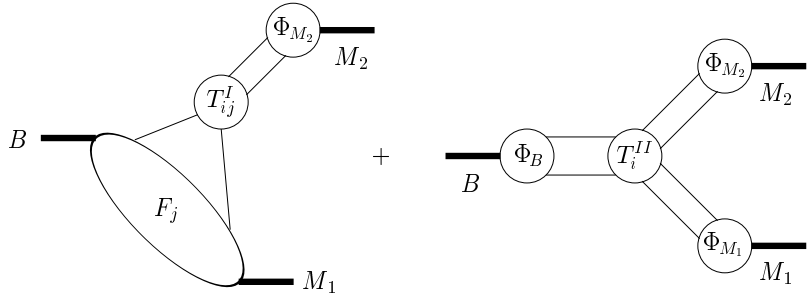}}
\centerline{\parbox{14cm}{\caption{\label{fig1}
Graphical representation of the factorization formula. Only one of the 
two form-factor terms in (\protect\ref{fact}) is shown for simplicity.}}}
\end{figure}

Based on the underlying physical principle of colour transparency (see 
\cite{Bj89,DG91,PW91} for early discussions in the context of decays 
to heavy--light final states), supported by a detailed diagrammatic 
analysis of infrared cancellations at leading power in the heavy-quark 
expansion, we have shown in previous work that the complexity of the 
hadronic matrix elements governing energetic, two-body hadronic decays 
of $B$ mesons simplifies greatly in the heavy-quark limit 
$m_b\gg\Lambda_{\rm QCD}$ \cite{BBNS1,BBNS2}. Consider $B\to\pi K$ 
decays as an example. To leading power in $\Lambda_{\rm QCD}/m_b$, but 
to all orders in perturbation theory, the matrix elements of the local 
operators $Q_i$ in the effective weak Hamiltonian in (\ref{Heff}) obey 
the factorization formula
\begin{eqnarray}\label{fact}
   \langle\pi K|Q_i|B\rangle
   &=& F_0^{B\to\pi}\,T_{K,i}^{\rm I}*f_K\Phi_K
    + F_0^{B\to K}\,T_{\pi,i}^{\rm I}*f_\pi\Phi_\pi
    \nonumber\\
   &&\mbox{}+ T_i^{\rm II}*f_B\Phi_B*f_K\Phi_K*f_\pi\Phi_\pi \,,
\end{eqnarray}
where $\Phi_M$ are leading-twist light-cone distribution amplitudes, 
and the $*$-products imply an integration over the light-cone momentum 
fractions of the constituent quarks inside the mesons. A graphical 
representation of this result is shown in Figure~\ref{fig1}. Because 
the energetic, collinear light-quark pair that ultimately evolves into 
the emission particle at the ``upper vertex'' is created by a 
point-like source, soft gluon exchange between this pair and the other 
quarks in the decay is power suppressed in the heavy-quark limit 
(colour transparency). In other words, whereas the hadronic physics 
governing the semileptonic $B\to M_1$ transition and the formation of 
the emission particle $M_2$ is genuinely nonperturbative, 
``nonfactorizable'' interactions connecting the two systems are 
dominated by hard gluon exchange.

\begin{figure}[t]
\epsfxsize=10cm
\centerline{\epsffile{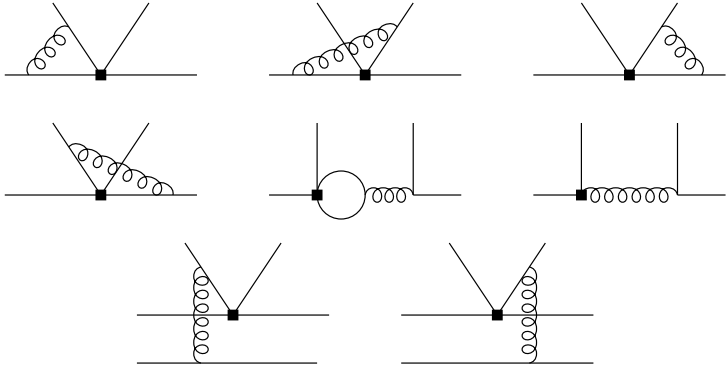}}
\centerline{\parbox{14cm}{\caption{\label{fig:graphs}
Order $\alpha_s$ corrections to the hard-scattering kernels 
$T_{M,i}^{\rm I}$ (first two rows) and $T_i^{\rm II}$ (last row). 
In the case of $T_{M,i}^{\rm I}$, the spectator quark does not 
participate in the hard interaction and is not drawn. The two lines 
directed upwards represent the quarks that make up one of the light 
mesons (the emission particle) in the final state.}}}
\end{figure}

The hard-scattering kernels $T_i^{\rm I,II}$ in (\ref{fact}) are
calculable in perturbation theory. $T_{M,i}^{\rm I}$ starts at tree 
level and, at higher order in $\alpha_s$, contains ``nonfactorizable'' 
corrections from hard gluon exchange or light-quark loops (penguin 
topologies). Hard, ``nonfactorizable'' interactions involving the 
spectator quark are part of $T_i^{\rm II}$. The relevant Feynman 
diagrams contributing to these kernels at next-to-leading are shown in 
Figure~\ref{fig:graphs}. Although individually these graphs contain 
infrared-sensitive regions at leading power, all soft and collinear 
divergences cancel in their sum, thus yielding a calculable 
short-distance contribution. Annihilation topologies are not included 
in (\ref{fact}) and Figure~\ref{fig:graphs}, because they do not 
contribute at leading order in $\Lambda_{\rm QCD}/m_b$. These 
power-suppressed contributions will be discussed separately in 
Section~\ref{subsec:annihilation}.

We stress that the factorization formula does not imply that hadronic 
$B$ decays are perturbative in nature. Dominant soft contributions to 
the decay amplitudes exist, which cannot be controlled in a 
hard-scattering approach. However, at leading power all these 
nonperturbative effects are contained in the semileptonic form factors 
and light-cone distribution amplitudes. Once these quantities are 
given, the nonleptonic decay amplitudes can be derived using 
perturbative approximations to the hard-scattering kernels. This allows 
us to compute perturbative corrections to ``naive factorization'' 
estimates of nonleptonic amplitudes, which is crucial for obtaining 
results that are independent of the renormalization scheme adopted in 
the calculation of the effective weak Hamiltonian. The hard-scattering 
kernels also contain imaginary parts, which determine the strong 
rescattering phases of the decay amplitudes. At leading power in 
$\Lambda_{\rm QCD}/m_b$ these imaginary parts are of perturbative 
origin.

In the remainder of this section, we discuss in detail the various 
issues to be addressed in the evaluation of the factorization formula. 
In Section~\ref{subsec:wilson}, a modified renormalization-group 
treatment of electroweak penguin effects is introduced, which is more 
appropriate than the standard scheme as far as applications to rare 
hadronic $B$ decays are concerned. Each of the diagrams in 
Figure~\ref{fig:graphs} contains a leading-power contribution relevant
to (\ref{fact}) and power-suppressed terms, which do not factorize in 
general. An important class of such power-suppressed effects is related 
to certain higher-twist meson distribution amplitudes. These amplitudes 
are defined in Section~\ref{subsec:power}, and their leading, 
chirally-enhanced contributions to the nonleptonic decay amplitudes
are evaluated. Section~\ref{subsec:airesults} contains a compendium of 
the relevant formulae for the calculation of the parameters $a_i$. 
Section~\ref{subsec:annihilation} is devoted to annihilation topologies 
and the definition of the parameters $b_i$.

\subsection{Wilson coefficients of electroweak penguin operators}
\label{subsec:wilson}

In the conventional treatment of the effective weak Hamiltonian
(\ref{Heff}), the initial conditions for the electroweak penguin 
coefficients at the scale $\mu=M_W$ are considered a next-to-leading 
order effect, because they are proportional to the electroweak gauge 
coupling $\alpha$ (see \cite{BJLW,CFMR} for a detailed discussion). For
our purposes, however, it is preferable to deviate from this standard 
power counting and introduce a modified approximation scheme. The 
reason is that the electroweak penguin contributions in $B\to\pi K$ 
decays and other rare processes based on $b\to s\bar q q$ transitions 
are important only because they compete with strongly CKM-suppressed 
tree topologies. Electroweak penguins and tree topologies together are 
responsible for the isospin-violating contributions to the decay 
amplitudes \cite{DeHe}. Therefore, their effects are important even 
though they are suppressed with respect to the leading QCD penguin 
amplitude, which conserves isospin. The ratio of electroweak penguin 
to tree amplitudes scales like $\alpha/\lambda^2\sim 1$, where 
$\lambda=0.22$ is the Wolfenstein parameter. Moreover, the dominant 
electroweak penguin effects are enhanced by a factor of $(m_t/M_W)^2$ 
and $1/\sin^2\!\theta_W$. Hence it is not appropriate to count $\alpha$ 
as a small parameter in the renormalization-group evolution, if the 
effect of interest is related to isospin breaking. 

We now describe a systematic modification of the usual leading and 
next-to-leading approximations, in which the dominant part of the 
electroweak penguin coefficients at the scale $\mu=M_W$ is treated as a 
leading-order effect. It is then consistent to include the QCD radiative 
corrections to the enhanced terms in the initial conditions for the 
electroweak penguin coefficients and, at the same time, the corrections 
of order $\alpha_s$ to the matrix elements of the electroweak penguin 
operators, which represent the next-to-leading order corrections to the 
hard-scattering kernels in the factorization formula. 

Using a compact matrix notation, the solution to the 
renormalization-group equation for the Wilson coefficients 
$C_1,\dots,C_{10}$ in (\ref{Heff}) can be written as
\begin{equation}\label{crge}
   \vec C(\mu) = \left[ \bm{U_0}
   + \frac{\alpha_s(\mu)}{4\pi}\,\bm{J\,U_0}
   - \frac{\alpha_s(M_W)}{4\pi}\,\bm{U_0\,J} 
   + \frac{\alpha}{4\pi} \left(
   \frac{4\pi}{\alpha_s(\mu)}\,\bm{R_0} + \bm{R_1} \right)
   \right] \vec C(M_W) \,.
\end{equation}
The matrices $\bm{U_0}$, $\bm{J}$, $\bm{R_0}$, and $\bm{R_1}$ depend on
the ratio $\alpha_s(\mu)/\alpha_s(M_W)$ and on the anomalous dimensions 
and $\beta$-function. At leading order, the evolution matrix reduces to 
$\bm{U_0}+(\alpha/\alpha_s) \bm{R_0}$. The remaining terms shown above 
are the next-to-leading corrections.

We now expand the coefficients $\vec C(M_W)$ at the weak scale as
\begin{equation}\label{cnlobar}
   \vec C(M_W)
   = \vec C_s^{(0)} + \frac{\alpha_s(M_W)}{4\pi}\,\vec C_s^{(1)}
   + \frac{\alpha}{4\pi} \left( \vec C_e^{(0)}
   + \frac{\alpha_s(M_W)}{4\pi}\,\vec C_e^{(1)}
   + \vec R_e^{(0)} \right) ,
\end{equation}
where superscripts indicate the order in the strong coupling constant 
$\alpha_s(M_W)$. The term proportional to $\alpha$ represents the 
electroweak contribution originating from photon-penguin, $Z$-penguin 
and box diagrams. We split this term into a contribution $\vec C_e$ 
containing all terms enhanced by the large top-quark mass and/or a 
factor of $1/\sin^2\!\theta_W$, and a remainder $\vec R_e$. As explained 
above, we treat $\vec C_e$ as a leading effect and hence include the 
first two terms in its expansion in powers of $\alpha_s(M_W)$. The 
remainder $\vec R_e$ is considered a next-to-leading effect, and so we 
only keep the first term in its perturbative expansion. Explicitly, the 
nonvanishing contributions to the initial conditions in the electroweak 
sector ($i=7,8,9,10$) are
\begin{equation}
   C_{e,7}^{(0)} = \frac{x_t}{3} \,, \qquad
   C_{e,9}^{(0)} = \frac{x_t}{3} + \frac{2}{3\sin^2\!\theta_W}
    \Big[ 10 B_0(x_t) - 4 C_0(x_t) \Big] \,,
\end{equation}
and
\begin{equation}
   R_{e,7}^{(0)} = R_{e,9}^{(0)} = \frac{8}{3}\,C_0(x_t)
   + \frac{2}{3}\,\tilde D_0(x_t) - \frac{x_t}{3} \,,
\end{equation}
where $x_t=m_t^2/M_W^2$ with $m_t=\bar m_t(m_t)$. The Inami--Lim 
functions $B_0(x)$, $C_0(x)$ and $\tilde D_0(x)$ can be found, e.g., in 
\cite{BBL}. Numerically, the remainder $\vec R_e^{(0)}$ is indeed much
smaller than $\vec C_e^{(0)}$, justifying our approximation scheme. 
Note that $\vec C_e^{(0)}$ is gauge and renormalization-scheme 
independent. The remainder $\vec R_e^{(0)}$ is gauge-independent, but 
it carries the usual next-to-leading order scheme dependence of the 
electroweak coefficients. Explicit expressions for the QCD corrections 
contributing to $\vec C_e^{(1)}$ have been obtained in \cite{BGH}. 
Using these results, we obtain the approximate expressions (valid for 
a high-energy matching scale $\mu_W=M_W$)
\begin{eqnarray}
   C^{(1)}_{e,7} &\simeq& -29.60\,x_t^{1.142} + 28.52\,x_t^{1.148}
    \,, \nonumber\\
   C^{(1)}_{e,8} &\simeq& 0.94\,x_t^{0.661}
    \,, \nonumber\\
   C^{(1)}_{e,9} &\simeq& -571.62\,x_t^{0.580} + 566.40\,x_t^{0.590}
    \,, \nonumber\\
   C^{(1)}_{e,10} &\simeq& -5.51\,x_t^{1.107} \,.
\end{eqnarray}
In the conventional treatment $\vec C_e^{(1)}$ would be absent, while 
the sum $(\vec C_e^{(0)}+\vec R_e^{(0)})$ would be the usual initial 
condition for the electroweak coefficients at the scale $\mu=M_W$, 
counted as a next-to-leading order effect. 

In addition to the modified counting scheme for powers of coupling 
constants, we make a further, numerically excellent approximation, 
which greatly simplifies the systematic evaluation of the Wilson 
coefficients. In essence, it amounts to neglecting QED effects in the
calculation of the Wilson coefficients $C_1,\dots,C_6$. The values of
these coefficients are obtained using the standard next-to-leading
order approximation including only strong-interaction effects. At the
same time, we neglect QED corrections to the matrix elements of the
operators $Q_1,\dots,Q_6$. (In fact, the virtual corrections of order 
$\alpha$ are infrared divergent and require the inclusion of photon 
bremsstrahlung contributions in order to obtain physical results. Our 
approximation scheme avoids this complication.) This treatment can be
justified by noting that QED and electroweak contributions to the decay 
amplitudes in (\ref{para}) are only important if they contribute to the
parameters $q\,e^{i\omega}$ and $q_C\,e^{i\omega_C}$. From 
(\ref{params}), it follows that the terms of order $\alpha$ contained 
in the coefficients $a_{7,\dots,10}$ are enhanced by the prefactor 
$1/\epsilon_{\rm KM}$. It is thus sufficient for all practical purposes 
to only include the order $\alpha$ corrections from the coefficients 
$a_{7,\dots,10}$. On the contrary, QED corrections to the other 
amplitude parameters can be safely neglected (i.e., the coefficients 
$a_{1,\ldots,6}$ do not contain terms proportional to $\alpha$). 
Precisely this is achieved by our approximation scheme. 

\begin{table}[t]
\centerline{\parbox{14cm}{\caption{\label{tab:wilco}
Wilson coefficients $C_i$ in the NDR scheme and based on our modified 
approximation scheme (see text). Input parameters are 
$\Lambda^{(5)}_{\overline{\rm MS}}=0.225$\,GeV, $m_t(m_t)=167$\,GeV, 
$m_b(m_b)=4.2$\,GeV, $M_W=80.4$\,GeV, $\alpha=1/129$, and 
$\sin^2\!\theta_W=0.23$.}}}
\begin{center}
\begin{tabular}{|l|c|c|c|c|c|c|}
\hline\hline
NLO & $C_1$ & $C_2$ & $C_3$ & $C_4$ & $C_5$ & $C_6$ \\
\hline
$\mu=m_b/2$ & 1.137 & $-0.295$ & 0.021 & $-0.051$ & 0.010 & $-0.065$ \\
$\mu=m_b$   & 1.081 & $-0.190$ & 0.014 & $-0.036$ & 0.009 & $-0.042$ \\
$\mu=2 m_b$ & 1.045 & $-0.113$ & 0.009 & $-0.025$ & 0.007 & $-0.027$ \\
\hline
 & $C_7/\alpha$ & $C_8/\alpha$ & $C_9/\alpha$ & $C_{10}/\alpha$
 & $C_{7\gamma}^{\rm eff}$ & $C_{8g}^{\rm eff}$ \\
\hline
$\mu=m_b/2$ & $-0.024$           & 0.096 & $-1.325$ & 0.331
 & --- & --- \\
$\mu=m_b$   & $-0.011$           & 0.060 & $-1.254$ & 0.223
 & --- & --- \\
$\mu=2 m_b$ & $\phantom{-}0.011$ & 0.039 & $-1.195$ & 0.144
 & --- & --- \\
\hline\hline
LO & $C_1$ & $C_2$ & $C_3$ & $C_4$ & $C_5$ & $C_6$ \\
\hline
$\mu=m_b/2$ & 1.185 & $-0.387$ & 0.018 & $-0.038$ & 0.010 & $-0.053$ \\
$\mu=m_b$   & 1.117 & $-0.268$ & 0.012 & $-0.027$ & 0.008 & $-0.034$ \\
$\mu=2 m_b$ & 1.074 & $-0.181$ & 0.008 & $-0.019$ & 0.006 & $-0.022$ \\
\hline
 & $C_7/\alpha$ & $C_8/\alpha$ & $C_9/\alpha$ & $C_{10}/\alpha$
 & $C_{7\gamma}^{\rm eff}$ & $C_{8g}^{\rm eff}$ \\
\hline
$\mu=m_b/2$ & $-0.012$           & 0.045 & $-1.358$ & 0.418
 & $-0.364$ & $-0.169$ \\
$\mu=m_b$   & $-0.001$           & 0.029 & $-1.276$ & 0.288
 & $-0.318$ & $-0.151$ \\
$\mu=2 m_b$ & $\phantom{-}0.018$ & 0.019 & $-1.212$ & 0.193
 & $-0.281$ & $-0.136$ \\
\hline\hline
\end{tabular}
\end{center}
\end{table}

At the technical level, the approximation described above can be 
explained in terms of the $10\times 10$ anomalous-dimension matrix for 
the operators $Q_1,\dots,Q_{10}$, written in block form as
\begin{equation}
   \bm{\gamma} = \left(
   \begin{array}{c|c}
   \bm{A}_{6\times 6}~ & ~\bm{B}_{6\times 4} \\ \hline
   \bm{C}_{4\times 6}~ & ~\bm{D}_{4\times 4}
   \end{array} \right) .
\end{equation}
We set $\bm{C}=\bm{0}$, thereby neglecting the mixing of the electroweak 
penguin operators into the operators $Q_1,\dots,Q_6$, and ignore 
contributions of order $\alpha$ to $\bm{A}$. At the same time, we drop 
the terms of order $\alpha$ in the matching conditions (\ref{cnlobar}) 
for $C_1,\dots,C_6$. We also omit terms of order $\alpha$ in $\bm{D}$, 
which would yield second-order corrections in $\alpha$. For the matrix 
$\bm{B}$, we use the complete next-to-leading order result including 
terms of order $\alpha$.

Numerical results for the Wilson coefficients obtained at leading and
next-to-leading order in our modified approximation scheme are given in 
Table~\ref{tab:wilco}. Throughout this work, we use the ``naive 
dimensional regularization'' (NDR) scheme with anticommuting $\gamma_5$, 
as defined in \cite{BJLW}. The matrix elements of the dipole operators 
$Q_{7\gamma}$ and $Q_{8g}$ enter the decay amplitudes only at 
next-to-leading order. Consequently, the standard leading-logarithmic 
approximation is sufficient for the coefficients $C_{7\gamma}$ and 
$C_{8g}$. In practice, it is advantageous to work with so-called
``effective'' coefficients, which in the NDR scheme are defined as 
$C_{7\gamma}^{\rm eff}=C_{7\gamma}-\frac13\,C_5-C_6$ and 
$C_{8g}^{\rm eff}=C_{8g}+C_5$. In the numerical analysis of (\ref{crge}) 
we consistently drop all terms of higher than next-to-leading order 
according to our modified counting scheme. Throughout we use the 
two-loop expression for the running coupling $\alpha_s(\mu)$ evaluated 
with $n_f=5$ light quark flavours.

\subsection{Meson distribution amplitudes and twist-3 projections}
\label{subsec:power}

Referring to the factorization formula shown graphically in 
Figure~\ref{fig1}, we denote by $x$ the longitudinal momentum fraction 
of the constituent quark in the emission meson $M_2$  
(the meson at the ``upper vertex''), and by $y$ the momentum fraction 
of the quark in the meson $M_1$. For a $\bar B$ meson 
decaying into two light mesons, we define light-cone distribution 
amplitudes by choosing the $+$ direction along the decay path of the 
light emission particle and denote by $\xi$ the light-cone momentum 
fraction of the light spectator antiquark. A massive pseudoscalar meson
has two leading-twist light-cone distribution amplitudes 
\cite{BBNS2,GrNe}, but only one of them enters our results. This 
amplitude is called $\Phi_B(\xi)$ and coincides with the function 
$\Phi_{B1}(\xi)$ defined in Section~2.3.3 of \cite{BBNS2}. The meson 
distribution amplitudes are normalized to 1 once the decay constants 
are factored out as in (\ref{fact}). For a light meson, we define the 
leading-twist amplitude $\Phi(x)$ in the usual way and assume that 
$\Phi(x)=O(1)$ if both $x$ and $(1-x)$ are of order unity, and 
$\Phi(x)=O(x)$ for $x\to 0$ (and similarly for $x\to 1$). For the $B$ 
meson, almost all momentum is carried by the heavy quark, and hence 
$\Phi_B(\xi)=O(m_b/\Lambda_{\rm QCD})$ and 
$\xi=O(\Lambda_{\rm QCD}/m_b)$.

Higher-twist light-cone distribution amplitudes for the light mesons 
give power-sup\-pressed contributions in the heavy-quark limit. 
However, as has been explained in Section~\ref{sec:2}, these can 
sometimes be large if they appear in conjunction with the chiral 
enhancement factors $r_\chi^M(\mu)$ defined in (\ref{rKdef}) and 
(\ref{rpidef}). The corresponding terms are associated with twist-3 
quark--antiquark distribution amplitudes and can be identified 
completely. Since the calculation of these contributions in momentum 
space is less straightforward than for the leading-twist contributions, 
we summarize the relevant projection operators below, following the 
discussion in \cite{BF00}.

The relevant definitions of the light-cone distribution amplitudes of 
a light pseudoscalar meson $P$ in terms of bilocal operator matrix 
elements are \cite{BraF}
\begin{eqnarray}\label{pidadef}
  \langle P(p)|\bar q(z_2)\gamma_\mu\gamma_5 q(z_1)|0\rangle
  &=& - i f_P\,p_\mu \int_0^1\! dx\,
   e^{i(x\,p\cdot z_2+\bar x\,p\cdot z_1)}\,\Phi(x) \,, \nonumber\\
  \langle P(p)|\bar q(z_2) i\gamma_5 q(z_1)|0\rangle
  &=& f_P\mu_P \int_0^1\! dx\, 
   e^{i(x\,p\cdot z_2+\bar x\,p\cdot z_1)}\,\Phi_p(x) \,, \nonumber\\
  \langle P(p)|\bar q(z_2)\sigma_{\mu\nu}\gamma_5 q(z_1)|0\rangle
  &=& i f_P\mu_P\,(p_\mu z_\nu - p_\nu z_\mu)
   \int_0^1\! dx\,e^{i(x\,p\cdot z_2+\bar x\,p\cdot z_1)}\,
   \frac{\Phi_\sigma(x)}{6} \,,
\end{eqnarray}
where $f_P$ is the decay constant, and we have defined $z=z_2-z_1$ and 
$\bar x=1-x$. The parameter $\mu_P=m_P^2/(m_1+m_2)$, where $m_{1,2}$ 
are the current quark masses of the meson constituents, is proportional 
to the chiral quark condensate. (This definition does not hold for the 
$\pi^0$ meson, in which case $\mu_{\pi^0}=m_\pi^2/(m_u+m_d)$ as for the 
charged pions \cite{BBNS2}.) $\Phi(x)$ is the leading-twist (twist-2)
distribution amplitude, whereas $\Phi_p(x)$ and $\Phi_\sigma(x)$ have 
subleading twist (twist-3). All three distribution amplitudes are 
normalized to 1, as follows by taking the limit $z_1\to z_2$. The above 
definitions can be combined into the matrix 
\begin{eqnarray}\label{pispacedef}
   &&\langle P(p)|\bar q_\beta(z_2)\,q_\alpha(z_1)|0\rangle
    \nonumber\\
   &&= \frac{i f_P}{4} \int_0^1\! dx\,
    e^{i(x\,p\cdot z_2+\bar x\,p\cdot z_1)}
    \left\{ \pslash\,\gamma_5\,\Phi(x)
    - \mu_P\gamma_5 \left( \Phi_p(x) - \sigma_{\mu\nu}\,p^\mu z^\nu\,
    \frac{\Phi_\sigma(x)}{6} \right) \right\}_{\alpha\beta} . \quad
\end{eqnarray}
We implicitly assume that the bilocal matrix elements are supplied with 
the appropriate path-ordered exponentials of gluon fields so as to 
make the definitions of the light-cone distribution amplitudes gauge 
invariant. (These exponentials are absent in light-cone gauge.) The 
distribution amplitudes depend on the renormalization scale $\mu$. This 
scale dependence is compensated by higher-order corrections to the 
hard-scattering kernels, which however are beyond the accuracy of the 
present calculation. We thus suppress the argument $\mu$ in the 
distribution amplitudes, because it is irrelevant to our discussion.

To obtain the corresponding projector of the quark--antiquark amplitude 
in momentum space, the transverse components of the coordinate $z$ must
be taken into account. The collinear approximation can be taken only 
after the projection has been applied. We therefore assign momenta 
\begin{equation}\label{momenta2}
   k_1^\mu = x p^\mu + k_\perp^\mu
    + \frac{\vec k_\perp^2}{2x\,p\cdot\bar p}\,\bar p^\mu \,, \qquad
   k_2^\mu = \bar x p^\mu - k_\perp^\mu
    + \frac{\vec k_\perp^2}{2\bar x\,p\cdot\bar p}\,\bar p^\mu
\end{equation}
to the quark and antiquark in the light meson, where $\bar p$ is a 
light-like vector whose 3-components point into the opposite direction
of $\vec{p}$. Then the exponential in 
(\ref{pispacedef}) becomes $e^{i(k_1\cdot z_2 + k_2\cdot z_1)}$. (Meson 
mass effects are neglected, so that $p$ and $\bar p$ can be considered 
as light-like.) The transverse components $k_\perp^\mu$ are defined with 
respect to the vectors $p$ and $\bar p$. Note that $k_1^2=k_2^2=0$. In 
general, the projector (\ref{pispacedef}) is part of a diagram expressed 
in configuration space. Transformation to momentum space is achieved by 
performing the integrations over $z_i$, which reduce to momentum-space 
$\delta$-functions after substituting
\begin{equation}\label{derivative}
   z^\nu \to (-i)\,\frac{\partial}{\partial k_{1\nu}}  
   = (-i) \left( \frac{\bar p^\nu}{p\cdot\bar p}\,
   \frac{\partial}{\partial x}
   + \frac{\partial}{\partial k_{\perp\nu}} + \dots \right) . 
\end{equation}
The ellipses denote a term proportional to $p^\nu$, which does not 
contribute to the result. We also omit terms of order $\vec k_\perp^2$, 
which cannot contribute in the limit $k_\perp\to 0$. As written above, 
the derivative acts on the hard-scattering amplitude in the 
momentum-space representation. Using an integration by parts, the 
derivative with respect to $x$ can be made to act on the light-cone 
distribution amplitude. The second term, which involves the derivative 
with respect to the transverse momentum, must be evaluated before the 
collinear limit $k_1\to x p$, $k_2\to\bar x p$ is taken. The light-cone 
projection operator of a light pseudoscalar meson in momentum space, 
including twist-3 two-particle contributions, then reads
\begin{equation}\label{pimeson2}
   M_{\alpha\beta}^P = \frac{i f_P}{4} \Bigg\{
   \pslash\,\gamma_5\,\Phi(x) - \mu_P\gamma_5 \left(
   \Phi_p(x) - i\sigma_{\mu\nu}\,\frac{p^\mu\bar p^\nu}{p\cdot\bar p}\,
   \frac{\Phi'_\sigma(x)}{6}
   + i\sigma_{\mu\nu}\,p^\mu\,\frac{\Phi_\sigma(x)}{6}\,
   \frac{\partial}{\partial k_{\perp\nu}} \right)
   \Bigg\}_{\alpha\beta}.
\end{equation}
It is understood that, after the derivative is taken, the momenta $k_1$ 
and $k_2$ are set equal to $x p$ and $\bar x p$, respectively. A 
complete description of the pseudoscalar meson at the twist-3 level 
would also include three-particle quark--antiquark--gluon contributions 
(see \cite{BraF} for a detailed discussion), which do not involve the 
large normalization factor $\mu_P$ and thus are omitted here. (Note
that the overall sign of (\ref{pimeson2}) as well as of
(\ref{geshpro}) below depends on the convention of ordering the quark fields
in (\ref{pispacedef}).)

The asymptotic limit of the leading-twist distribution amplitude, valid
for $\mu\to\infty$, is $\Phi(x)=6x(1-x)$. For finite value of the 
renormalization scale, it is convenient and conventional to employ an 
expansion in Gegenbauer polynomials of the form
\begin{equation}\label{gegenbauer}
   \Phi_M(x,\mu) = 6x(1-x) \left[ 1 + \sum_{n=1}^\infty 
   \alpha_n^M(\mu)\,C_n^{(3/2)}(2x-1) \right] .
\end{equation}
In numerical evaluations it will be sufficient to truncate this 
expansion at $n=2$, using $C_1^{(3/2)}(u)=3u$ and 
$C_2^{(3/2)}(u)=\frac32(5u^2-1)$. The Gegenbauer moments 
$\alpha_n^M(\mu)$ are multiplicatively renormalized. The scale 
dependence of these coefficients enters our results only 
at order $\alpha_s^2$, which is beyond the accuracy of a next-to-leading 
order calculation. 

The twist-3 two-particle distribution amplitudes are determined by the 
three-particle distributions via the equations of motion, except for a 
single term \cite{BraF}. In the approximation adopted here, where only 
terms proportional to $\mu_P$ are kept and all three-particle 
distributions are neglected, the twist-3 amplitudes must obey the 
equations of motion
\begin{equation}\label{eompi}
   \frac{x}{2} \left( \Phi_p(x) + \frac{\Phi'_\sigma(x)}{6} \right)
   = \frac{\Phi_\sigma(x)}{6} \,, \qquad
   \frac{1-x}{2} \left( \Phi_p(x) - \frac{\Phi'_\sigma(x)}{6} \right)
   = \frac{\Phi_\sigma(x)}{6} \,. 
\end{equation}
These equations enforce that we must use the asymptotic forms 
$\Phi_p(x)=1$ and $\Phi_\sigma(x)=6x(1-x)$. It will be important below 
that $\Phi_p(x)$ and $\Phi'_\sigma(x)$ do not vanish at the endpoints 
$x=0$ or 1. We finally observe that the $k_\perp$-derivative in 
(\ref{pimeson2}) can be substituted by
\begin{equation}
   \frac{\partial}{\partial k_{\perp\nu}}
   \to \frac{2 k_\perp^\nu}{k_\perp^2} \,.
\end{equation}
This is because we may expand the amplitude to first order in $k_\perp$ 
(higher powers do not contribute in the $k_\perp\to 0$ limit), and use
\begin{equation}
   \frac{\partial}{\partial k_{\perp\nu}} k_\perp^\lambda
   = \frac{\langle 2 k_\perp^\nu k_\perp^\lambda\rangle}{k_\perp^2}
   = g_\perp^{\nu\lambda} \,,
\end{equation}
which holds after averaging $k_\perp$ over the azimuthal angle
(denoted by $\langle\dots\rangle$). We use the definition 
$g_\perp^{\nu\lambda}={\rm diag}(0,-1,-1,0)$. The twist-3 terms in 
(\ref{pimeson2}) can now be combined into the projector \cite{Gesh}
\begin{equation}
\label{geshpro}
   - \frac{i f_P\mu_P}{4}\,\gamma_5\, 
   \frac{\kslash_2\,\kslash_1}{k_2\cdot k_1}\,\Phi_p(x) \,, 
\end{equation}
where $k_{1,2}$ are the quark and antiquark momenta defined in 
(\ref{momenta2}), and the factor of $\Phi_p(x)=1$ is simply there to 
remind us that this is a twist-3 projection. In our analysis below, we 
will quote the results of the twist-3 projections in this form, i.e., 
after eliminating $\Phi_\sigma^{(\prime)}(x)$ using the equations of 
motion. Expressions in terms of two functions $\Phi_p(x)$ and 
$\Phi_\sigma(x)$ are ambiguous, but reduce to the same expression upon 
substituting the asymptotic forms of the distribution amplitudes. 

\subsection{Comments on the calculation}
\label{subsec:comments}

We now describe some technical aspects of the calculation of the 
various diagrams in more detail. The complete results for the parameters 
$a_i$ will be given in Section~\ref{subsec:airesults}.

\subsubsection*{Vertex corrections}

The calculation of the four one-loop vertex diagrams (first four 
diagrams in Figure~\ref{fig:graphs}) involves the trace
\begin{equation}\label{traceM2}
   \mbox{tr} \left( M^{M_2} \!\left[
   \frac{(2k_1^\rho+\gamma^\rho\lslash)\Gamma}{(xq+\ell)^2}
   - \frac{\Gamma(2k_2^\rho+\lslash\gamma^\rho)}{(\bar x q+\ell)^2}
   \right] \right) ,
\end{equation}
where $M^{M_2}$ is the projector from (\ref{pimeson2}), $q$ the momentum 
of the meson $M_2$, $k_1$ ($k_2$) the momentum of the quark (antiquark) 
in this meson, and $\ell$ the momentum of the gluon. We have used that 
$M^{M_2}\kslash_1=\kslash_2 M^{M_2}=0$ by the equations of motion, and 
that we can put $k_1=x q$ and $k_2=\bar x q$ in the denominator. A
Fierz transformation may be necessary in order to arrive at the trace 
(\ref{traceM2}). The possible $\Gamma$ structures are therefore $V-A$ 
(contributing to $a_{1,\dots,4,9,10}$), $V+A$ (contributing to 
$a_{5,7}$), and $S+P$ (contributing to $a_{6,8}$).

If $\Gamma=V\pm A$, only the leading-twist light-cone distribution 
amplitude $\Phi(x)$ in (\ref{pimeson2}) contributes under the trace. We 
then recover the results of \cite{BBNS1,BBNS2} and find that there 
exist no chirally-enhanced power corrections to $a_{1,\dots,5,7,9,10}$ 
from the vertex diagrams. If $\Gamma=S+P$, the leading-twist 
contribution to the trace vanishes, while the twist-3 result is 
proportional to $r_\chi^{M_2}=2\mu_{M_2}/m_b$. This gives an order 
$\alpha_s$ correction to the coefficients $a_{6,8}$, which are 
multiplied by $r_\chi^{M_2}$ already in naive factorization. We can now 
exploit the fact that $\mbox{tr}\,\sigma_{\mu\nu}=0$ to show that the 
term involving the transverse-momentum derivative in (\ref{pimeson2}) 
does not contribute to the trace. At this stage, the collinear limit can
be taken to compute the kernel in the usual way. The projection on 
$\Phi_p(x)$ yields a result containing symmetric and antisymmetric parts 
under the exchange of $x$ and $(1-x)$. As explained above, in the 
approximation of keeping only chirally-enhanced terms we are forced to 
assume the asymptotic form for $\Phi_p(x)$, so that the antisymmetric 
part of the kernel integrates to zero. The symmetric part turns out to 
be a scheme-dependent constant and is responsible for the ``$-6$'' in 
the expressions for $a_{6,8}$ in (\ref{ai}) below. The kernel resulting 
from the $\Phi'_\sigma(x)$ projection is symmetric under
$x\leftrightarrow(1-x)$ and thus vanishes after integration with 
$\Phi'_\sigma(x)$.

\subsubsection*{Penguin diagrams}

We now consider the penguin contractions (fifth diagram in 
Figure~\ref{fig:graphs}), restricting our attention first to the twist-2 
part of the projector (\ref{pimeson2}). The corresponding contributions 
to the hard-scattering kernels have been given in our previous work 
\cite{BBNS1}. Because there have been conflicting results for the 
penguin terms in the recent literature (see Section~\ref{sec:comp}), we 
wish to clarify the origin of the discrepancies here.

\begin{figure}[t]
\epsfxsize=8cm
\centerline{\epsffile{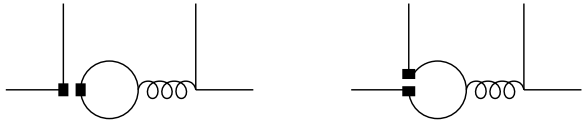}}
\centerline{\parbox{14cm}{\caption{\label{fig:penguin} 
The two different penguin contractions.}}}
\end{figure}

The point to note is that, depending on the structure of the 
$(V-A)\otimes(V\pm A)$ four fermion operators in the effective weak 
Hamiltonian (without Fierz transformation!), there exist two distinct 
penguin contractions with different contractions of spinor indices, as 
indicated in Figure~\ref{fig:penguin}. Although the two diagrams are 
related by Fierz transformations in four dimensions, in dimensional
regularization they give different results in the NDR scheme with 
anticommuting $\gamma_5$, because this scheme does not preserve the 
Fierz identities in $d$ dimensions \cite{BJLW}. The results for the two 
contractions shown in the figure involve 
\begin{equation}
   \mbox{left:} \quad
   \frac{2}{3}\,\ln\frac{m_b^2}{\mu^2} - G(s) \,, \qquad
   \mbox{right: } \quad
   \frac{2}{3} \left( \ln\frac{m_b^2}{\mu^2} + 1 \right) - G(s) \,,
\end{equation}
where the function $G(s)$ is given in (\ref{GK}) below. The first
contraction appears in matrix elements of the operators $Q_{4,6}$, while
the second one enters in matrix elements of $Q_{1,3}$. This assumes the 
standard Fierz-form of the effective Hamiltonian, which is employed in 
the calculation of the Wilson coefficients in the NDR scheme. The 
operator $Q_5$ is special, because its contribution is a pure 
ultraviolet effect which, by definition, is absorbed into the definition 
of the ``effective'' Wilson coefficient $C_{8g}^{\rm eff}=C_{8g}+C_5$ of 
the chromomagnetic dipole operator \cite{BBL}. The discrepancies between 
our results and some of the papers discussed in Section~\ref{sec:comp} 
seem to arise from the fact that $Q_4$ and $Q_6$ are treated like 
$Q_{1,3}$. 

Including now the twist-3 part of the projector (\ref{pimeson2}), we
still find that the penguin contractions can be straightforwardly
evaluated in terms of four-quark operators, before the actual projection 
is made. The on-shell conditions for the external quarks connected to 
the gluon need to be used in this step. The projection is then very 
simple and only the $\gamma_5$ projector multiplied by $\Phi_p(x)$ 
contributes. The resulting kernel is identical to that obtained with the 
twist-2 projection. 

The calculation of the matrix element of the chromomagnetic dipole 
operator (sixth diagram in Figure~\ref{fig:graphs}) is more interesting 
in this respect. In this case it is no longer possible to reduce the 
amplitude to the usual structures involving four-quark operators. This 
is related to the fact that at the twist-3 level the $k_\perp$ momenta 
cannot immediately be put to zero. The complete twist-3 projection 
(\ref{pimeson2}) has to be performed to evaluate the diagram, which 
leads to the expression
\begin{equation}
   \frac{1}{(1-x)m_b^2}\,\bar u_q(p_q)\gamma^\mu M^{M_2} 
   (\Pslash\gamma_\mu-\gamma_\mu\Pslash)(1+\gamma_5)u_b(p_b) \,,
\end{equation}
where $P=p_q+k_2=p_b-k_1$ denotes the momentum of the gluon, and the 
prefactor comes from the gluon propagator. $M^{M_2}$ projects on the 
emission particle $M_2$ with momentum $q$, and $p_b$, $p_q$ denote the 
momenta of the $b$ quark and the quark in $M_1$, respectively. We find 
that all four terms of the projector contribute when evaluated on this 
expression, giving the result
\begin{eqnarray}\label{chromo3}
  &&i f_P\,\bar u_q(p_q) \left[
   (1+\gamma_5)\,\frac{\Phi(x)}{\bar x}
   + \frac{\mu_P}{m_b}\,(1-\gamma_5)\!\left( \frac{3}{2}\,\Phi_p(x)
   + \frac{1}{2}\,\frac{\Phi'_\sigma(x)}{6}
   + \frac{1}{\bar x}\,\frac{\Phi_\sigma(x)}{6}\right)\right] u_b(p_b)
   \nonumber\\
  &&\hspace{2cm}= i f_P\,\bar u_q(p_q) \left[ 
   (1+\gamma_5)\,\frac{\Phi(x)}{\bar x}
   + \frac{2\mu_P}{m_b}\,(1-\gamma_5)\,\Phi_p(x) \right] u_b(p_b) \,.
\end{eqnarray}
The second line is obtained after using the equations of motion 
(\ref{eompi}) for the twist-3 distribution amplitudes, so that the 
asymptotic form $\Phi_p(x)=1$ is understood. We observe that the factor 
$1/\bar x$ from the gluon propagator is cancelled in the twist-3 term, 
and so the convolution integral has no endpoint divergence.

One can see from (\ref{chromo3}) that the matrix element of the 
chromomagnetic operator is obtained incorrectly at the twist-3 level 
if the incomplete projector containing only $\Phi_p(x)$ is used. Note 
that, in general, this leads to gauge-dependent results, since the 
equations of motion are not respected. 

\subsubsection*{Hard spectator interaction}

The calculation of the two diagrams in the third row of 
Figure~\ref{fig:graphs} leads to the same trace as in (\ref{traceM2}). 
It therefore follows that, for $\Gamma=S+P$, the projection on $M_2$ 
can result at most in a constant multiplying the distribution amplitude 
$\Phi_p(x)$. However, the constant obtained for the vertex diagrams 
resulted entirely from a term of order $\epsilon=(2-d/2)$ in the trace 
multiplying the ultraviolet-divergent loop diagram. Since the hard 
spectator contributions result from tree diagrams, the trace can be 
evaluated in four dimensions, and this constant is absent. We thus 
conclude that there is no hard spectator correction to the parameters 
$a_{6,8}$ at order $\alpha_s$. 

If $\Gamma=V\pm A$, only the twist-2 distribution amplitude contributes 
for the emission particle $M_2$, but all four terms in the projector 
for $M_1$ contribute to the result. We also find that both terms in the 
$B$-meson projection (as given in \cite{BBNS2}) contribute, but one 
of the two $B$-meson light-cone distribution amplitudes drops out after 
implementing the equations of motion (\ref{eompi}). Contrary to the 
other corrections, the kernels $T_i^{\rm II}$ resulting from hard 
spectator interactions have logarithmic endpoint singularities at 
twist-3 level. They arise from integrals of the form 
$\int_0^1\!dy/\bar y$, where $\bar y$ is the momentum of the antiquark 
in the meson that picks up the spectator antiquark from the $B$ meson. 
These endpoint singularities prevent a reliable perturbative calculation 
of the chirally-enhanced power corrections to the hard spectator 
interactions. (The endpoint singularities are missed if the incomplete 
light-meson projector with only the $\gamma_5$ projection at twist-3 is 
employed.)

\subsection{\boldmath Results for the parameters $a_i$\unboldmath}
\label{subsec:airesults}

After these preliminaries, we now present the results for the 
coefficients $a_i$ obtained at next-to-leading order in $\alpha_s$, and 
including the complete set of chirally-enhanced power corrections to 
the heavy-quark limit. We focus on the case of $B\to\pi K$ decays, but 
similar results (with obvious substitutions) hold for all other $B$ 
decays into two flavour-nonsinglet pseudoscalar mesons. For later 
convenience, every coefficient $a_i(\pi K)$ is split into two terms: 
$a_i(\pi K)=a_{i,\rm I}(\pi K)+a_{i,\rm II}(\pi K)$. The first term 
contains the naive factorization contribution and the sum of vertex and 
penguin corrections (the form-factor terms in the factorization formula 
(\ref{fact})), while the second one arises from the hard spectator 
interactions (the hard-scattering term in the factorization formula).
Weak annihilation effects are not included here; they will be discussed 
separately in Section~\ref{subsec:annihilation}. The calculation of the 
kernels described above results in
\begin{eqnarray}\label{ai}
   a_{1,\rm I} &=& C_1 + \frac{C_2}{N_c} \left[ 1
    + \frac{C_F\alpha_s}{4\pi}\,V_K \right] \,,
    \nonumber\\[-1.25cm]
   &&\hspace{7.7cm}
    a_{1,\rm II} = \frac{C_2}{N_c}\,\frac{C_F\pi\alpha_s}{N_c}\,
    H_{K\pi} \,,
    \nonumber\\
   a_{2,\rm I} &=& C_2 + \frac{C_1}{N_c} \left[ 1
    + \frac{C_F\alpha_s}{4\pi}\,V_\pi \right] \,,
    \nonumber\\[-1.25cm]
   &&\hspace{7.7cm}
    a_{2,\rm II} = \frac{C_1}{N_c}\,\frac{C_F\pi\alpha_s}{N_c}\,
    H_{\pi K} \,,
    \nonumber\\
   a_{3,\rm I} &=& C_3 + \frac{C_4}{N_c} \left[ 1
    + \frac{C_F\alpha_s}{4\pi}\,V_\pi \right] \,,
    \nonumber\\[-1.25cm]
   &&\hspace{7.7cm} 
    a_{3,\rm II} = \frac{C_4}{N_c}\,\frac{C_F\pi\alpha_s}{N_c}\,
    H_{\pi K} \,,
    \nonumber\\
   a_{4,\rm I}^p &=& C_4 + \frac{C_3}{N_c} \left[ 1
    + \frac{C_F\alpha_s}{4\pi}\,V_K \right]
    + \frac{C_F\alpha_s}{4\pi}\,\frac{P_{K,2}^p}{N_c} \,, 
    \nonumber\\[-1.25cm]
   &&\hspace{7.7cm} 
    a_{4,\rm II} = \frac{C_3}{N_c}\,\frac{C_F\pi\alpha_s}{N_c}\,
    H_{K\pi} \,,
    \nonumber\\
   a_{5,\rm I} &=& C_5 + \frac{C_6}{N_c} \left[ 1
    + \frac{C_F\alpha_s}{4\pi}\,(-V_\pi') \right] \,,
    \nonumber\\[-1.25cm]
   &&\hspace{7.7cm} 
    a_{5,\rm II} = \frac{C_6}{N_c}\,\frac{C_F\pi\alpha_s}{N_c}\,
    (-H'_{\pi K})
    \,, \nonumber\\
   a_{6,\rm I}^p &=& C_6 + \frac{C_5}{N_c} \left( 1
    - 6\cdot\frac{C_F\alpha_s}{4\pi} \right)
    + \frac{C_F\alpha_s}{4\pi}\,\frac{P_{K,3}^p}{N_c} \,, 
    \nonumber\\[-1cm]
   &&\hspace{7.7cm} 
    a_{6,\rm II} = 0 \,,
    \nonumber\\[0.25cm]
   a_{7,\rm I} &=& C_7 + \frac{C_8}{N_c} \left[ 1
    + \frac{C_F\alpha_s}{4\pi}\,(-V_\pi') \right] \,,
    \nonumber\\[-1.25cm]
   &&\hspace{7.7cm} 
    a_{7,\rm II} = \frac{C_8}{N_c}\,\frac{C_F\pi\alpha_s}{N_c}\,
    (-H'_{\pi K})
    \,, \nonumber\\
   a_{8,\rm I}^p &=& C_8 + \frac{C_7}{N_c} \left( 1
    - 6\cdot\frac{C_F\alpha_s}{4\pi} \right)
    + \frac{\alpha}{9\pi}\,\frac{P_{K,3}^{p,{\rm EW}}}{N_c} \,, 
    \nonumber\\[-1cm]
   &&\hspace{7.7cm} 
    a_{8,\rm II} = 0 \,,
    \nonumber\\[0.25cm]
   a_{9,\rm I} &=& C_9 + \frac{C_{10}}{N_c} \left[ 1
    + \frac{C_F\alpha_s}{4\pi}\,V_\pi \right] \,,
    \nonumber\\[-1.25cm]
   &&\hspace{7.7cm} 
    a_{9,\rm II} = \frac{C_{10}}{N_c}\,\frac{C_F\pi\alpha_s}{N_c}\,
    H_{\pi K} \,,
    \nonumber\\
   a_{10,\rm I}^p &=& C_{10} + \frac{C_9}{N_c} \left[ 1
    + \frac{C_F\alpha_s}{4\pi}\,V_K \right]
    + \frac{\alpha}{9\pi}\,\frac{P_{K,2}^{p,{\rm EW}}}{N_c} \,,
    \nonumber\\[-1.25cm]
   &&\hspace{7.7cm} 
    a_{10,\rm II} = \frac{C_9}{N_c}\,\frac{C_F\pi\alpha_s}{N_c}\,
    H_{K\pi} \,,
\end{eqnarray}
where $C_i\equiv C_i(\mu)$, $\alpha_s\equiv\alpha_s(\mu)$, 
$C_F=(N_c^2-1)/(2N_c)$, and $N_c=3$. The quantities $V_M^{(\prime)}$, 
$H_{M_2 M_1}^{(\prime)}$, $P_{K,2}^p$, $P_{K,3}^p$, 
$P_{K,2}^{p,{\rm EW}}$, and $P_{K,3}^{p,{\rm EW}}$ are hadronic 
parameters that contain all nonperturbative dynamics. These quantities 
consist of convolutions of hard-scattering kernels with meson 
distribution amplitudes. Specifically, the terms $V_M^{(\prime)}$ result 
from the vertex corrections (first four diagrams in 
Figure~\ref{fig:graphs}), $P_{K,2}^p$ and $P_{K,3}^p$ 
($P_{K,2}^{p,{\rm EW}}$ and $P_{K,3}^{p,{\rm EW}}$) arise from QCD 
(electroweak) penguin contractions and the contributions from the dipole 
operators (fifth and sixth diagrams in Figure~\ref{fig:graphs}), and 
$H_{M_2 M_1}^{(\prime)}$ are due hard gluon exchange involving the 
spectator quark in the $B$ meson (last two diagrams in 
Figure~\ref{fig:graphs}). For the penguin terms, the subscript 2 or 3 
indicates the twist of the corresponding projection.

In the numerical evaluation of these expressions we consistently drop 
higher-order terms in the products of the Wilson coefficients with the 
next-to-leading order corrections. Also, in the computation of the 
$O(\alpha)$ corrections we confine ourselves to the penguin contractions 
of the current--current operators $Q_1$ and $Q_2$ and to the 
contribution of the electromagnetic dipole operator, which have the 
largest Wilson coefficients. These contributions, which are contained 
in the quantities $P_{K,2}^{p,{\rm EW}}$ and $P_{K,3}^{p,{\rm EW}}$, 
suffice to cancel the renormalization-scheme dependence of the 
electroweak penguin coefficients $C_{7,\ldots,10}$ at next-to-leading 
order. Additional $O(\alpha)$ corrections proportional to the Wilson 
coefficients $C_{3,\ldots,6}$ of the QCD penguin operators exist, but 
because these coefficients are very small their effects can be safely 
neglected.

\subsubsection*{Vertex and penguin contributions}

We now collect the relevant formulae needed for the calculation of the 
coefficients $a_{i,\rm I}$. The vertex corrections are given by 
($M=\pi,K$)
\begin{eqnarray}\label{FM}
   V_M &=& 12\ln\frac{m_b}{\mu} - 18 + \int_0^1\! dx\,g(x)\,\Phi_M(x)
    \,, \nonumber\\
   V_M' &=& 12\ln\frac{m_b}{\mu} - 6 + \int_0^1\! dx\,g(1-x)\,\Phi_M(x)
    \,, \nonumber\\
   g(x) &=& 3\left( \frac{1-2x}{1-x}\ln x-i\pi \right) \nonumber\\
   &&\mbox{}+ \left[ 2 \,\mbox{Li}_2(x) - \ln^2\!x + \frac{2\ln x}{1-x}
    - (3+2i\pi)\ln x - (x\leftrightarrow 1-x) \right] ,
\end{eqnarray}
where $\mbox{Li}_2(x)$ is the dilogarithm. The constants $18$ and $6$ 
are specific to the NDR scheme. The function $g(x)$ can be obtained from
the corresponding function relevant to, e.g., $\bar B^0\to D^+ K^-$ 
decays \cite{BBNS2} by taking the limit $m_c\to 0$. (Recently, a partial 
two-loop result for the vertex corrections has been obtained in 
\cite{BBN01}, where the terms of order $\beta_0\alpha_s^2$ were 
calculated analytically. To be consistent with the next-to-leading 
order analysis of this paper, we will not make use of this result.)  

If the leading-twist light-cone distribution amplitudes are expanded in 
Gegenbauer polynomials as shown in (\ref{gegenbauer}), the relevant 
convolution integral can be evaluated analytically, giving
\begin{equation}\label{vgegenb}
   \int_0^1\!dx\,g(x)\,\Phi_M(x)
   = -\frac12 - 3i\pi + \left( \frac{11}{2} - 3i\pi \right)
   \alpha_1^M - \frac{21}{20}\,\alpha_2^M + \dots \,.
\end{equation}
The integral with $g(x)\to g(1-x)$ is obtained by changing the sign of 
the odd Gegenbauer coefficients. Since the pion distribution amplitude 
$\Phi_\pi(x)$ is symmetric under the exchange $x\leftrightarrow(1-x)$, 
it follows that $V_\pi'=V_\pi+12$. 

Next, the penguin contributions are
\begin{eqnarray}\label{PK}
   P_{K,2}^p &=& C_1 \left[ \frac43\ln\frac{m_b}{\mu}
    + \frac23 - G_K(s_p) \right]
    + C_3 \left[ \frac83\ln\frac{m_b}{\mu} + \frac43
    - G_K(0) - G_K(1) \right] \nonumber\\
   &&\mbox{}+ (C_4+C_6) \left[ \frac{4n_f}{3}\ln\frac{m_b}{\mu}
    - (n_f-2) G_K(0) - G_K(s_c) - G_K(1) \right] \nonumber\\
   &&\mbox{}- 2 C_{8g}^{\rm eff} \int_0^1 \frac{dx}{1-x}\,
    \Phi_K(x) \,, \nonumber\\
   P_{K,2}^{p,{\rm EW}} &=& (C_1+N_c C_2) \left[
    \frac43\ln\frac{m_b}{\mu} + \frac23 - G_K(s_p) \right]
    - 3\,C_{7\gamma}^{\rm eff} \int_0^1 \frac{dx}{1-x}\,\Phi_K(x) \,,
\end{eqnarray}
where $n_f=5$ is the number of light quark flavours, and $s_u=0$, 
$s_c=(m_c/m_b)^2$ are mass ratios involved in the evaluation of the 
penguin diagrams. Small electroweak corrections from $C_7,\ldots,C_{10}$ 
are consistently neglected in $P^p_{K,2}$ within our approximations.
In principle, in $P_{K,2}^{p,{\rm EW}}$ also contributions from
$C_3,\ldots,C_6$ appear. Their impact is extremely small numerically
and we drop them for simplicity. Similar comments apply to (\ref{hatPK}) 
below. The function $G_K(s)$ is given by
\begin{eqnarray}\label{GK}
   G_K(s) &=& \int_0^1\!dx\,G(s-i\epsilon,1-x)\,\Phi_K(x) \,, \\
   G(s,x) &=& -4\int_0^1\!du\,u(1-u) \ln[s-u(1-u)x] \nonumber\\
   &=& \frac{2(12s+5x-3x\ln s)}{9x}
    - \frac{4\sqrt{4s-x}\,(2s+x)}{3x^{3/2}}
    \arctan\sqrt{\frac{x}{4s-x}} \,. 
\end{eqnarray}
Its expansion in terms of Gegenbauer moments reads 
\begin{eqnarray}
   G_K(s_c) &=& \frac53 - \frac23\ln s_c + \frac{\alpha_1^K}{2}
    + \frac{\alpha_2^K}{5} + \frac43 \left( 8 + 9 \alpha_1^K
    + 9 \alpha_2^K \right) s_c \nonumber\\
   &&\mbox{}+ 2(8 + 63 \alpha_1^K + 214 \alpha_2^K) s_c^2
    - 24 (9 \alpha_1^K + 80 \alpha_2^K) s_c^3
    + 2880 \alpha_2^K s_c^4 \nonumber\\
   &&\mbox{}- \frac23\sqrt{1-4s_c}\,\bigg[ 1 + 2 s_c
    + 6 (4 + 27 \alpha_1^K + 78 \alpha_2^K) s_c^2 \nonumber\\
   &&\quad\mbox{}- 36 (9 \alpha_1^K + 70 \alpha_2^K) s_c^3 
    + 4320 \alpha_2^K s_c^4 \bigg] \left(
    2 \,\mbox{arctanh}\sqrt{1-4 s_c} - i\pi \right) \nonumber\\
   &&\mbox{}+ 12 s_c^2\,\bigg[ 1 + 3 \alpha_1^K + 6 \alpha_2^K
    - \frac43 \left( 1 + 9 \alpha_1^K + 36 \alpha_2^K \right) s_c
    \nonumber\\
   &&\quad\mbox{}+ 18 (\alpha_1^K + 10 \alpha_2^K) s_c^2 
    - 240 \alpha_2^K s_c^3 \bigg] \left(
    2 \,\mbox{arctanh}\sqrt{1-4 s_c} - i\pi \right)^2 + \dots \,,
    \nonumber\\
   G_K(0) &=& \frac53 + \frac{2i\pi}{3} + \frac{\alpha_1^K}{2}
   + \frac{\alpha_2^K}{5} + \dots \,, \nonumber\\
   G_K(1) &=& \frac{85}{3} - 6\sqrt3\,\pi + \frac{4\pi^2}{9}
    - \left( \frac{155}{2} - 36\sqrt3\,\pi+12\pi^2 \right)
    \alpha_1^K \nonumber\\
   &&\mbox{}+ \left( \frac{7001}{5} - 504\sqrt3\,\pi
    + 136\pi^2 \right) \alpha_2^K + \dots \,.
\end{eqnarray}
The contribution of the dipole operators in (\ref{PK}) involve the 
integral
\begin{equation}\label{eq54}
   \int_0^1 \frac{dx}{1-x}\,\Phi_M(x)
   = 3(1+\alpha_1^M+\alpha_2^M+\dots) \,.
\end{equation}

The twist-3 terms from the penguin diagrams are related to the twist-2
terms by the simple replacement $\Phi_K(x)\to\Phi_p^K(x)=1$. For the 
terms proportional to $C_{7\gamma}^{\rm eff}$ and $C_{8g}^{\rm eff}$, 
however, the twist-3 projection yields an additional factor of $(1-x)$,
which cancels the denominator in (\ref{eq54}). We therefore find
\begin{eqnarray}\label{hatPK}
   P_{K,3}^p &=& C_1 \left[ \frac43\ln\frac{m_b}{\mu}
    + \frac23 - \hat G_K(s_p) \right]
    + C_3 \left[ \frac83\ln\frac{m_b}{\mu} + \frac43
    - \hat G_K(0) - \hat G_K(1) \right] \nonumber\\
   &+& (C_4+C_6) \left[ \frac{4n_f}{3}\ln\frac{m_b}{\mu}
    - (n_f-2) \hat G_K(0) - \hat G_K(s_c) - \hat G_K(1) \right]
    - 2 C_{8g}^{\rm eff} \,, \nonumber\\
   P_{K,3}^{p,{\rm EW}} &=& (C_1+N_c C_2) \left[
    \frac43\ln\frac{m_b}{\mu} + \frac23 - \hat G_K(s_p) \right]
    - 3\,C_{7\gamma}^{\rm eff} \,,
\end{eqnarray}
with 
\begin{equation}\label{penfunction1}
   \hat G_K(s) = \int_0^1\!dx\,G(s-i\epsilon,1-x)\,\Phi_p^K(x) \,. 
\end{equation}
Inserting $\Phi_p^K(x)=1$, this integral is evaluated to
\begin{eqnarray}\label{penfunction2}
   \hat G_K(s_c) &=& \frac{16}{9}\,(1-3s_c) - \frac23 \left[
    \ln s_c + (1-4s_c)^{3/2} \left( 2\,\mbox{arctanh}\sqrt{1-4 s_c}
    - i\pi \right) \right] \,, \nonumber\\
   \hat G_K(0) &=& \frac{16}{9} + \frac{2\pi}{3}\,i \,, \qquad
    \hat G_K(1) = \frac{2\pi}{\sqrt3} - \frac{32}{9} \,. 
\end{eqnarray}

The typical parton off-shellness in the loop diagrams contributing to 
the vertex and penguin contributions to the hard-scattering kernels is 
of order $m_b$, and hence it is appropriate to choose a value 
$\mu\sim m_b$ for the renormalization scale in the Wilson coefficients 
$C_i$ and in the kernels $T_i^{\rm I}$ when evaluating the quantities 
$a_{i,\rm I}$. The $\mu$-dependent terms in expressions (\ref{FM}), 
(\ref{PK}) and (\ref{hatPK}) cancel the renormalization-scale dependence 
of the Wilson coefficients $C_i(\mu)$ at next-to-leading order. 
Similarly, the constants accompanying the various logarithms in 
these expressions are renormalization-scheme dependent and combine with 
scheme-dependent constants in the expressions for the Wilson 
coefficients to give renormalization-group invariant results.

The coefficients $a_{i,\rm I}$ contain strong-interaction phases via the
imaginary parts of the functions $g(x)$ and $G(s,x)$. At next-to-leading 
order and to leading power in $\Lambda_{\rm QCD}/m_b$, these phases 
yield the asymptotic contributions to the final-state rescattering 
phases of the $B\to\pi K$ decay amplitudes. The presence of a 
strong-interaction phase in the penguin function $G(s,x)$ is well known 
and commonly referred to as the Bander--Silverman--Soni mechanism 
\cite{BSS}. It was included in many phenomenological investigations of 
nonleptonic $B$ decays; however, there was always an argument as to how 
to choose the gluon momentum, $k_g^2$, in the fourth diagram in 
Figure~\ref{fig:graphs}. In our approach there is no ambiguity to this 
choice, because the distribution of $k_g^2=(1-x)m_b^2$ is determined by 
the kaon distribution amplitudes. The imaginary part of the function 
$g(x)$ is another source of rescattering phases, which arises from hard 
gluon exchanges between the two outgoing mesons \cite{BBNS1,BBNS2}. The
appearance of this phase is a new element of the QCD factorization 
approach. 

\subsubsection*{Hard-scattering contributions}

The hard spectator interactions shown in the last two diagrams in 
Figure~\ref{fig:graphs} give leading-twist and chirally-enhanced twist-3 
contributions to the kernels $T_i^{\rm II}$. We include these 
hard-scattering contributions as parts of the coefficients $a_i$, 
although they are not related to factorized matrix elements in the usual 
sense. Only the twist-2 terms are dominated by hard gluon exchange and 
thus calculable. Nevertheless, for consistency with our treatment of the 
penguin coefficients we should also include the chirally-enhanced terms 
of subleading twist, which however have logarithmic endpoint 
singularities.

At tree level, the integrals over meson distribution amplitudes for the
hard spectator contributions factorize. We find (the momentum
fractions $x$ and $y$ are defined in Section~\ref{subsec:power})
\begin{eqnarray}\label{Hiexpr}
   H_{K\pi} &=& \frac{f_B f_\pi}{m_B^2\,F_0^{B\to\pi}(0)}\,
    \int_0^1 \frac{d\xi}{\xi}\,\Phi_B(\xi)
    \int_0^1 \frac{dx}{\bar x}\,\Phi_K(x)
    \int_0^1 \frac{dy}{\bar y} \left[ \Phi_\pi(y)
    + \frac{2\mu_\pi}{m_b}\,\frac{\bar x}{x}\,\Phi_p^\pi(y) \right]
    \nonumber\\
   &=& \frac{f_B f_\pi}{m_B\lambda_B\,F_0^{B\to\pi}(0)} \left[
    \langle\bar x^{-1}\rangle_K\,\langle\bar y^{-1}\rangle_\pi
    + r_\chi^\pi\,\langle x^{-1}\rangle_K\,X_H^\pi \right] \,, \nonumber\\
   H_{\pi K} &=& \frac{f_B f_K}{m_B^2\,F_0^{B\to K}(0)}\,
    \int_0^1 \frac{d\xi}{\xi}\,\Phi_B(\xi)
    \int_0^1 \frac{dx}{\bar x}\,\Phi_\pi(x)
    \int_0^1 \frac{dy}{\bar y} \left[ \Phi_K(y)
    + \frac{2\mu_K}{m_b}\,\frac{\bar x}{x}\,\Phi_p^K(y) \right]
    \nonumber\\
   &=& \frac{f_B f_K}{m_B\lambda_B\,F_0^{B\to K}(0)} \left[
    \langle\bar x^{-1}\rangle_\pi\,\langle\bar y^{-1}\rangle_K
    + r_\chi^K\,\langle x^{-1}\rangle_\pi\,X_H^K \right] \,, \nonumber\\
   H'_{\pi K} &=& \frac{f_B f_K}{m_B^2\,F_0^{B\to K}(0)}\,
    \int_0^1 \frac{d\xi}{\xi}\,\Phi_B(\xi)
    \int_0^1 \frac{dx}{x}\,\Phi_\pi(x)
    \int_0^1 \frac{dy}{\bar y} \left[ \Phi_K(y)
    + \frac{2\mu_K}{m_b}\,\frac{x}{\bar x}\,\Phi_p^K(y) \right]
    \nonumber\\
   &=& \frac{f_B f_K}{m_B\lambda_B\,F_0^{B\to K}(0)} \left[
    \langle x^{-1}\rangle_\pi\,\langle\bar y^{-1}\rangle_K
    + r_\chi^K\,\langle\bar x^{-1}\rangle_\pi\,X_H^K \right] \,,
\end{eqnarray}
where we have defined the moments
\begin{equation}
   \int_0^1 \frac{d\xi}{\xi}\,\Phi_B(\xi)
   \equiv \frac{m_B}{\lambda_B} \,, \qquad
   \int_0^1\!dx\,x^n\,\Phi_M(x) \equiv \langle x^n\rangle_M \,.
\end{equation}
The quantity $\lambda_B$ parameterizes our ignorance about the 
$B$-meson distribution amplitude \cite{BBNS1}. Not much is known about 
this parameter except for the upper bound $3\lambda_B\le 4\bar\Lambda$
\cite{Korch}, where $\bar\Lambda=m_B-m_b$ is a scheme-dependent 
parameter. In practice, this just means that $\lambda_B$ is expected to 
be less than 600\,MeV or so. The ratios $2\mu_K/m_b$ and $2\mu_\pi/m_b$ 
multiplying the twist-3 terms coincide with the parameters $r_\chi^K$ 
and $r_\chi^\pi$ introduced in (\ref{rKdef}) and (\ref{rpidef}), 
respectively. The twist-3 contribution involves the logarithmically 
divergent integral ($M=\pi$ or $K$)
\begin{equation}\label{Xdef}
   X_H^M\equiv \int_0^1 \frac{dy}{1-y}\,\Phi_p^M(y)
   = \int_0^1 \frac{dy}{1-y} \,.
\end{equation}
As previously we have to use the asymptotic form for $\Phi_p^M(y)$,
so no distinction between pion and kaon is necessary. (Consequently, the
superscript ``$M$'' will often be dropped from now on.) The divergence 
results from the region where the spectator quark in the $B$ meson 
enters the light final-state meson at the ``lower vertex'' in 
Figure~\ref{fig1} as a soft quark. The twist-3 hard-scattering kernels 
do not provide sufficient endpoint suppression to render this 
contribution subleading. In practice, the singularity will be smoothed
out by soft physics related to the intrinsic transverse momentum and
off-shellness of the partons, which unfortunately does not admit a 
perturbative treatment (see also the discussion in 
Section~\ref{sec:comp}). In particular, the resulting contribution may 
be complex due to soft rescattering in higher orders. For the purpose 
of power counting, we note that the effect of transverse momentum and 
off-shellness would be to modify $(1-y)\to(1-y)+\epsilon$ with 
$\epsilon=O(\Lambda_{\rm QCD}/m_b)$ in the denominator in (\ref{Xdef}). 
We thus expect that $X_H^M\sim\ln(m_b/\Lambda_{\rm QCD})$, however, with 
a potentially complex coefficient.

Considering the off-shellness of the gluon in the last two diagrams 
in Figure~\ref{fig:graphs}, it is natural to associate a scale 
$\mu_h\sim(\Lambda_{\rm QCD}\,m_b)^{1/2}$, rather than $\mu\sim m_b$, 
with the hard-scattering contributions of leading power. Hence, we set 
$\alpha_s=\alpha_s(\mu_h)$ (and also evaluate the Wilson coefficients 
at the scale $\mu_h$) for the twist-2 contributions to the quantities 
$a_{i,\rm II}$ in (\ref{ai}). Specifically, we use 
$\mu_h=\sqrt{\Lambda_h\,\mu}$ with $\Lambda_h=0.5$\,GeV for the scale 
in the hard-scattering diagrams. However, because of the endpoint 
divergence the contributions proportional to $X_H^M$ must be considered 
as nonperturbative effects dominated by small-momentum interactions. 
For this reason, we should more properly write
\begin{equation}\label{better}
   H_{K\pi} = \frac{f_B f_\pi}{m_B\lambda_B\,F_0^{B\to\pi}(0)}
   \left[ \langle\bar x^{-1}\rangle_K\,\langle\bar y^{-1}\rangle_\pi
   + \frac{\alpha_s(\mu_s)\,r_\chi^\pi(\mu_s)}{\alpha_s(\mu_h)}\,
   \langle x^{-1}\rangle_K\,X_H^\pi \right] ,
\end{equation}
and similarly for the other two quantities in (\ref{Hiexpr}). Here 
$\mu_s$ may be a soft scale. In other words, in principle there is no 
reason to expect that the twist-3 contributions should be governed by a
perturbative coupling constant. However, it turns out that the product
$\alpha_s(\mu_s)\,r_\chi^\pi(\mu_s)$ is almost renormalization-group
invariant (it scales only as $[\alpha_s(\mu)]^{1/25}$ with $N_c=3$ and
four flavours), and therefore we may evaluate it at the scale $\mu_h$, 
so that the ratio of running couplings in (\ref{better}) equals 1.

The expressions for the hard-scattering contributions in (\ref{Hiexpr}) 
contain many poorly known parameters. Besides the divergent quantity 
$X_H$ and the wave-function parameter $\lambda_B$, they depend on the 
$B$-meson decay constant and heavy-to-light form factors. Fortunately, 
it turns out that ratios of the different hard-scattering contribution 
have very small uncertainties. Using the symmetry of the pion 
distribution amplitude, we find that
\begin{equation}\label{Hrela}
   H_{\pi K}' = H_{\pi K} \,, \qquad
   H_{K\pi} \simeq R_{\pi K}\,H_{\pi K} \,,
\end{equation}
where $R_{\pi K}=A_{K\pi}/A_{\pi K}$ is the ratio of factorized matrix 
elements defined in (\ref{ratios}). As argued in 
Section~\ref{subsec:power}, in the approximation where only 
chirally-enhanced power corrections are included one is forced to 
employ the asymptotic forms of the twist-3 distribution amplitudes 
$\Phi_p(x)$ and $\Phi_\sigma(x)$. It is then not necessary to keep other 
nonasymptotic effects in the twist-3 terms. That is, we are free to 
replace the moments of twist-2 amplitudes multiplying $X_H$ by their 
asymptotic values. In this approximation, the twist-3 contributions in 
(\ref{Hiexpr}) reduce to a universal, multiplicative correction of the 
twist-2 term, and we obtain the approximate form of the second relation 
in (\ref{Hrela}). In other words, up to small SU(3) violations the main
uncertainties in the description of the hard-scattering terms combine
into a single poorly-known quantity $H_{\pi K}$.

Finally, note that in (\ref{Hiexpr}) we have not assumed any symmetry 
properties of the twist-2 light-meson distribution amplitudes. 
Therefore, with obvious substitutions our results can be applied 
directly to other decays, such as $B\to\pi\pi$ and $B_s\to K^+ K^-$.

\subsection{\boldmath Weak annihilation contributions $b_i$\unboldmath}
\label{subsec:annihilation}

Weak annihilation contributions to charmless hadronic $B$ decays are
power suppressed in the heavy-quark limit and hence do not appear in
the factorization formula (\ref{fact}). Nevertheless, as emphasized 
in \cite{KLS00}, these contributions may be numerically important for
realistic $B$-meson decays. Besides their power suppression, weak 
annihilation effects differ from the hard spectator interactions 
discussed earlier in that they exhibit endpoint singularities even at 
twist-2 order in the light-cone expansion for the final-state mesons, 
and therefore cannot be computed self-consistently in the context of 
a hard-scattering approach. In the following discussion, we will ignore 
the soft endpoint divergences and derive results for the annihilation 
contributions in terms of convolutions of ``hard-scattering'' kernels 
with light-cone distribution amplitudes, including again the 
chirally-enhanced twist-3 projections. Despite the fact that such a 
treatment is not entirely self-consistent, it is nevertheless useful to 
estimate the importance of annihilation for particular final states.

\begin{figure}[t]
\epsfxsize=15cm
\centerline{\epsffile{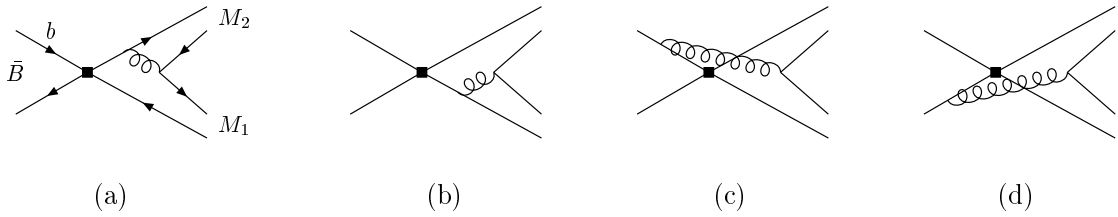}}
\centerline{\parbox{14cm}{\caption{\label{fig_annh} 
Annihilation diagrams.}}}
\end{figure}

At leading order in $\alpha_s$, the annihilation kernels follow from 
the diagrams shown in Figure~\ref{fig_annh}. They result in a further 
contribution to the hard-scattering term in the factorization 
formula, in addition to the hard spectator scattering discussed in 
Section~\ref{subsec:power}. It will be convenient for future 
applications to keep the discussion of annihilation contributions 
general. We thus consider a generic $b$-quark decay and use the 
convention that $M_2$ contains a quark from the weak decay vertex with 
longitudinal momentum fraction $x$, and $M_1$ contains an antiquark 
from the weak vertex with momentum fraction $\bar y$. The four-quark 
operators in the effective weak Hamiltonian are Fierz-transformed into 
the form $(\bar q_1 b)_{\Gamma_1}(\bar q_2 q_3)_{\Gamma_2}$, such that 
the quarks in the first bracket refer to the constituents of the 
$\bar B$ meson. If the colour indices of this bracket are of the form 
$\bar q_{1i} b_i$, diagrams (c) and (d) do not contribute, while 
diagrams (a) and (b) give rise to a colour factor $C_F/N_c$. If the 
colour indices are of the form $\bar q_{1i} b_j$, then the colour factor 
of all four diagrams is $C_F/N_c^2$. The projections onto the light-cone
distribution amplitudes are done in the same manner as for the hard 
spectator scattering described in Section~\ref{subsec:power}. At leading
power (and assuming that $x,y\gg\xi$) the integration over the $B$-meson 
distribution amplitude is trivial and yields the $B$-meson decay 
constant, since the kernels are $\xi$ independent. The remainders of the 
diagrams can be expressed in terms of the following building blocks:
\begin{eqnarray}
   A_1^i &=& \pi\alpha_s \int_0^1\! dx dy\, 
    \left\{ \Phi_{M_2}(x)\,\Phi_{M_1}(y)
    \left[ \frac{1}{y(1-x\bar y)} + \frac{1}{\bar x^2 y} \right]
    + \frac{4\mu_{M_1}\mu_{M_2}}{m_b^2}\,\frac{2}{\bar x y} \right\} ,
    \nonumber\\
   A_1^f &=& 0 \,, \nonumber\\ 
   A_2^i &=& \pi\alpha_s \int_0^1\! dx dy\, 
    \left\{ \Phi_{M_2}(x)\,\Phi_{M_1}(y)
    \left[ \frac{1}{\bar x(1-x\bar y)} + \frac{1}{\bar x y^2} \right]
    + \frac{4\mu_{M_1}\mu_{M_2}}{m_b^2}\,\frac{2}{\bar x y} \right\} ,
    \nonumber\\
   A_2^f &=& 0 \,, \nonumber\\ 
   A_3^i &=& \pi\alpha_s \int_0^1\! dx dy\,
    \left\{ \frac{2\mu_{M_1}}{m_b}\,\Phi_{M_2}(x)\,
    \frac{2\bar y}{\bar x y(1-x\bar y)}
    - \frac{2\mu_{M_2}}{m_b}\,\Phi_{M_1}(y)\,
    \frac{2x}{\bar x y(1-x\bar y)} \right\} , \nonumber\\
   A_3^f &=& \pi\alpha_s \int_0^1\! dx dy\,
    \left\{ \frac{2\mu_{M_1}}{m_b}\,\Phi_{M_2}(x)\,
    \frac{2(1+\bar x)}{\bar x^2 y}
    + \frac{2\mu_{M_2}}{m_b}\,\Phi_{M_1}(y)\,
    \frac{2(1+y)}{\bar x y^2} \right\} .
\end{eqnarray}
Here the superscripts $i$ and $f$ refer to gluon emission from the 
initial- and final-state quarks, respectively. The subscript $k$ on 
$A_k^{i,f}$ refers to one of the three possible Dirac structures 
$\Gamma_1\otimes\Gamma_2$, namely $k=1$ for $(V-A)\otimes(V-A)$, $k=2$ 
for $(V-A)\otimes(V+A)$, and $k=3$ for $(-2)(S-P)\otimes(S+P)$. As 
always, $\Phi_M(x)$ denotes the leading-twist light-cone distribution 
amplitude of a pseudoscalar meson $M$, and the asymptotic forms of the 
twist-3 amplitudes have been used. Note that in the limit of symmetric 
(under $x\leftrightarrow\bar x$) distribution amplitudes, and assuming 
SU(3) flavour symmetry, we have $A_1^i=A_2^i$ and $A_3^i=0$. In this 
approximation the annihilation contributions can be parameterized by 
only two quantities ($A_1^i$ and $A_3^f$). For an estimate of these 
annihilation contributions we use the asymptotic form of the 
leading-twist distribution amplitudes to obtain
\begin{eqnarray}\label{XAdef}
   A_1^i &\approx& \pi\alpha_s \left[
    18 \left( X_A -4 + \frac{\pi^2}{3} \right)
    + 2 r_\chi^2\,X_A^2 \right] \,, \nonumber\\
   A_3^f &\approx & 12\pi\alpha_s\,r_\chi\,(2 X_A^2 - X_A) \,,
\end{eqnarray}
where $X_A=\int_0^1 dy/y$ parameterizes the divergent endpoint 
integrals. Similar to the case of the twist-3 hard-scattering 
contributions parameterized by $X_H$, we will treat the quantity $X_A$ 
as a phenomenological parameter. Clearly, taking the same value of 
$X_A$ for all annihilation terms is a crude model. We shall see below, 
however, that the parameter $A_3^f$ (contributing to penguin 
annihilation topologies) gives the dominant contribution. Therefore, 
our treatment effectively amounts to defining a model for this 
particular parameter.

To complete the calculation we need to account for the flavour 
structure of the various operators. It is convenient to introduce the 
compact notation
\begin{equation}
   \langle M_1 M_2| j_1\times j_2 |\bar B_q\rangle 
   \equiv i c f_{B_q} f_{M_1} f_{M_2} \,,
\end{equation}
where the constant $c$ takes into account factors of $(-1)$ or 
$1/\sqrt{2}$ appearing in the quark wave-functions of some of the 
mesons, and $c\ne 0$ only if the flavours of the ``currents'' $j_1$ and 
$j_2$ match those of the mesons $M_1$ and $M_2$, respectively. It is 
apparent from Figure~\ref{fig_annh} that the products $j_1\times j_2$ 
have the flavour structure
\begin{equation}
   \sigma_{q_1}^{q_2} = \sum_{q'}\,(\bar q' q_2)\times(\bar q_1 q') \,,
\end{equation}
where the sum over $q'=u,d,s$ arises from the $g\to q'\bar q'$ 
vertex in the annihilation diagrams. Effectively, the flavour structure
$\sigma_{q_1}^{q_2}$ ``creates'' a quark $q_1$ (lower index) and an 
antiquark $\bar q_2$ (upper index), together with a flavour-singlet 
$q'\bar q'$ pair. It is then straightforward to find the set of meson 
final states to which a given flavour structure contributes. All 
$\sigma$ operators contributing to charged $B^-$ decays have the 
structure $\sigma_d^u$ or $\sigma_s^u$. The corresponding two-particle 
final states with light (flavour-nonsinglet) pseudoscalar mesons are
\begin{equation}\label{chargedB}
   \sigma_d^u: ~~\pi^0\pi^- ,\, \pi^-\pi^0 ,\, K^- K^0 \,;
   \hspace{1.2cm}
   \sigma_s^u: ~~\pi^0 K^- ,\, \pi^-\bar K^0 \,. 
\end{equation}
Operators contributing to neutral $\bar B_r$ decays (with $r=d,s$) have
the structure $\sigma_d^r$, $\sigma_s^r$, $\sigma_u^u$, or one of the 
``penguin structures'' $\mbox{tr}(\sigma)\equiv\sum_q\,\sigma_q^q$ and
$\mbox{tr}(Q\,\sigma)\equiv\sum_q\,e_q\,\sigma_q^q$. Here $Q$ is the 
charge operator for the quarks, and the sum (trace) over $q$ is 
inherited from the QCD and electroweak penguin operators in the 
effective weak Hamiltonian. The final states to which these operators 
contribute are
\begin{eqnarray}\label{neutralBs}
   \sigma_d^d:&& ~~\pi^+\pi^- ,\, \pi^0\pi^0 ,\,\bar K^0 K^0 \,;
    \hspace{1.2cm}
    \sigma_s^d: ~~\pi^+ K^- ,\, \pi^0\bar K^0 \,; 
    \nonumber\\
   \sigma_d^s:&& ~~K^+\pi^- ,\, K^0\pi^0 \,; 
    \hspace{1.2cm}
    \sigma_s^s: ~~K^+ K^- ,\, K^0\bar K^0 \,; 
    \nonumber\\
   \sigma_u^u:&& ~~\pi^0\pi^0 ,\, \pi^-\pi^+ ,\, K^- K^+ \,;
    \nonumber\\
   \mbox{tr}(\sigma) ,\,
   \mbox{tr}(Q\,\sigma):&&
    ~~\pi^0\pi^0 ,\, \pi^-\pi^+ ,\, \pi^+\pi^- ,\, K^- K^+ ,\,
    K^+ K^- ,\, K^0\bar K^0 ,\, \bar K^0 K^0 \,. 
\end{eqnarray}

It follows from the above discussion that the weak annihilation 
contributions to the decay amplitudes can be parameterized in terms of 
the coefficients
\begin{eqnarray}\label{bidef}
   b_1 &=& \frac{C_F}{N_c^2}\,C_1 A_1^i \,, \qquad
    b_3 = \frac{C_F}{N_c^2} \Big[ C_3 A_1^i + C_5 (A_3^i+A_3^f)
    + N_c C_6 A_3^f \Big] \,, \nonumber\\
   b_2 &=& \frac{C_F}{N_c^2}\,C_2 A_1^i \,, \qquad
    b_4 = \frac{C_F}{N_c^2}\,\Big[ C_4 A_1^i + C_6 A_2^i \Big] \,,
    \nonumber\\
   b_3^{\rm EW} &=& \frac{C_F}{N_c^2} \Big[ C_9 A_1^i
    + C_7 (A_3^i+A_3^f) + N_c C_8 A_3^f \Big] \,, \nonumber\\
   b_4^{\rm EW} &=& \frac{C_F}{N_c^2}\,\Big[ C_{10} A_1^i
    + C_8 A_2^i \Big] \,.
\end{eqnarray}
They correspond to current--current annihilation ($b_1,b_2$), penguin  
annihilation ($b_3,b_4$), and electroweak penguin annihilation 
($b_3^{\rm EW},b_4^{\rm EW}$), where within each pair the two 
coefficients correspond to different flavour structures. The quantities
$b_i$ depend on the final-state mesons through the light-cone 
distribution amplitudes entering the expressions for $A_k^{i,f}$ and 
thus should be written as $b_i(M_1 M_2)$. We suppress this notation 
when confusion cannot arise. As for the hard spectator terms, we will
evaluate the various quantities in (\ref{bidef}) at the scale 
$\mu_h=\sqrt{\Lambda_h\,\mu}$.

The effective weak Hamiltonian for $\bar B$-meson decays contains a 
strangeness-conserving part (${\cal H}_{\Delta S=0}$) and a 
strangeness-changing part (${\cal H}_{\Delta S=1}$). Using the above
definitions, the annihilation contributions to the matrix elements of 
${\cal H}_{\Delta S=1}$ can be written as 
\begin{equation}\label{aanham}
   \langle M_1 M_2|{\cal H}_{\Delta S=1}|\bar B\rangle
   = \frac{G_F}{\sqrt 2} \sum_{p=u,c} \lambda_p\,
   \langle M_1 M_2|{\cal T}_p^{\rm ann}|\bar B\rangle \,,
\end{equation}
where $\lambda_p=V_{pb} V^*_{ps}$, and 
\begin{eqnarray}
   {\cal T}_p^{\rm ann}
   &=& \delta_{up}\,(\delta_{rn}\,b_1\,\sigma_u^u
    + \delta_{ru}\,b_2\,\sigma_s^u) + b_3\,\sigma_s^r
    + \delta_{rn}\,b_4\,\mbox{tr}(\sigma) \nonumber\\
   &&\mbox{}+ \frac{3}{2}\,b_3^{\rm EW}\,e_r\,\sigma_s^r
    + \frac{3}{2}\,\delta_{rn}\,b_4^{\rm EW}\,\mbox{tr}(Q\,\sigma) \,.
\end{eqnarray}
The index $r$ refers to the flavour of the spectator quark inside the 
$B$ meson in (\ref{aanham}), and $\delta_{rn}=\delta_{rd}+\delta_{rs}$
equals 1 for neutral $\bar{B}$ mesons and 0 for $B^-$. 
The matrix elements of 
${\cal H}_{\Delta S=0}$ take an identical form, except that $\lambda_p$ 
is replaced with $\lambda_p'=V_{pb} V^*_{pd}$ in this case, and 
$\sigma_s^q$ must be replaced with $\sigma_d^q$. 

It is now straightforward to derive the annihilation contribution to 
a particular final state in terms of the coefficients $b_i(M_1 M_2)$. 
The expressions for the decay modes discussed in this paper have been 
given earlier in (\ref{BKPiann}) and (\ref{Bpipiann}). We will later 
use the approximation (\ref{XAdef}) for the quantities $A_k^{i,f}$ to 
estimate the annihilation coefficients $b_i$ numerically. We should 
recall, however, that the annihilation kernels have been derived under 
the assumption of hard scattering. Specifically, we have neglected the 
momentum fraction $\xi$ of the spectator quark in the $B$ meson compared 
to $x$, $\bar x$, $y$, $\bar y$ in deriving the kernels. This is the
reason why the results for $A_k^{i,f}$ turned out to be independent of 
the form of the $B$-meson distribution amplitude. In the endpoint 
regions, one or two of the variables $x$, $\bar x$, $y$, $\bar y$ can 
be of order $\xi$, invalidating this approximation. Therefore, our
numerical results for the weak annihilation contributions presented in 
the next section must be considered as model-dependent estimates.

\section{Numerical analysis of amplitude parameters}
\label{sec:numerics}

\begin{table}[t]
\centerline{\parbox{14cm}{\caption{\label{tab:inputs}
Summary of theoretical input parameters.}}}
\begin{center}
\begin{tabular}{|c|c|c|c|c|}
\hline\hline
\multicolumn{5}{|c|}{QCD Scale and Running Quark Masses} \\
\hline
$\Lambda_{\overline{\rm MS}}^{(5)}$ & $m_b(m_b)$ & $m_c(m_b)$
 & $m_s(2\,\mbox{GeV})$ & $(m_u+m_d)(2\,\mbox{GeV})$ \\
\hline
225\,MeV & ~4.2\,GeV~ & $(1.3\pm 0.2)$\,GeV & $(110\pm 25)$\,MeV
 & $(9.1\pm 2.1)$\,MeV \\
\hline
\end{tabular}
{\tabcolsep=0.479cm\begin{tabular}{|c|c|c|c|c|}
\hline
\multicolumn{5}{|c|}{Parameters Related to Hadronic Matrix Elements} \\
\hline
$f_\pi$ & $f_K$ & $f_B$ & $F_0^{B\to\pi}(0)$ & $R_{\pi K}$ \\
\hline
131\,MeV & 160\,MeV & $(180\pm 40)$\,MeV & $0.28\pm 0.05$
 & $0.9\pm 0.1$ \\
\hline
\end{tabular}}
{\tabcolsep=0.552cm\begin{tabular}{|c|c|c|c|c|}
\hline
\multicolumn{5}{|c|}{Parameters of Distribution Amplitudes} \\
\hline
$\alpha_1^K$ & $\alpha_2^K$ & $\alpha_1^\pi$ & $\alpha_2^\pi$
 & $\lambda_B$ \\
\hline
$0.3\pm 0.3$ & $0.1\pm 0.3$ & $\quad 0\quad$ & $0.1\pm 0.3$
 & $(350\pm 150)$\,MeV \\
\hline\hline
\end{tabular}}
\end{center}
\end{table}

In this section we summarize the numerical values of the parameters 
$a_i$ and $b_i$ entering the $B\to\pi K$ decay amplitudes and perform
detailed estimates of various sources of theoretical uncertainties. 
Other decays such as $B\to\pi\pi$ will be discussed later. 

The theoretical input parameters used in our analysis, together with 
their respective ranges of uncertainty, are summarized in 
Table~\ref{tab:inputs}. The quark masses are running masses in the 
$\overline{\mbox{MS}}$ scheme. Note that the value of the charm-quark 
mass is given at $\mu=m_b$. The ratio $s_c=(m_c/m_b)^2$ needed for the 
calculation of the penguin contributions is scale independent. The 
values of the light quark masses are such that $r_\chi^K=r_\chi^\pi$. 
We hold $(m_u+m_d)/m_s$ fixed and use $m_s$ as an input parameter. 
(This implies that in our error estimation procedure 
$|P_{\pi\pi}/T_{\pi\pi}|$ indirectly depends on $m_s$.) The value of the 
QCD scale parameter corresponds to $\alpha_s(M_Z)=0.118$ in the 
$\overline{\rm MS}$ scheme. The values for the $B$-meson decay constant 
$f_B$, the semileptonic form factor $F_0^{B\to\pi}(0)$, and the hadronic 
parameter $R_{\pi K}$ in (\ref{ratios}) are consistent with recent 
determinations of these quantities using light-cone QCD sum rules 
\cite{Khod,Patr}, form-factor models (see, e.g., \cite{Stech00}), and 
lattice gauge theory (see, e.g., \cite{Bpilat}). The last row in the 
table contains our values for the Gegenbauer moments of the pion and 
kaon light-cone distribution amplitudes, and for the 
$B$-meson wave-function parameter $\lambda_B$. The Gegenbauer moments 
for the light mesons are adopted with a conservative error estimate 
that encompasses most of the parameter ranges obtained from 
phenomenological or QCD sum rule determinations of these quantities. 
The value of $\lambda_B$ is an 
educated guess guided by the model determinations 
$\lambda_B\approx\frac23\bar\Lambda\approx 300$\,MeV \cite{GrNe} and 
$\lambda_B=(380\pm 120)$\,MeV \cite{Korch}. For comparison, using the 
models of \cite{KLS00} we obtain $\lambda_B=(410\pm 170)$\,MeV.

\subsection{Vertex and penguin contributions}

We begin the discussion with the vertex and penguin contributions, i.e., 
the terms $a_{i,\rm I}(\pi K)$ in (\ref{ai}). In this case all 
convolution integrals are finite, even for the power-suppressed 
coefficients $r_\chi^K a_{6,\rm I}(\pi K)$ and 
$r_\chi^K a_{8,\rm I}(\pi K)$. Table~\ref{tab:ai} contains the values 
of the various coefficients for three different values of the
renormalization scale, and with theoretical uncertainties due to the 
variation of the pion and kaon light-cone distribution amplitudes and
the charm-quark mass, as specified in Table~\ref{tab:inputs}. We vary 
the Gegenbauer moments independently and quote the maximal variation in 
the table. The main characteristics of theoretical uncertainties are as
follows:

{\tabcolsep=0.09cm
\begin{table}[t]
\centerline{\parbox{14cm}{\caption{\label{tab:ai}
Next-to-leading order results for the coefficients $a_{i,\rm I}(\pi K)$ 
for three different choices of the renormalization scale. Numbers in 
parentheses show the maximal change in the last digit(s) under variation 
of the Gegenbauer moments of the light-cone distribution amplitudes; if 
present, numbers in square brackets show the change under variation of 
the charm-quark mass.}}}
\begin{center}
\begin{tabular}{|c|ccc|ccc|}
\hline\hline
 & \multicolumn{3}{c|}{Real Part}
 & \multicolumn{3}{c|}{Imaginary Part} \\ 
\hline
$\mu$ & $m_b/2$ & $m_b$ & $2 m_b$ & $m_b/2$ & $m_b$ & $2 m_b$ \\
\hline
$a_{1,\rm I}$ & 
 $1.073(8)$ & $1.054(4)$ & $1.037(2)$
 & $0.048(11)$ & $0.026(6)$ & $0.015(3)$ \\
$a_{2,\rm I}$ & 
 $-0.039(4)$ & $0.005(3)$ & $0.045(2)$ & 
 $-0.113$ & $-0.084$ & $-0.066$ \\
$a_{3,\rm I}$ & 
 $0.008$ & $0.006$ & $0.004$ & 
 $0.004$ & $0.002$ & $0.001$ \\
$-a_{4,\rm I}^u$ & 
 $0.031(4)$ & $0.029(3)$ & $0.027(2)$ & 
 $0.023(0)$ & $0.017$ & $0.014$ \\
$-a_{4,\rm I}^c$ & 
 \,$0.036(9)[2]$ & $0.033(6)[1]$ & $0.030(4)[1]$ & 
 \,$0.005(3)[4]$ & $0.004(3)[3]$ & $0.004(2)[2]$ \\
$-a_{5,\rm I}$ &  
 $0.011$ & $0.007$ & $0.004$ & 
 $0.005$ & $0.003$ & $0.001$ \\
$-r_\chi^K a_{6,\rm I}^u$ & 
 $0.052$ & $0.052$ & $0.052$ & 
 $0.017$ & $0.018$ & $0.019$ \\
$-r_\chi^K a_{6,\rm I}^c$ & 
 $0.056$ & $0.056$ & $0.056$ & 
 $0.005[3]$ & $0.007[3]$ & $0.008[3]$ \\
$a_{7,\rm I}/\alpha$ & 
 $0.007$ & $0.011$ & $0.025$ & 
 $0.004$ & $0.002$ & $0.001$ \\
$r_\chi^K a_{8,\rm I}^u/\alpha$ &  
 $0.090$ & $0.077$ & $0.059$ & 
 $-0.001$ & $-0.009$ & $-0.020$ \\
~$r_\chi^K a_{8,\rm I}^c/\alpha$~ & 
 $0.090$ & $0.075$ & $0.055$ &
 $-0.000$ & $-0.005[1]$ & $-0.010[3]$ \\
$-a_{9,\rm I}/\alpha$ & 
 $1.258(1)$ & $1.222(1)$ & $1.181$ & 
 $0.040$ & $0.022$ & $0.012$ \\
$a_{10,\rm I}^u/\alpha$ & 
 $0.062(27)$ & $0.020(20)$ & $-0.025(16)$ & 
 $0.168(39)$ &$0.116(29)$ & $0.084(22)$ \\
$a_{10,\rm I}^c/\alpha$ & 
 $0.062(27)$ & $0.018(21)[1]$ & $\!-0.028(17)[1]$
 & $0.168(39)$ & $0.121(30)[1]$ & $0.093(25)[2]$ \\
\hline\hline
\end{tabular}
\end{center}
\end{table}}

\noindent
{\em Renormalization-scale dependence:\/} The residual scale dependence 
at next-to-leading order depends on the size of the leading-order 
Wilson coefficients and the magnitude of the Wilson coefficient that 
multiplies the next-to-leading order correction. In general, we find a 
significant reduction of scale dependence compared to the $a_i$ 
parameters obtained at leading order (corresponding to naive 
factorization). Only for the parameters $a_2$ and $a_{10}$ a sizeable 
scale dependence remains at next-to-leading order, even though it is 
reduced by about a factor of 2 relative to the leading order. In
general, the imaginary parts of the coefficients $a_{i,\rm I}$, which
occur first at order $\alpha_s$, have a larger scale dependence than the 
real parts. 

\noindent
{\em Light-cone distribution amplitudes:\/} The explicit expressions in 
Section~\ref{subsec:airesults} show that the second Gegenbauer moments 
$\alpha_2^K$ and $\alpha_2^\pi$ enter the results for the vertex and 
penguin contributions typically with small coefficients. Therefore, the 
main uncertainty comes from the first moment $\alpha_1^K$ of the kaon 
distribution amplitude, which affects only $a_1$, $a_4$ and $a_{10}$. 
The real part of $a_{10}$ is particularly uncertain. There is also some 
uncertainty in $a_2$, because the dependence on the second moment of 
the pion light-cone distribution amplitude is amplified by the large 
Wilson coefficient $C_1$. From Table~\ref{tab:ai}, we conclude that the 
dependence on the distribution amplitudes is almost always smaller than 
the scale dependence. Since, for consistency, we must use the 
asymptotic twist-3 distribution amplitudes, the coefficients $a_6$ and 
$a_8$ show no dependence on the Gegenbauer moments. This is clearly an 
approximation, which is valid only if the chirally-enhanced power 
corrections dominate over the remaining power corrections. (This is a 
questionable approximation, because the quark--antiquark--gluon 
distribution amplitude impacts on $\Phi_p(x)$ with a large numerical 
coefficient \cite{BraF}.) As a consequence, we cannot control 
SU(3)-breaking effects at twist-3 order. In practice, we expect such 
SU(3) violations to have a similar (hence, small) effect as those at 
leading twist.

\noindent
{\em Charm and strange-quark masses:\/} The value of the charm-quark 
mass affects the penguin contributions with a charm-quark loop. This 
leads to a significant uncertainty in the imaginary parts of $a_4^c$ 
and $a_6^c$. The real parts of the coefficients are much less affected. 
Note that the chiral enhancement factor $r_\chi^K$ multiplying the 
coefficients $a_6$ and $a_8$ in Table~\ref{tab:ai} is inversely 
proportional to the strange-quark mass. The $\pm 25$\,MeV uncertainty 
in the value of $m_s$ leads to a sizeable uncertainty in the values of 
the products $r_\chi^K a_{6,8}$ which, except for the imaginary parts 
of $r_\chi^K a_{6,8}^c$, is a much larger effect than the dependence on 
the charm-quark mass.  

\noindent
{\em Unknown power corrections:\/} A source of theoretical uncertainty 
that is difficult to estimate arises from power corrections which cannot 
be computed using the QCD factorization approach. Naive dimensional 
analysis suggests that such corrections are of order 
$\Lambda_{\rm QCD}/m_b\sim(10\mbox{--}20)\%$, but they could be 
enhanced, e.g., by large Wilson coefficients. This situation may 
potentially be realized for the parameters $a_2$ and $a_{10}$. Recently,
some power corrections to QCD factorization for the $\pi\pi$ final state
have been investigated in the framework of QCD sum rules \cite{AKH} and
using the renormalon calculus \cite{BBN01}. (Power corrections for 
final states with one heavy meson were also considered in 
\cite{BBNS2,Burr01}.) In none of these cases particularly large 
corrections have been identified.

\subsection{Hard spectator interactions}
\label{subsec:hardspec}

\begin{table}[t]
\centerline{\parbox{14cm}{\caption{\label{tab:aiII}
Coefficients $a_{i,\rm II}(\pi K)$ for three different choices of the 
renormalization scale and fixed default values of the quantity 
$H_{\pi K}$. All scale-dependent quantities are evaluated at the scale 
$\mu_h=\sqrt{\Lambda_h\,\mu}$ with $\Lambda_h=0.5$\,GeV. Numbers in 
parentheses show the maximal change in the last digit(s) under 
variation of the Gegenbauer moments; numbers in square brackets show 
the dependence under variation of $R_{\pi K}$.}}}
\begin{center}
\begin{tabular}{|c|ccc|}
\hline\hline
$\mu$ & $m_b/2$ & $m_b$ & $2 m_b$ \\
$\mu_h$ & 1.02\,GeV & 1.45\,GeV & 2.05\,GeV \\
\hline
$H_{\pi K}^{\rm default}$ & 0.92 & 0.99 & 1.05 \\
\hline
$a_{1,\rm II}$ & $-0.087(16)[10]$ & $-0.061(14)[7]$ & $-0.045(12)[5]$ \\
$a_{2,\rm II}$ & $\phantom{-}0.231$ & $\phantom{-}0.192$
 & $\phantom{-}0.167$ \\
$a_{3,\rm II}$ & $-0.010$ & $-0.007$ & $-0.005$ \\
$a_{4,\rm II}$ & $0.004(1)$ & $0.003(1)$ & $0.002(1)$ \\
$a_{5,\rm II}$ & $\phantom{-}0.016$ & $\phantom{-}0.010$
 & $\phantom{-}0.008$ \\
$a_{7,\rm II}/\alpha$ & $-0.014$ & $-0.009$ & $-0.006$ \\
$a_{9,\rm II}/\alpha$ & $\phantom{-}0.112$ & $\phantom{-}0.080$
 & $\phantom{-}0.060$ \\
$a_{10,\rm II}/\alpha$ & $-0.221(42)[25]$ & $-0.182(42)[20]$
 & $-0.157(42)[17]$ \\
\hline\hline
\end{tabular}
\end{center}
\end{table}

We have argued in Section~\ref{subsec:airesults} that the description 
of the hard-scattering contributions to the coefficients $a_i$ suffers 
from large theoretical uncertainties, which however can be parameterized 
in terms of a single (complex) quantity $H_{\pi K}$ defined in 
(\ref{Hiexpr}). If this quantity is fixed, then the hard-scattering 
contributions $a_{i,\rm II}$ can be calculated with relatively small 
uncertainties. In Table~\ref{tab:aiII}, we show the results for these 
coefficients obtained by keeping $H_{\pi K}$ fixed at its central value 
(using central values for all input parameters and setting 
$X_H=\ln(m_B/\Lambda_h)\approx 2.4$). Because the hard-scattering terms 
arise first at order $\alpha_s$, they exhibit a relatively strong scale 
dependence. In addition, the coefficients $a_1$, $a_4$ and $a_{10}$ have 
some dependence on the Gegenbauer moments and on the value of the ratio 
$R_{\pi K}$. Although the twist-3 correction is sizeable, it does not 
dominate the result for the hard spectator term. With our default value 
for $X_H$ we obtain an enhancement of the leading twist-2 term by about 
40\%.

Comparison of the results for $a_{i,\rm I}$ and $a_{i,\rm II}$ in
Tables~\ref{tab:ai} and \ref{tab:aiII} shows that in most cases the hard 
spectator terms are of a similar magnitude as the vertex corrections. 
Notable exceptions are the coefficients $a_2$ and $a_{10}$, for which 
the hard-scattering contributions are the dominant effects. The 
predictions for these coefficients are correspondingly uncertain. On the 
other hand, the hard spectator contributions are very small (or absent) 
in the case of the coefficients $a_4$, $a_6$ and $a_8$.

\begin{figure}[t]
\epsfxsize=7cm
\centerline{\epsffile{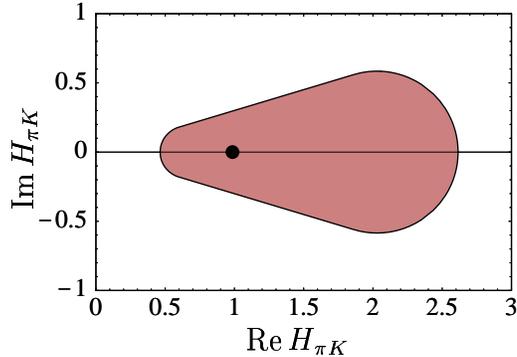}}
\centerline{\parbox{14cm}{\caption{\label{fig:Hpik} 
Ranges for the complex parameter $H_{\pi K}$. The dot shows the default 
value used in obtaining the results in Table~\protect\ref{tab:aiII}.}}}
\end{figure}

So far we have ignored the large overall uncertainty in the 
hard-scattering terms resulting from the uncertainty in the value of 
$H_{\pi K}$. For an estimate of this quantity, we parameterize the 
divergent integral $X_H$ in (\ref{Xdef}) in the form
\begin{equation}\label{XHparam}
   X_H = \left( 1 + \varrho_H\,e^{i\varphi_H} \right)
   \ln\frac{m_B}{\Lambda_h} \,; \qquad
   \varrho_H \le 1
\end{equation}
with an arbitrary phase $\varphi_H$, which may be caused by soft 
rescattering. In other words, we assign a 100\% uncertainty to the 
``default value'' $X_H=\ln(m_M/\Lambda_h)\approx 2.4$. If, in addition,
the parameters $\lambda_B$, $f_B$ and $F_0^{B\to\pi}(0)$ are varied
within the ranges shown in Table~\ref{tab:inputs}, the result for
$H_{\pi K}$ is confined to the interior of a region in the complex 
plane shown in Figure~\ref{fig:Hpik}. (Variations of the Gegenbauer 
moments or the parameter $R_{\pi K}$ have a minor effect and can be 
safely neglected in this plot.) The obtained values for $H_{\pi K}$ are 
of order unity, but with an uncertainty of at least a factor 2 and a 
potentially significant strong-interaction phase (of up to about 
$\pm 17^\circ$ with our choice of parameters).

\subsection{Annihilation contributions}
\label{subsec:annihilation_numerics}

As emphasized earlier, the results for the weak annihilation 
contributions derived in Section~\ref{subsec:annihilation} are based on
the assumption of hard scattering, which is invalidated by the presence 
of endpoint singularities. Nevertheless, eqs.~(\ref{XAdef}) and 
(\ref{bidef}) can be employed as a model for the annihilation terms, 
which we expect to give the correct order of magnitude of the effects. 
In analogy with the previous section, we parameterize the divergent 
integral $X_A$ in the form
\begin{equation}\label{XAparam}
   X_A = \left( 1 + \varrho_A\,e^{i\varphi_A} \right)
   \ln\frac{m_B}{\Lambda_h} \,; \qquad
   \varrho_A\le 1
\end{equation}
with an arbitrary phase $\varphi_A$. Table~\ref{tab:bi} shows the 
results for the annihilation contributions obtained with the default 
value $X_A=\ln(m_B/\Lambda_h)$. They have an overall uncertainty of 
about 30\% due to the error in the value of the ratio 
$r_A=(3.0\pm 0.9)\cdot 10^{-3}$ defined in (\ref{rAdef}). 

\begin{table}[t]
\centerline{\parbox{14cm}{\caption{\label{tab:bi}
Annihilation coefficients $r_A b_i^{({\rm EW})}$ for three different 
choices of the renormalization scale and fixed default values of all 
input parameters.}}}
\begin{center}
\begin{tabular}{|c|ccc|}
\hline\hline
$\mu$ & $m_b/2$ & $m_b$ & $2 m_b$ \\
$\mu_h$ & 1.02\,GeV & 1.45\,GeV & 2.05\,GeV \\
\hline
$r_A b_1$ & $\phantom{-}0.025$ & $\phantom{-}0.021$
 & $\phantom{-}0.018$ \\
$r_A b_2$ & $-0.011$ & $-0.008$ & $-0.006$ \\
$r_A b_3$ & $-0.008$ & $-0.006$ & $-0.005$ \\
$r_A b_4$ & $-0.003$ & $-0.002$ & $-0.001$ \\
$r_A b_3^{\rm EW}/\alpha$ & $-0.021$ & $-0.018$ & $-0.016$ \\
$r_A b_4^{\rm EW}/\alpha$ & $\phantom{-}0.014$ & $\phantom{-}0.010$
 & $\phantom{-}0.007$ \\
\hline\hline
\end{tabular}
\end{center}
\end{table}

We observe that the default values for the annihilation contributions 
are rather small, compatible with being first-order power corrections 
of a canonical size. Specifically, from the relations for the amplitude
parameters in (\ref{params}) it follows that for an estimate of the most 
important annihilation effects in $B\to\pi K$ decays we should compare 
$r_A b_3$ with $(a_4^c+r_\chi^K a_6^c)$ (denominator of 
$\varepsilon_{3/2}$, $\varepsilon_T$, $\varepsilon_a$), $r_A(b_2+b_3)$ 
with $(a_4^u+r_\chi^K a_6^u)$ (numerator of $\varepsilon_a$), and
$r_A b_3^{\rm EW}$ with $(a_{10}^c+r_\chi^K a_8^c)$ (numerator of 
$q_C$). Similarly, from (\ref{TPpipi}) it follows that in $B\to\pi\pi$ 
decays we should compare $r_A(b_3+2b_4)$ with $(a_4^c+r_\chi^\pi a_6^c)$
(numerator of $P_{\pi\pi}/T_{\pi\pi}$). In all cases, annihilation 
effects can be neglected in comparison with $a_1\approx 1$. With the 
default values from Table~\ref{tab:bi} the annihilation contributions 
are always a moderate correction of less than 25\% to the leading terms 
obtained from the QCD factorization formula. However, these estimates 
have a large uncertainty.

\begin{figure}[t]
\epsfxsize=14cm
\centerline{\epsffile{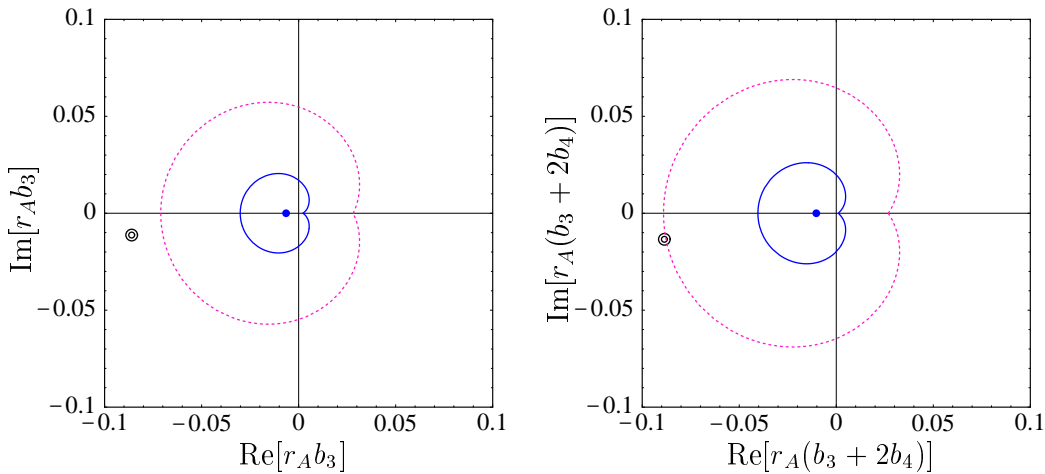}}
\centerline{\parbox{14cm}{\caption{\label{fig:bi2}
Ranges for $r_A b_3$ (left) and $r_A(b_3+2b_4)$ (right) parameterizing, 
respectively, penguin annihilation effects in $B\to\pi K$ and 
$B\to\pi\pi$ decays. Solid lines refer to $\varrho_A=1$, dashed ones to 
$\varrho_A=2$. The dots show the default values. For comparison, the 
central values for the leading penguin coefficients 
$(a_4^c+r_\chi a_6^c)$ are shown by the double circles.}}}
\end{figure}

Phenomenologically most relevant are the penguin annihilation effects
parameterized by $b_3$ and $b_4$. They tend to increase the penguin 
amplitudes $P$ in $B\to\pi K$ decays and $P_{\pi\pi}$ in $B\to\pi\pi$
decays, thereby reducing the values of the tree-to-penguin ratios
$\varepsilon_{3/2}$ and $\varepsilon_T$, and increasing the value of the 
penguin-to-tree ratio $P_{\pi\pi}/T_{\pi\pi}$. The fact that penguin 
annihilation graphs can significantly enhance the penguin amplitude has 
been noted first in \cite{KLS00}. We confirm this effect; however, with
our default parameter variations we find a more moderate enhancement 
than these authors. In order to illustrate the effect and its dependence 
on the value of the quantity $X_A$, we show in Figure~\ref{fig:bi2} the 
combinations $r_A b_3$ and $r_A(b_3+2b_4)$ parameterizing the penguin 
annihilation contributions in $B\to\pi K$ and $B\to\pi\pi$ decays. The 
default values for the leading penguin coefficients 
$(a_4^c+r_\chi a_6^c)$ are shown for comparison. The regions bounded by 
the solid lines refer to our standard choice $\varrho_A\le 1$ in 
(\ref{XAparam}). In this case, the annihilation contribution can 
increase the penguin amplitudes by up to 30--40\%. The dashed curves 
show the accessible parameter space in the more extreme case where we 
let $\varrho_A\le 2$ (corresponding to a 200\% uncertainty in the value 
of the divergent integral $X_A$). Then the annihilation contributions 
can be almost as large as the leading penguin terms. We will see later 
that there are no experimental indications of such large annihilation 
effects.

\subsection{\boldmath Amplitude parameters for $B\to\pi K,\pi\pi$ 
decays\unboldmath}
\label{sec:num}

We are now in a position to combine the results for the parameters
$a_{i,\rm I}$, $a_{i,\rm II}$ and $b_i$ discussed in the previous 
subsections into complete predictions for the decay amplitudes. We 
focus first on the amplitude parameters defined in (\ref{params}). 
They are sufficient to calculate any ratio of 
$B\to\pi K$ decay amplitudes, such as CP asymmetries and ratios of 
CP-averaged branching fractions. We also study the ratio 
$P_{\pi\pi}/T_{\pi\pi}$ in $B\to\pi\pi$ decays defined in 
(\ref{TPpipi}). A more extensive 
phenomenological analysis will be performed in Section~\ref{sec:4}. 

It will be important to distinguish two types of theoretical 
uncertainties: those arising from the variation of input parameters to 
the factorization formula, and those associated with power corrections 
to factorization. Uncertainties of the first kind have a well-defined 
meaning and can, at least in principle, be reduced in a systematic way. 
They include the dependence on the renormalization scale, quark masses, 
moments of light-cone distribution amplitudes, and hadronic quantities 
such as $f_B$ and $F_0^{B\to\pi}(0)$. The errors in the input 
parameters can be reduced, e.g., by using experimental data or lattice 
calculations. The residual dependence on the renormalization scale can 
be reduced by calculating higher-order corrections to the 
hard-scattering kernels. Our predictions also depend on the value of 
$|V_{ub}/V_{cb}|$; however, this should not be considered a theoretical 
uncertainty. Ultimately, the goal is to use hadronic $B$ decays to 
learn about such CKM parameters. 

Theoretical uncertainties related to power corrections to the 
factorization formula are of a different quality. Since factorization
does, in general, not hold beyond leading power, there is no systematic
formalism known that would allow us to analyze power corrections in a
model-independent way. This is a general problem of QCD factorization 
theorems in cases where no operator product expansion can be applied. 
(Another familiar example are event-shape variables in $e^+ e^-$ 
annihilation to hadrons.) In the present work, we have identified 
sources of potentially large power corrections (chirally-enhanced 
contributions and weak annihilation terms) and estimated their effects. 
These estimates are uncertain due to logarithmically divergent endpoint 
contributions, which indicate the dominance of soft gluon exchange. 
Without significant conceptual progress in the understanding of power 
corrections to observables that do not admit a local operator product 
expansion, it will be difficult to reduce these uncertainties in a 
systematic way.

\begin{table}[p]
\centerline{\parbox{14cm}{\caption{\label{tab:pars} 
Predictions for the amplitude parameters including all theoretical 
uncertainties of the first kind (see text), but using default values 
for the power corrections to QCD factorization. The first error on the 
central value is the sum of all theoretical uncertainties added in 
quadrature. The second error (if present) shows the dependence on 
$|V_{ub}/V_{cb}|$. The last column indicates the two most important
contributions to the theoretical uncertainty. For each quantity, the 
second line shows the result without weak annihilation contributions
(except for $q$ and $\omega$, which do not receive annihilation 
terms).}}}
\begin{center}
\begin{tabular}{|c|c|ll|}
\hline\hline
Parameter & Central Value & \multicolumn{2}{c|}{Dominant Errors} \\
\hline
$\varepsilon_{3/2}$~(\%) & 
 $23.9\pm 4.5\pm 4.8$ & $\pm 3.5\,(m_s)$ & $\pm 1.4\,(\mu)$ \\
& $25.7\pm 4.8\pm 5.1$ & $\pm 3.6\,(m_s)$ & $\pm 1.6\,(\alpha_2^K)$ \\
$\phi$~(deg) & 
 $-9.6\pm 3.8$ & $\pm 3.5\,(m_c)$ & $\pm 1.4\,(\alpha_1^K)$ \\
& $-10.2\pm 4.1$ & $\pm 3.7\,(m_c)$ & $\pm 1.5\,(\alpha_1^K)$ \\
\hline
$\varepsilon_T$~(\%) & 
 $20.6\pm 3.5\pm 4.1$ & $\pm 3.2\,(m_s)$ & $\pm 0.9\,(\mu)$ \\
& $22.0\pm 3.6\pm 4.4$ & $\pm 3.3\,(m_s)$ & $\pm 0.8\,(\alpha_2^K)$ \\
$\phi_T$~(deg) & 
 $-5.7\pm 4.4$ & $\pm 3.5\,(m_c)$ & $\mp 2.3\,(\mu)$ \\
& $-6.2\pm 4.6$ & $\pm 3.7\,(m_c)$ & $\mp 2.2\,(\mu)$ \\
\hline
$\varepsilon_a$~(\%) & 
 $2.0\pm 0.1\pm 0.4$ & $\pm 0.1\,(m_c)$ & $\mp 0.1\,(\mu)$ \\
& $1.9\pm 0.1\pm 0.4$ & $\pm 0.1\,(m_c)$ & ~~~ --- \\
$\phi_a$~(deg) & 
 $13.6\pm 4.4$ & $\pm 3.7\,(m_c)$ & $\pm 1.7\,(\alpha_1^K)$ \\
& $16.6\pm 5.2$ & $\pm 3.9\,(m_c)$ & $\mp 2.8\,(\mu)$ \\
\hline
$q$~(\%) & 
 $58.8\pm 6.7\mp 11.8$ & $\pm 6.4\,(R_{\pi K})$ & $\pm 1.3\,(\mu)$ \\
$\omega$~(deg) & 
 $-2.5\pm 2.8$ & $\pm 1.9\,(\mu)$ & $\mp 1.8\,(\alpha_1^K)$ \\
\hline
$q_C$~(\%) & 
 $8.3\pm 4.5\mp 1.7$ & $\mp 2.7\,(\lambda_B)$
 & $\pm 2.3\,(\alpha_1^K)$ \\
& $8.9\pm 4.9\mp 1.8$ & $\mp 3.1\,(\lambda_B)$
 & $\pm 2.3\,(\alpha_1^K)$ \\
$\omega_C$~(deg) & 
 $-60.2\pm 49.5$ & $\pm 31.7\,(\mu)$ & $\mp 27.9\,(\lambda_B)$ \\
& $-54.2\pm 44.2$ & $\pm 29.5\,(\mu)$ & $\mp 24.1\,(\lambda_B)$ \\
\hline
$|P_{\pi\pi}/T_{\pi\pi}|$~(\%) & 
 $28.5\pm 5.1\mp 5.7$ & $\mp 4.6\,(m_s)$ & $\mp 1.8\,(\mu)$ \\
& $25.9\pm 4.3\mp 5.2$ & $\mp 4.1\,(m_s)$ & $\mp 0.8\,(\mu)$ \\
$\mbox{arg}(P_{\pi\pi}/T_{\pi\pi})\!$ & 
 $8.2\pm 3.8$ & $\mp 3.3\,(m_c)$ & $\pm 2.0\,(\mu)$ \\
(deg) &
 $9.0\pm 4.1$ & $\mp 3.6\,(m_c)$ & $\pm 1.8\,(\mu)$ \\
\hline\hline
\end{tabular}
\end{center}
\end{table}

Our results for the amplitude parameters including all theoretical 
uncertainties of the first kind are shown in Table~\ref{tab:pars}. They
are obtained by keeping the parameters $X_H$ and $X_A$ entering the 
power corrections fixed at the default value $X_H=X_A
=\ln(m_B/\Lambda_h)$. All other input parameters are varied within the 
ranges shown in Table~\ref{tab:inputs}. Following common practice, we 
vary the renormalization scale $\mu$ between $m_b/2$ and $2 m_b$. The 
individual contributions to the error are then added in quadrature to 
obtain the total theoretical uncertainty of the first kind. The two 
most important contributions to the total error are shown in the last 
column of the table. The sign convention is such that the upper (lower) 
sign corresponds to increasing (decreasing) the value of an input 
parameter. Finally, the second error on the central value (if present) 
indicates the sensitivity to the uncertainty in the ratio 
$|V_{ub}/V_{cb}|=0.085\pm 0.017$, which we assume to be 20\%. This is 
not a hadronic uncertainty and therefore should not be combined with 
the first error. The amplitude parameters are either proportional (or 
inversely proportional) to $|V_{ub}/V_{cb}|$ or independent of this 
parameter, so that this error can easily be readjusted if needed.

In almost all cases the theoretical uncertainty is dominated by a 
single source. With the exception of the strong-interaction phase
$\omega_C$, our next-to-leading order results for the amplitude 
parameters are very stable under variation of the renormalization scale. 
The uncertainty in the values of the tree-to-penguin ratios 
$\varepsilon_{3/2}$, $\varepsilon_T$, and $|P_{\pi\pi}/T_{\pi\pi}|$ is 
dominated by the error on the strange-quark mass, whereas the 
corresponding strong-interaction phases $\phi$, $\phi_T$, and 
$\mbox{arg}(P_{\pi\pi}/T_{\pi\pi})$, as well as the phase $\phi_a$, are 
most sensitive to the error on the charm-quark mass. The largest 
uncertainty in the electroweak-penguin parameter $q$ comes from the 
SU(3) violations parameterized by the amplitude ratio $R_{\pi K}$ in 
(\ref{ratios}). The uncertainty in the value of the wave-function 
parameter $\lambda_B$ is the dominant source of uncertainty for the 
electroweak-penguin parameter $q_C$. Note that the Gegenbauer moments 
of the pion and kaon light-cone distribution amplitudes are never the 
dominant contribution to the error. This shows that the precise shapes 
of these amplitudes are of minor importance for phenomenological 
applications of QCD factorization.

\begin{figure}[t]
\epsfxsize=16.3cm
\centerline{\epsffile{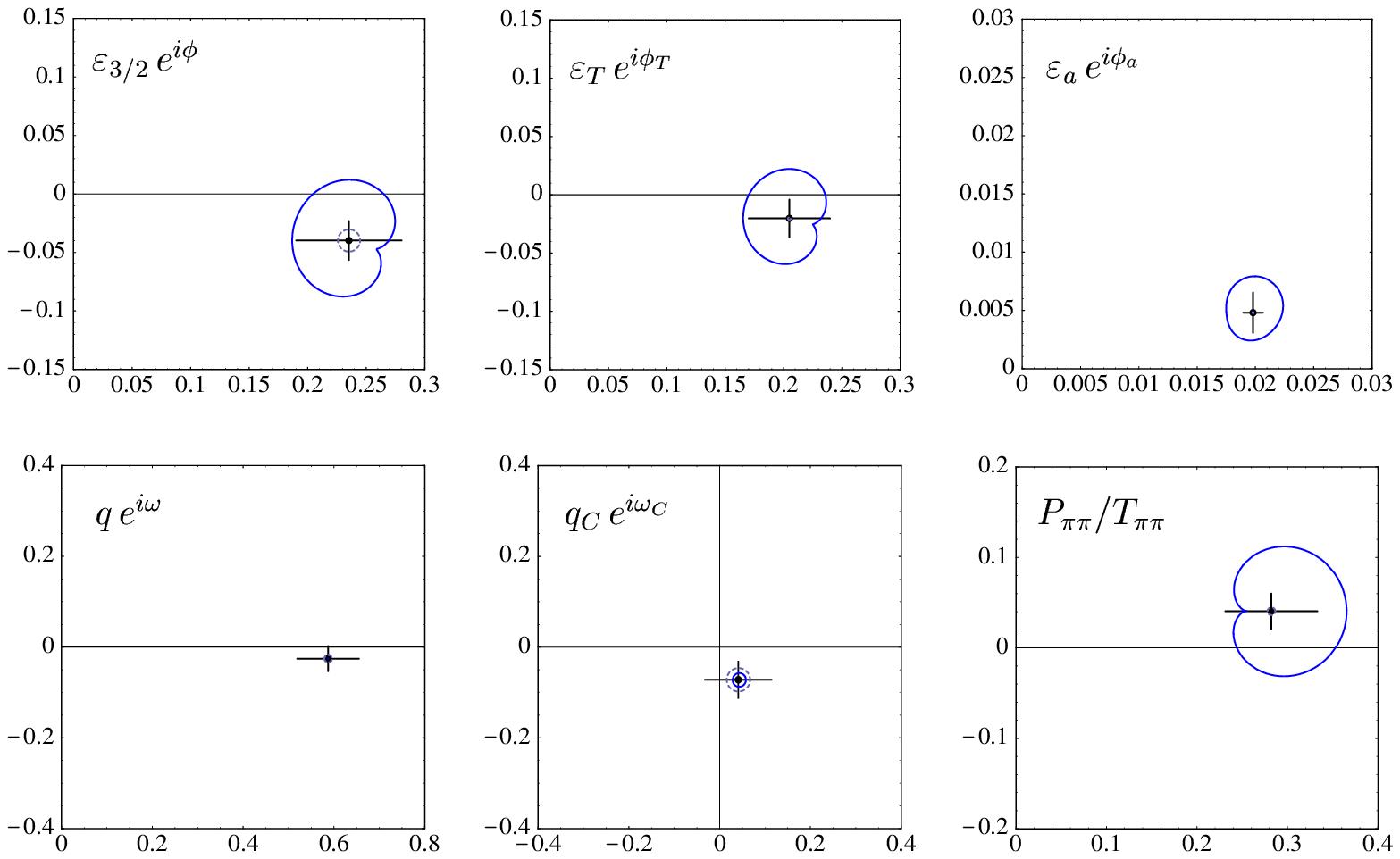}}
\centerline{\parbox{14cm}{\caption{\label{fig:pars}
Results for the amplitude parameters. Dots with error bars show the 
default values and uncertainties of the first kind, while the regions
bounded by the curves show the variation of the central values under
variation of $X_H$ with $\varrho_H=1$ (dashed curves), and variation 
of $X_A$ with $\varrho_A=1$ (solid curves).}}}
\end{figure}

In order to study the theoretical uncertainties of the second kind, 
related to our estimate of potentially large power corrections to the 
factorization formula, we show in Figure~\ref{fig:pars} results for 
amplitude parameters obtained using central values for all input 
parameters, but varying the complex quantities $X_H$ and $X_A$ 
according to the parameterizations (\ref{XHparam}) and (\ref{XAparam}).
The dots with error bars show the central results for the amplitude 
parameters as given in Table~\ref{tab:pars} (for fixed 
$|V_{ub}/V_{cb}|=0.085$). The dashed curves (visible only for 
$\varepsilon_{3/2}$ and $q_C$) bound the parameter space obtained by 
variation of $X_H$, whereas the solid curves (present for all parameters 
except $q$) define the region obtained by variation of $X_A$. It is
evident that the chirally-enhanced twist-3 corrections to the hard 
spectator interactions do not lead to a dominant uncertainty. Their 
effect is always smaller than the uncertainty due to parameter 
variations of the first kind. The uncertainty in the description of weak 
annihilation is significant for the tree-to-penguin ratios 
$\varepsilon_{3/2}$, $\varepsilon_T$, $P_{\pi\pi}/T_{\pi\pi}$ and their 
phases. The resulting variation is typically as large as the total 
theoretical uncertainty of the first kind. Although the modelling of weak 
annihilation thus introduces sizeable uncertainties, we stress that 
their impact on the values of the amplitude parameters is still a
moderate correction (see also Table~\ref{tab:pars}). Moreover, it is 
important that the dominant effect is a universal contribution to the 
leading penguin amplitudes $P$ and $P_{\pi\pi}$, which leads to 
correlations between the various amplitude parameters. Ultimately, this
will help to constrain the annihilation contributions using experimental 
data (see Section~\ref{sec:NR} below).

\section{Phenomenological applications}
\label{sec:4}

This section, which can be read without studying the technical details 
of our analysis, illustrates several applications of our result to the 
phenomenology of $B\to\pi K$ and $B\to\pi\pi$ decays. The main goal 
will be to obtain information about the Wolfenstein parameters 
$\bar\rho$ and $\bar\eta$ defining the unitarity triangle. After some 
general remarks in Section~\ref{sec:intro}, we start by considering 
methods for constraining $\bar\rho$ and $\bar\eta$ that depend on a 
minimal amount of theoretical input about QCD dynamics. The simplest 
such strategy is the Fleischer--Mannel bound \cite{FM97} discussed in 
Section~\ref{sec:FM}. In the following Section~\ref{sec:NR}, we review 
analysis strategies developed by Rosner and one of us 
\cite{NR2,NR1,Mat98}, which are based on CP-averaged rate measurements 
for the charged modes $B^\pm\to\pi K,\pi\pi$. These methods provide 
powerful constraints in the $(\bar\rho,\bar\eta)$ plane with minimal 
theoretical uncertainties. The third method, the determination of 
$\sin2\alpha$ from the time-dependent CP asymmetry in 
$B^0,\bar B^0\to\pi^+\pi^-$ decays, is explored in 
Section~\ref{sec:sin2a}. Towards the end of our discussion we will then 
rely more heavily on the new theoretical results obtained in the present 
work. In Section~\ref{sec:BRs}, we give predictions for the absolute 
values of branching fractions and results for various ratios of 
CP-averaged branching ratios as a function of $\gamma$. We then perform
a global fit to the experimental data on the branching ratios and extract
the corresponding allowed region in the $(\bar\rho,\bar\eta)$ plane. In 
the final Section~\ref{sec:CP}, we show predictions for the direct CP 
asymmetries in the various $B\to\pi K$ and $B\to\pi\pi$ decay modes. 
Because of their sensitivity to strong-interaction phases, these results 
have the largest theoretical uncertainties.

\subsection{General observations}
\label{sec:intro}

Several points following from the numerical analysis in 
Section~\ref{sec:num} are worth repeating here, because they will have 
a direct impact on some of the analysis strategies mentioned below:

\vspace*{-0.3cm}
\paragraph{\rm 1.} 
The rescattering effects parameterized by $\varepsilon_a$ are very small
and not much affected by theoretical uncertainties. Therefore, the decay
amplitude for $B^-\to\pi^-\bar K^0$ in (\ref{para}) is, to a very good
approximation, given by the pure penguin amplitude $P$. This has two 
important consequences: first, it is a safe approximation to neglect
terms of order $\varepsilon_a^2$ in the squared decay amplitudes; 
secondly, it follows that the direct CP asymmetry in the decays 
$B^\pm\to\pi^\pm K^0$, 
\begin{equation}
   A_{\rm CP}(\pi^+ K^0)
   = - 2\varepsilon_a\sin\phi_a\sin\gamma + O(\varepsilon_a^2)
   \approx -1\% \times \sin\gamma \,,
\end{equation}
is tiny and unobservable in the foreseeable future. (We define the CP 
asymmetries as the difference of the $B$-meson minus $\bar B$-meson 
decay rates divided by their sum.) An experimental finding of a sizeable 
asymmetry in this decay mode would have to be interpreted as a sign of 
physics beyond the Standard Model, or as a gross failure of the QCD 
factorization formula, indicating the presence of large, uncontrollable 
power corrections. For completeness, we note that the prediction that
$\varepsilon_a$ is small can be tested experimentally by measuring the
CP-averaged branching ratio for the decays $B^\pm\to K^\pm K^0$ 
\cite{Mat98,Falk}.

\vspace*{-0.3cm}
\paragraph{\rm 2.} 
The strong-interaction phase $\omega=-(2.5\pm 2.8)^\circ$ is accurately 
predicted and tiny, consistent with zero within errors. Also, the 
value of $q$ is not affected by annihilation contributions. These 
observations confirm a theoretical argument presented in \cite{Fl96,NR1},
which uses Fierz identities and top-quark dominance in the electroweak 
penguin diagrams to show that $\omega\approx 0$, and that $q$ can be 
calculated in a model-independent way up to SU(3)-breaking corrections. 
The argument is based on the fact that in the limit of V-spin 
($s\leftrightarrow u$) symmetry the leading electroweak penguin 
contributions parameterized by $q\,e^{i\omega}$ can be related to the 
current--current contributions from the operators $Q_1^u$ and $Q_2^u$ 
using Fierz identities. Neglecting the small contributions from the 
operators $Q_7$, $Q_8$, and $Q_{7\gamma}$, one then obtains
\begin{equation}\label{qsimple}
   q\,e^{i\omega} \simeq 
   - \frac{3}{2\epsilon_{\rm KM}}\,\frac{C_9+C_{10}}{C_1+C_2}
   \simeq \frac{1}{\epsilon_{\rm KM}}\,
   \frac{\alpha}{8\pi}\,\frac{x_t}{\sin^2\!\theta_W}
   \left( 1 + \frac{3\ln x_t}{x_t-1} \right) ,
\end{equation}
where the ratio of Wilson coefficients is renormalization-scheme 
invariant and can thus be evaluated at the electroweak scale. The 
strong-interaction phase $\omega$ vanishes in this approximation. When 
SU(3)-breaking corrections and small electromagnetic contributions 
are included, the above result is rescaled by a factor 
$R_q=(0.84\pm 0.10)\,e^{-i(2.5\pm 2.8)^\circ}$. About half of the 
deviation from 1 is due to (mostly ``factorizable'') SU(3) violations. 
The important point to note is that the smallness of $\omega$ is a 
model-independent result that does not rely on the QCD factorization 
formula. It follows that terms of second order in $\omega$ can be 
safely neglected.

\vspace*{-0.3cm}
\paragraph{\rm 3.} 
The electroweak penguin contributions to the $\bar B^0\to\pi^+ K^-$
decay amplitude, parameterized by $q_C$ in (\ref{para}), are small and 
can also be treated to first order to good approximation. In the 
literature, these effects are sometimes referred to as 
``colour-suppressed'' electroweak penguin contributions.

\vspace*{-0.3cm}
\paragraph{\rm 4.} 
The strong-interaction phases $\phi$, $\phi_T$, and 
$\mbox{arg}(P_{\pi\pi}/T_{\pi\pi})$ are small. We find central values 
of order $10^\circ$ or less in magnitude, with an uncertainty of about 
a factor 2 due to potentially large annihilation contributions and 
higher-order perturbative corrections to the hard-scattering kernels. It 
follows that the cosines of these phases deviate from 1 by only a few 
percent, and the direct CP asymmetries are suppressed by a factor 
$|\sin\phi_i|\sim 0.1$--0.3. We note, in this context, that the 
smallness of the phase difference $\phi_T-\phi\approx 4^\circ$ (see
Table~\ref{tab:pars}) can be understood based on simple physical 
arguments and implies a strong correlation between the direct CP 
asymmetries in the decays $B^\pm\to\pi^0 K^\pm$ and 
$B^0\to\pi^\mp K^\pm$ \cite{GR}.

\vspace*{0.2cm}
\noindent
In the following sections, we explain how these general observations
can be put to work in different analysis strategies. We start with 
those strategies that avoid theoretical input on tree-to-penguin
ratios such as $\varepsilon_{3/2}$ and $\varepsilon_T$. This eliminates
the sensitivity to weak annihilation effects and hence the main 
uncertainty of the QCD factorization approach. As a result, these
strategies are particularly clean from a theoretical point of view.

{\tabcolsep=0.2cm
\begin{table}[t]
\centerline{\parbox{14cm}{\caption{\label{tab:data}
Experimental results for the CP-averaged $B\to\pi K$ and $B\to\pi\pi$ 
branching ratios in units of $10^{-6}$. The BaBar and Belle results
are preliminary. Our averages ignore correlations.}}}
\begin{center}
\begin{tabular}{|l|c|c|c|r|}
\hline\hline
Decay Mode & CLEO \protect\cite{CLEO00} & BaBar \protect\cite{BaBar01}
 & Belle \protect\cite{Belle01} & Average\hspace{0.15cm} \\
\hline
$B^0\to\pi^+\pi^-$ & $4.3_{\,-1.4}^{\,+1.6}\pm 0.5$
 & $4.1\pm 1.0\pm 0.7$ & $5.9_{\,-2.1}^{\,+2.4}\pm 0.5$ 
 & $4.4\pm 0.9$ \\
$B^\pm\to\pi^\pm\pi^0$ & $5.6_{\,-2.3}^{\,+2.6}\pm 1.7$
 & $5.1_{\,-1.8}^{\,+2.0}\pm 0.8$ & $7.1_{\,-3.0\,-1.2}^{\,+3.6\,+0.9}$
 & $5.6\pm 1.5$ \\
 & $< 12.7~(90\%~\mbox{C.L.})$ & $< 9.0~(90\%~\mbox{C.L.})$
 & $< 12.6~(90\%~\mbox{C.L.})$ & \\
\hline
$B^0\to\pi^\mp K^\pm$ & $17.2_{\,-2.4}^{\,+2.5}\pm 1.2$
 & $16.7\pm 1.6_{\,-1.7}^{\,+1.2}$ & $18.7_{\,-3.0}^{\,+3.3}\pm 1.6$
 & $17.2\pm 1.5$ \\
$B^\pm\to\pi^0 K^\pm$ & $11.6_{\,-2.7\,-1.3}^{\,+3.0\,+1.4}$
 & $10.8_{\,-1.9\,-1.2}^{\,+2.1\,+1.0}$
 & $17.0_{\,-3.3\,-2.2}^{\,+3.7\,+2.0}$
 & $12.1\pm 1.7$ \\
$B^\pm\to\pi^\pm K^0$ & $18.2_{\,-4.0}^{\,+4.6}\pm 1.6$
 & $18.2_{\,-3.0\,-2.0}^{\,+3.3\,+1.6}$
 & $13.1_{\,-4.6}^{\,+5.5}\pm 2.6$
 & $17.2\pm 2.5$ \\
$B^0\to\pi^0 K^0$ & $14.6_{\,-5.1\,-3.3}^{\,+5.9\,+2.4}$
 & $8.2_{\,-2.7\,-1.2}^{\,+3.1\,+1.1}$
 & $14.6_{\,-5.1}^{\,+6.1}\pm 2.7$
 & $10.3\pm 2.5$ \\
\hline\hline
\end{tabular}
\end{center}
\end{table}}

Experimental data for the CP-averaged $B\to\pi K$ and $B\to\pi\pi$
branching fractions as reported by several experimental groups are 
collected in Table~\ref{tab:data}. The last column shows our naive 
averages of these results neglecting correlations. We use the average
result for the $B^\pm\to\pi^\pm\pi^0$ branching ratio in despite of the 
fact that the individual measurements of this mode have less than 
$3\sigma$ significance.

\subsection{Fleischer--Mannel bound}
\label{sec:FM}

Based on a few plausible assumptions, Fleischer and Mannel have 
proposed a very simple method for obtaining a bound on the weak phase
$\gamma$ from the measurement of a single ratio 
\begin{equation}
   R = \frac{\tau(B^+)}{\tau(B^0)}\,  
   \frac{\mbox{Br}(B^0\to\pi^- K^+)+\mbox{Br}(\bar B^0\to\pi^+ K^-)}
        {\mbox{Br}(B^+\to\pi^+ K^0)+\mbox{Br}(B^-\to\pi^-\bar K^0)}
\end{equation}
of CP-averaged branching ratios \cite{FM97} (see \cite{Fl98} for more 
sophisticated generalizations of this method). The current value of this
ratio obtained from Table~\ref{tab:data} is $R=1.06\pm 0.18$. Neglecting 
the small rescattering effects parameterized by $\varepsilon_a$ and the 
colour-suppressed electroweak penguin contribution $q_C$ (observations~1
and 3), it follows from (\ref{para}) that
\begin{equation}
\label{fmbound}
   R \simeq 1 - 2\varepsilon_T\cos\phi_T\cos\gamma + \varepsilon_T^2
   \ge \sin^2\gamma \,.
\end{equation}
The bound excludes a region near $|\gamma|=90^\circ$ provided that 
$R<1$. It is valid for any values of $\varepsilon_T$ and the 
strong-interaction phase $\phi_T$ and is thus independent of theoretical 
assumptions about the tree-to-penguin ratio in these decays. 

Our calculations confirm the dynamical assumptions that go into the
derivation of this bound. The neglected terms in the exact formula for 
$R$ are of order $\varepsilon_a\,\varepsilon_T$ or $q_C\,\varepsilon_T$. 
Both are at the few percent level and can be safely neglected until the
experimental error on $R$ is much below 10\%. (More accurately, we find 
that the bound (\ref{fmbound}) is violated by at most 1.5\% and only in 
the region $72^\circ<\gamma<86^\circ$.)

\subsection{Strategies based on charged modes}
\label{sec:NR}

The theoretical analysis of the charged decays $B^\pm\to\pi K$ profits 
from the fact that, with the exception of the strong-interaction phases 
$\phi$ and $\phi_a$, the hadronic parameters entering the 
parameterization of the corresponding decay amplitudes in (\ref{para}) 
can be constrained in the limit of SU(3) symmetry \cite{NR1,Mat98,Fl98}. 
Therefore, the QCD factorization formula is needed only to reduce the 
uncertainties in the estimate of SU(3)-breaking corrections. We have 
already seen above how SU(3) symmetry and Fierz identities help to 
calculate the quantity $q\,e^{i\omega}$ with small theoretical 
uncertainties. In addition, the decay amplitudes for the charged modes 
$B^\pm\to\pi K$ in (\ref{para}) depend on the tree-to-penguin ratio 
$\varepsilon_{3/2}$ and the very small rescattering parameter 
$\varepsilon_a$ (and the corresponding phases $\phi$ and $\phi_a$). It 
is apparent from Figure~\ref{fig:pars} that our prediction for 
$\varepsilon_{3/2}$ has a large uncertainty due to weak annihilation 
contributions as well as parameter variations. The advantage of the 
charged $B^\pm\to\pi K$ modes is that the tree-to-penguin ratio can be 
determined, up to small SU(3) violations, using experimental data. 
Specifically, a certain combination of the parameters 
$\varepsilon_{3/2}$ and $\varepsilon_a$ can be measured by comparing the 
CP-averaged branching fractions for the decays $B^\pm\to\pi^\pm\pi^0$ 
and $B^\pm\to\pi^\pm K^0$. The relation is
\begin{equation}\label{R1}
   \bar\varepsilon_{3/2} =
   \frac{\varepsilon_{3/2}}
    {\sqrt{1+2\varepsilon_a\cos\phi_a\cos\gamma+\varepsilon_a^2}}
   \equiv R_{\rm th}\,\varepsilon_{\rm exp} \,,
\end{equation}
where 
\begin{equation}\label{epsexp}
   \varepsilon_{\rm exp}
   = \tan\theta_C\,\frac{f_K}{f_\pi}\, \left[ 
   \frac{2[\mbox{Br}(B^+\to\pi^+\pi^0)+\mbox{Br}(B^-\to\pi^-\pi^0)]}
        {\mbox{Br}(B^+\to\pi^+ K^0)+\mbox{Br}(B^-\to\pi^-\bar K^0)}
   \right]^{1/2} 
\end{equation}
is an observable, and $R_{\rm th}=1$ in the limit of U-spin 
($d\leftrightarrow s$) symmetry \cite{NR1}. In the theoretical analysis 
of $B^\pm\to\pi K$ decays it is convenient to replace the parameter 
$\varepsilon_{3/2}$ by $\bar\varepsilon_{3/2}$. Since 
$\varepsilon_a=O(\epsilon_{\rm KM})$ is very small (observation~1), both 
quantities take very similar values. The experimental uncertainty in the 
current value $\varepsilon_{\rm exp}=0.223\pm 0.034$ is still large. 
Ultimately, however, the accuracy in the determination of 
$\bar\varepsilon_{3/2}$ is only limited by the theoretical uncertainty 
in the calculation of the SU(3)-breaking corrections to $R_{\rm th}$. 
The QCD factorization approach helps reducing the model dependence in 
this calculation. Neglecting the tiny contributions from electroweak 
penguins,
\begin{equation}\label{R1eq}
   R_{\rm th} = \left|\frac{a_1(\pi K)+R_{\pi K}\,a_2(\pi K)}
         {a_1(\pi\pi)+a_2(\pi\pi)}\right| \,
   \frac{F_0^{B\to\pi}(m_K^2)}{F_0^{B\to\pi}(m_\pi^2)} \,.
\end{equation}
The form-factor ratio can be safely set equal to 1. (Deviations
are of order $(m_K^2-m_\pi^2)/m_B^2\approx 1\%$.) Note that there are
no annihilation contributions to $R_{\rm th}$. If we keep the parameter 
$X_H$ governing the twist-3 contributions to the hard spectator 
interactions fixed vary all other input parameters over their respective 
ranges of uncertainty, we find that $R_{\rm th}=0.98\pm 0.02$. Next, 
if we keep the input parameters fixed but vary $X_H$ as shown in 
(\ref{XHparam}), we find a variation of about $\pm 0.01$. However, since 
our focus here is on SU(3) violations we should be more conservative 
and let the quantities $X_H^K$ and $X_H^\pi$ vary independently. This
gives a larger variation of about $\pm 0.03$. Taken altogether, we 
obtain
\begin{equation}\label{Rthval}
   R_{\rm th} = 0.98\pm 0.05 \,.
\end{equation}
Combining this with the experimental value of $\varepsilon_{\rm exp}$
gives $\bar\varepsilon_{3/2}=0.218\pm 0.034_{\rm exp}\pm 0.011_{\rm th}$.

Our results for the SU(3)-breaking corrections parameterized by 
$R_{\rm th}$ and $R_q$ (the quantity that corrects (\ref{qsimple})) are 
valid up to nonfactorizable corrections that simultaneously violate 
SU(3) symmetry and are power-suppressed in $\Lambda_{\rm QCD}/m_b$. The
potentially most important power corrections are included in the error
estimate. The remaining uncertainties are of order
\begin{equation}\label{parametr}
   O\!\left( \frac{1}{N_c}\cdot\frac{m_s-m_d}{m_b} \right) ,
\end{equation}
and, by naive power counting, should not amount to more than a few
percent.

Whereas our theoretical predictions for the parameter 
$\varepsilon_{3/2}$ were affected by large uncertainties (see 
Figure~\ref{fig:pars}), relations (\ref{R1}) and (\ref{Rthval}) can be 
combined to obtain a much more accurate value for the related parameter 
$\bar\varepsilon_{3/2}$ provided, of course, that the experimental 
value of $\varepsilon_{\rm exp}$ has a small error. This parameter can 
then be used in the phenomenological analysis of $B^\pm\to\pi K$ decays. 
Moreover, the comparison of the so-determined value of this parameter 
with our prediction for the tree-to-penguin ratio $\varepsilon_{3/2}$ 
provides a nontrivial test of the QCD factorization approach, and 
ultimately could help to reduce the uncertainties in our modelling of 
weak annihilation contributions. 

\begin{figure}[t]
\epsfxsize=7cm
\centerline{\epsffile{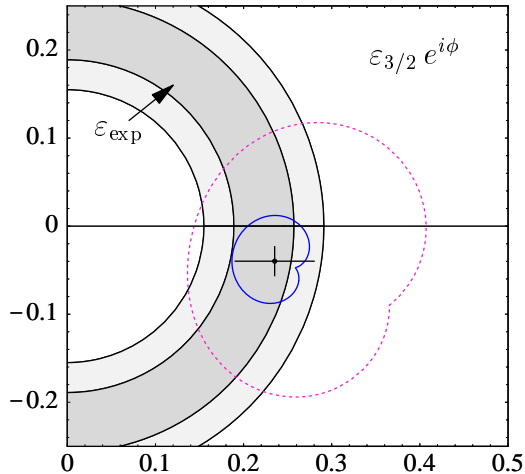}}
\centerline{\parbox{14cm}{\caption{\label{fig:eps32}
Comparison of the prediction for the tree-to-penguin ratio
$\varepsilon_{3/2}$ in $B^\pm\to\pi K$ decays with the experimental 
value (at $1\sigma$ and $2\sigma$) of the quantity 
$\varepsilon_{\rm exp}$ defined in (\protect\ref{epsexp}). The region 
bounded by the solid line refers to our default annihilation model with 
$\varrho_A=1$. The dashed line corresponds to $\varrho_A=2$.}}}
\end{figure}

To illustrate this last point, we show in Figure~\ref{fig:eps32} our
prediction for $\varepsilon_{3/2}\,e^{i\phi}$ (corresponding to the
first plot in Figure~\ref{fig:pars}) and underlay as a gray band the 
experimental value for $\varepsilon_{\rm exp}$ with its $1\sigma$ and
$2\sigma$ errors. (The two quantities should agree if we neglect the 
small deviation of $R_{\rm th}$ from 1, and the tiny contribution of 
$\varepsilon_a$ in (\ref{R1}).) It is pleasing that the central 
experimental value is in excellent agreement with our theoretical 
prediction. The ``smallness'' of the tree-to-penguin ratio in 
$B\to\pi K$ decays has often been interpreted as evidence for large, 
nonfactorizable contributions to the decay amplitudes. Here we find that 
this effect is reproduced in the QCD factorization approach without any 
tuning of parameters. This is an important result. A deviation of our 
prediction from the experimental result could have been considered as an 
indication of large corrections to QCD factorization, such as enhanced 
weak annihilation effects not covered by our simple model estimates. As 
an example, the dashed curve in the figure shows the allowed region 
obtained by increasing the value of the parameter $\varrho_A$ in 
(\ref{XAparam}) from 1 to 2. Fortunately, the data provide no evidence 
for the existence of such large deviations from our central
prediction. (However, the evidence from $\varepsilon_{\rm exp}$ alone 
does also not exclude a potentially large annihilation contribution, 
if it has a large phase.)

\subsubsection*{Model-independent bound on \boldmath$\gamma$\unboldmath}

A key observable in the study of the weak phase $\gamma$ is the ratio 
of the CP-averaged branching ratios in the two $B^\pm\to\pi K$ decay 
modes, defined as
\begin{equation}\label{Rst}
   R_* = 
   \frac{\mbox{Br}(B^+\to\pi^+ K^0)+\mbox{Br}(B^-\to\pi^-\bar K^0)}
        {2[\mbox{Br}(B^+\to\pi^0 K^+)+\mbox{Br}(B^-\to\pi^0 K^+)]}
   \,.
\end{equation}
Its current value is $R_*=0.71\pm 0.14$. The theoretical expression for 
this ratio obtained using the parameterization in (\ref{para}) 
is 
\begin{eqnarray}\label{expr}
   R_*^{-1} &=& 1 + 2\bar\varepsilon_{3/2}\cos\phi\,(q-\cos\gamma)
    + \bar\varepsilon_{3/2}^2 (1 - 2q\cos\gamma + q^2) \nonumber\\
   &&\mbox{}- 2\bar\varepsilon_{3/2}\,\varepsilon_a
    \Big[ \sin^2\!\gamma\cos\phi\cos\phi_a
    + (1-q\cos\gamma) \sin\phi\sin\phi_a \Big] \nonumber\\
   &&\mbox{}- 2\bar\varepsilon_{3/2}\,q\sin\phi\sin\omega
    + O(\varepsilon_a^2,\omega^2,\varepsilon_a\,\omega) \,.
\end{eqnarray}
(The parameter $q$ is also called $\delta_{\rm EW}$ in the literature 
on this ratio.) Note that the rescattering effects described by 
$\varepsilon_a$, as well as the terms linear in the small phase 
$\omega$, are suppressed by a factor of $\bar\varepsilon_{3/2}$ and 
thus reduced to the percent level. Higher-order terms in these small 
parameters can be safely neglected (observations~1 and 2). 

From a measurement of the ratio $R_*$ alone a bound on $\cos\gamma$ can 
be derived, which for $R_*<1$ implies a nontrivial constraint on the 
Wolfenstein parameters $\bar\rho$ and $\bar\eta$ \cite{NR1}. Only 
CP-averaged branching ratios are needed for this purpose. The idea is 
to allow the strong-interaction phases $\phi$ and $\phi_a$ in 
(\ref{expr}) to take any value between 0 and $2\pi$. (Of course, the 
QCD factorization approach predicts rather restricted ranges for these 
strong-interaction phases. For the moment, however, we will not make 
use of these predictions.) This gives \cite{NR1,Mat98}
\begin{equation}\label{Rstbound}
   R_*^{-1} < \left( 1 + \bar\varepsilon_{3/2}\,|q-\cos\gamma| \right)^2
   + \bar\varepsilon_{3/2} (\bar\varepsilon_{3/2} + 2\varepsilon_a)
   \sin^2\!\gamma + O(\varepsilon_a^2,\varepsilon_a\,\omega,\omega^2) \,.
\end{equation}
Note that there is no term linear in $\omega$ on the right-hand side. 
Provided $R_*$ is significantly smaller than 1, the bound implies an 
exclusion region for $\cos\gamma$, which becomes larger the smaller the 
values of $R_*$ and $\bar\varepsilon_{3/2}$ turn out to be. The effect 
of the rescattering contribution proportional to $\varepsilon_a$ on the
right-hand side of the bound is numerically very small. 

\begin{figure}[t]
\epsfxsize=14cm
\centerline{\epsffile{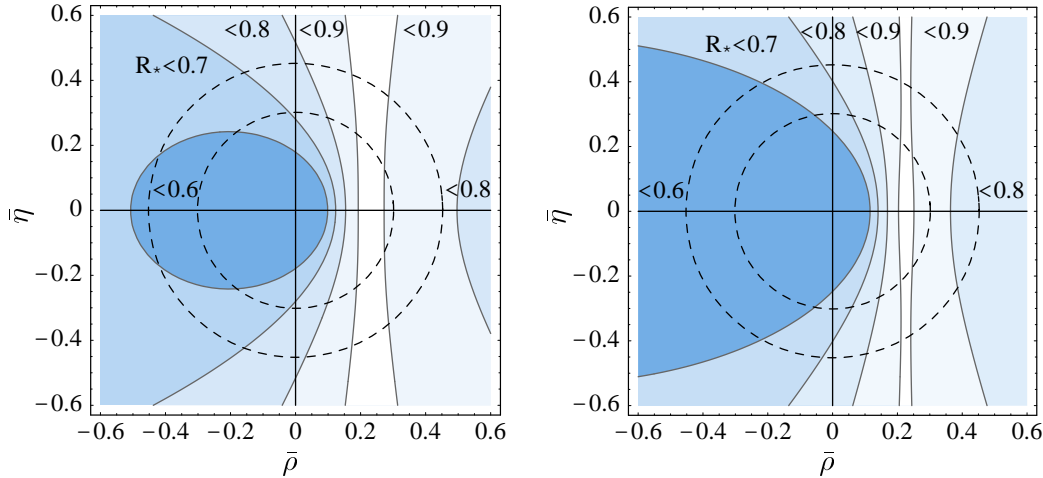}}
\centerline{\parbox{14cm}{\caption{\label{fig:CKMbound}
Theoretical constraints on the Wolfenstein parameters 
$(\bar\rho,\bar\eta)$ implied by (still hypothetical) experimental 
upper bounds on the ratios $R_*$ and $\varepsilon_{\rm exp}$. The
left-hand plot refers to $\varepsilon_{\rm exp}<0.19$, the right-hand
one to $\varepsilon_{\rm exp}<0.25$. For a given upper bound on $R_*$, 
$\bar\rho$ and $\bar\eta$ are restricted to lie inside the corresponding
shaded region. The dashed circles show the allowed region implied by
the measurement of $|V_{ub}/V_{cb}|$ in semileptonic $B$ decays.}}}
\end{figure}

For fixed value of $q$, eq.~(\ref{Rstbound}) excludes a region in
$\cos\gamma$ provided that $R_*<1$. However, we should take into account
that the value of $q$ itself depends on the Wolfenstein parameters 
$\bar\rho$ and $\bar\eta$. Hence, it is more appropriate to display the
relation (\ref{Rstbound}) as a constraint in the $(\bar\rho,\bar\eta)$ 
plane. We use
\begin{equation}
   \cos\gamma = \frac{\bar\rho}{\sqrt{\bar\rho^2+\bar\eta^2}} \,, \qquad
   q = \frac{\hat q}{R_b} 
   = \frac{0.222\pm 0.025}{\sqrt{\bar\rho^2+\bar\eta^2}} \,,
\end{equation}
where the numerical value $\hat q=q\,R_b=0.222\pm 0.025$ corresponds to 
$|V_{ub}/V_{cb}|=0.085\pm 0.017$. The experimental inputs to the bound 
are the measured ratios $R_*$ and $\varepsilon_{\rm exp}$ of CP-averaged 
branching ratios. The theoretical inputs are the value of $\hat q$, the
parameter $R_{\rm th}$ in the relation $\bar\varepsilon_{3/2}
=R_{\rm th}\,\varepsilon_{\rm exp}$, and a value for the rescattering 
parameter $\varepsilon_a$. The accuracy of our predictions for $\hat q$ 
and $R_{\rm th}$ is intrinsically limited only by effects that have a 
strong parametric suppression, as shown in (\ref{parametr}). As far 
as $\varepsilon_a$ is concerned, the bound becomes weaker the larger the 
value of $\varepsilon_a$. In our analysis we take $\varepsilon_a<0.04$, 
corresponding to an upper bound that is twice as large as predicted by 
the QCD factorization approach. 

Figure~\ref{fig:CKMbound} illustrates the resulting constraint in the 
$(\bar\rho,\bar\eta)$ plane obtained for some representative upper 
bounds on $R_*$ and $\varepsilon_{\rm exp}$. For comparison, the dashed 
circles show the constraint arising from the measurement of the ratio 
$|V_{ub}/V_{cb}|$ in semileptonic $B$ decays. (Note that the information 
from kaon CP violation excludes $\bar\eta<0$ in the Standard Model.) It 
is evident that, depending on the values of $R_*$ and 
$\varepsilon_{\rm exp}$, the constraint may be very nontrivial. If 
$R_*<0.7$, then only values $|\gamma|>90^\circ$ are allowed, which are 
significantly larger than those favoured by the global analysis of the 
unitarity triangle (see \cite{UT1,UT2,UT3,Hoecker} for some recent 
discussions of the standard analysis). For yet smaller values $R_*<0.6$, 
the arising constraint would become inconsistent with the global 
analysis. This would be an indication of some new flavour physics beyond 
the Standard Model. On the other hand, if $R_*>0.9$ then the excluded 
region is too small (or absent) to be of phenomenological significance. 
The present uncertainty in the value of $R_*$ is too large to tell which 
of these possibilities is realized.

\subsubsection*{\boldmath Determination of $\gamma$ and constraint in
the $(\bar\rho,\bar\eta)$ plane\unboldmath}
\label{sec:NRdet}

Ultimately, the goal is of course not only to derive a bound in the
$(\bar\rho,\bar\eta)$ plane, but to determine the Wolfenstein parameters
(and thus the unitarity triangle). This requires obtaining information 
about the strong-interaction phase $\phi$ in (\ref{expr}), which can 
be achieved either through the measurement of a CP asymmetry or with the 
help of theory. A strategy for a model-independent determination of 
$\gamma$ from $B^\pm\to\pi K,\pi\pi$ decays has been suggested in 
\cite{NR2}. It generalizes a method proposed by Gronau, Rosner and 
London \cite{GRL} to include the effects of electroweak penguins. The 
approach has later been refined to account for rescattering 
contributions to the $B^\pm\to\pi^\pm K^0$ decay amplitudes \cite{Mat98}.
However, this method relies on the measurement of a direct CP asymmetry 
in addition to $R_*$ and hence requires very high statistics. Here, we 
suggest an easier strategy for a theory-guided determination of 
$\bar\rho$ and $\bar\eta$, which does not require the measurement of 
a CP asymmetry. Instead, we will exploit the prediction of the QCD 
factorization approach that the strong-interaction phase $\phi$ is small,
i.e., 
\begin{equation}\label{phiest}
   \sin\phi = O[\alpha_s(m_b),\Lambda_{\rm QCD}/m_b] \,.
\end{equation}
This implies that the deviation of $\cos\phi$ from 1 is a second-order 
effect in $\alpha_s(m_b)$ and/or $\Lambda_{\rm QCD}/m_b$. More 
specifically, we found that $\phi\approx-(10\pm 15)^\circ$, where the 
uncertainty is dominated by the weak annihilation contribution. Taking 
this value literally we would conclude that $\cos\phi>0.9$; however, 
to be conservative we allow for a larger phase such that $\cos\phi>0.8$. 
(This prediction can be tested experimentally once the direct CP 
asymmetry in the decays $B^\pm\to\pi^0 K^\pm$ has been measured 
\cite{NR2,Mat98}.) With the help of this result, a measurement of the 
ratio $R_*$ can be used to obtain a narrow allowed region in the 
$(\bar\rho,\bar\eta)$ plane, which for fixed value of $|V_{ub}/V_{cb}|$ 
corresponds to a determination of $\gamma$ that is unique up to a sign. 
To this end, we rewrite (\ref{expr}) as
\begin{equation}\label{gamth}
   \cos\gamma = q
   + \frac{1-R_*^{-1} + \bar\varepsilon_{3/2}^2 (1-q^2)
           - 2\bar\varepsilon_{3/2}\,\delta}
          {2\bar\varepsilon_{3/2}(\cos\phi + \bar\varepsilon_{3/2}\,q)} 
   + O(\varepsilon_a^2,\varepsilon_a\,\omega,\omega^2) \,,
\end{equation}
where 
\begin{equation}
   \delta = \varepsilon_a \Big[ \sin^2\!\gamma\cos\phi\cos\phi_a 
   + (1-q\cos\gamma) \sin\phi\sin\phi_a \Big]
   + q\sin\phi\sin\omega \,.
\end{equation}
It is safe to take $0<\delta<0.05$ and treat this as an independent
parameter. Then, in addition to $\hat q$ and $R_{\rm th}$, we must 
specify a range for $\cos\phi$. 

\begin{figure}[t]
\epsfxsize=14cm
\centerline{\epsffile{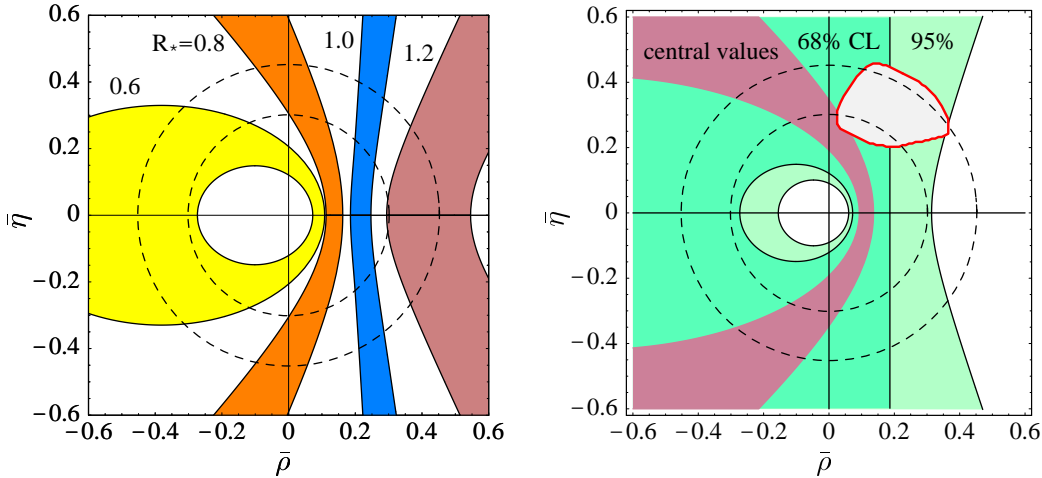}}
\centerline{\parbox{14cm}{\caption{\label{fig:CKMfit}
Left: Allowed regions in the $(\bar\rho,\bar\eta)$ plane corresponding 
to $\varepsilon_{\rm exp}=0.22$ and different values of $R_*$ as 
indicated. The widths of the bands reflect the total theoretical 
uncertainty, obtained by scanning the input parameters inside the ranges 
specified in the text. Right: Allowed regions at 68\% and 95\% 
confidence level obtained from the current experimental results on the 
branching ratios. The dark band shows the theoretical uncertainty for 
the central experimental values. The light gray area is the allowed 
region obtained from the standard global fit of the unitarity triangle
\protect\cite{Hoecker}.}}}
\end{figure}

In evaluating the result (\ref{gamth}), we scan the theory input
parameters in the ranges $0.197<\hat q<0.247$, $0.93<R_{\rm th}<1.03$, 
$0<\delta<0.05$, and $0.8<\cos\phi<1$ (corresponding to 
$|\phi|<37^\circ$). The left-hand plot in Figure~\ref{fig:CKMfit} shows 
the allowed regions in the $(\bar\rho,\bar\eta)$ plane obtained for 
some representative values $R_*$ and the current central value 
$\varepsilon_{\rm exp}=0.22$. We stress that with this 
method a useful constraint on the Wolfenstein parameters is obtained 
for any value of these parameters. When combined with a measurement of 
$|V_{ub}/V_{cb}|$, this determines the weak phase $\gamma$ up to a sign
ambiguity. Note that the theoretical accuracy of the method is high,
especially if the ratio $R_*$ turns out to be close to 1, corresponding 
to a value of $\gamma$ as suggested by the global analysis of the 
unitarity triangle. In that case the theoretical uncertainty in the
determination of $\gamma$ is about $10^\circ$ (for fixed value of 
$|V_{ub}/V_{cb}|$). Also, the resulting constraint is then very weakly
dependent on the value of the parameter $\varepsilon_{\rm exp}$. We 
stress that the width of the band in the $(\bar\rho,\bar\eta)$ 
plane corresponding to $R_*=1$ is narrower than the widths of the 
theoretical error bands corresponding to the ``standard'' constraints 
on the unitarity triangle derived from charmless semileptonic $B$ decays,
$B$--$\bar B$ mixing, and CP violation in $K$--$\bar K$ mixing. Only the
measurement of $\sin2\beta$ is theoretically cleaner. The uncertainty 
in the extraction of $\gamma$ increases as one considers values of $R_*$ 
significantly less than 0.8 or larger than 1.2. However, such values 
would be inconsistent with the global analysis of the unitarity triangle 
\cite{UT1,UT2,UT3,Hoecker} and thus provide evidence for physics beyond 
the Standard Model.

With the current values of the branching ratios collected in 
Table~\ref{tab:data}, the method proposed here is at the verge of 
providing a useful constraint in the $(\bar\rho,\bar\eta)$ plane. This 
is shown in the right-hand plot in Figure~\ref{fig:CKMfit}, where we 
indicate the resulting allowed regions at 68\% and 95\% confidence level 
and compare them with the allowed region (light gray area) obtained from the 
standard global fit of the unitarity triangle. Here and below we use the
most recent result for the standard fit obtained in \cite{Hoecker}, which
includes the measurements of $\sin2\beta$ at the $B$-factories and 
adopts a conservative treatment of theoretical uncertainties that is
similar in spirit to the one adopted here. In evaluating the $1\sigma$ 
and $2\sigma$ domains of the quantities $R_*$ and 
$\varepsilon_{\rm exp}$, we take into account the correlation implied by 
the fact that the $B^\pm\to\pi^\pm K^0$ branching ratio enters both 
quantities. Specifically, we vary the branching fractions for 
$B^\pm\to\pi^\pm\pi^0$, $\pi^\pm K^0$ and $\pi^0 K^\pm$ independently 
such that $\chi^2\le 1$ (for $1\sigma$) or 4 (for $2\sigma$). In this
way we find that the minimum and maximum values of $R_*$ at 95\% 
confidence level are 0.47 and 1.07, respectively. If in the future the 
upper value can be reduced, the resulting allowed region in the 
$(\bar\rho,\bar\eta)$ plane will no longer fully overlap with the 
standard domain.

\subsection{\boldmath Determination of $\sin 2\alpha$ from 
$B\to\pi^+\pi^-$ decays\unboldmath}
\label{sec:sin2a}

The methods described so far in this section provide constraints in the
$(\bar\rho,\bar\eta)$ plane that, in essence, correspond to a 
determination of the magnitude of $\gamma=\mbox{arg}(V_{ub}^*)$. 
Independent information about the unitarity triangle can be obtained 
from a measurement of the time-dependent CP asymmetry in the decays
$B^0,\bar B^0\to\pi^+\pi^-$, which is sensitive to the $B_d$--$\bar B_d$
mixing phase $e^{-2i\beta}$. We define
\begin{eqnarray}
   A_{\rm CP}^{\pi\pi}(t)
   &=& \frac{\mbox{Br}(B^0(t)\to\pi^+\pi^-)
             - \mbox{Br}(\bar B^0(t)\to\pi^+\pi^-)}
            {\mbox{Br}(B^0(t)\to\pi^+\pi^-)
             + \mbox{Br}(\bar B^0(t)\to\pi^+\pi^-)} \nonumber\\
   &=& - S_{\pi\pi} \sin(\Delta m_B\,t)
    + C_{\pi\pi} \cos(\Delta m_B\,t) \,,
\end{eqnarray}
where
\begin{equation}\label{Spipi}
   S_{\pi\pi} = \frac{2\,\mbox{Im}\,\lambda_{\pi\pi}}
                       {1+|\lambda_{\pi\pi}|^2} \,, \quad
   C_{\pi\pi} = \frac{1-|\lambda_{\pi\pi}|^2}
                       {1+|\lambda_{\pi\pi}|^2} \,, \quad
   \lambda_{\pi\pi} = e^{-2i\beta}\,
    \frac{e^{-i\gamma} + P_{\pi\pi}/T_{\pi\pi}}
         {e^{i\gamma} + P_{\pi\pi}/T_{\pi\pi}} \,.
\end{equation}
The coefficient $C_{\pi\pi}$, which is a function of the weak phase 
$\gamma$, coincides with the direct CP asymmetry to be discussed later.
The mixing-induced asymmetry $S_{\pi\pi}$ depends on $\gamma$ and 
$\beta$. In fact, in the limit where $P_{\pi\pi}/T_{\pi\pi}$ is set to 
zero it follows that $\lambda_{\pi\pi}=e^{-2i(\beta+\gamma)}
=e^{2i\alpha}$, and hence $S_{\pi\pi}=\sin2\alpha$. 

\begin{figure}[t]
\epsfxsize=7cm
\centerline{\epsffile{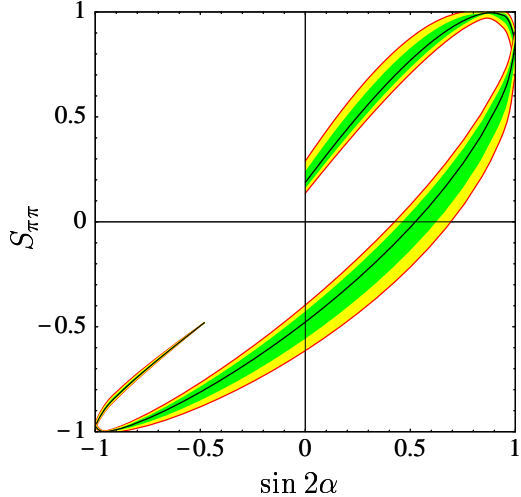}}
\centerline{\parbox{14cm}{\caption{\label{fig:sin2a}
Relation between $\sin2\alpha$ and the mixing-induced CP asymmetry
$S_{\pi\pi}$, assuming $\sin2\beta=0.48$. The dark band reflects 
parameter variations of the first kind, the light band shows the total 
theoretical uncertainty. The lower portion of the band refers to values 
$45^\circ<\alpha<135^\circ$, the upper one to $0<\alpha<45^\circ$ 
(right branch) or $135^\circ<\alpha<180^\circ-\beta$ (left branch).}}}
\end{figure}

When the penguin contributions to the $B\to\pi\pi$ decay amplitudes are
included, the relation between the coefficient $S_{\pi\pi}$ and 
$\sin2\alpha$ receives hadronic corrections \cite{GL90,SiWo94,GQ97,Ch98},
which can be calculated using the QCD factorization approach 
\cite{BBNS1}. To illustrate the effect, we first assume that 
$|V_{ub}/V_{cb}|$ and the weak phase $\beta$ have been determined 
accurately. Then using $\gamma=180^\circ-\alpha-\beta$ the expression 
for $\lambda_{\pi\pi}$ in (\ref{Spipi}) becomes a function of $\alpha$ 
and our prediction for the penguin-to-tree ratio $P_{\pi\pi}/T_{\pi\pi}$.
If we further assume that the unitarity triangle lies in the upper half 
of the $(\bar\rho,\bar\eta)$ plane, then a measurement of $S_{\pi\pi}$ 
determines $\sin2\alpha$ with at most a two-fold discrete ambiguity. 
Figure~\ref{fig:sin2a} shows the relation between the two quantities for 
the particular case where $|V_{ub}/V_{cb}|=0.085$ and 
$\beta=14.3^\circ$, corresponding to $\sin2\beta=0.48$ (the current 
world average is $\sin2\beta=0.48\pm 0.16$ \cite{sin2bref}). The dark 
band shows the theoretical uncertainty due to input parameter variations 
as specified in Table~\ref{tab:inputs}, whereas the light band indicates 
the total theoretical uncertainty including the effects of weak 
annihilation and twist-3 hard spectator interactions. We observe that
for negative values $\sin2\alpha$ as preferred by the global analysis of 
the unitarity triangle \cite{UT2,UT3,Hoecker}, a measurement of the 
coefficient $S_{\pi\pi}$ could be used to determine $\sin2\alpha$ with 
a theoretical uncertainty of about $\pm 0.1$. Interestingly, for such 
values of $\sin2\alpha$ the ``penguin pollution'' effect enhances the 
value of the mixing-induced CP asymmetry, yielding values of 
$S_{\pi\pi}$ between $-0.5$ and $-1$. Such a large asymmetry should be 
relatively easy to observe experimentally.

\begin{figure}[t]
\epsfxsize=14cm
\centerline{\epsffile{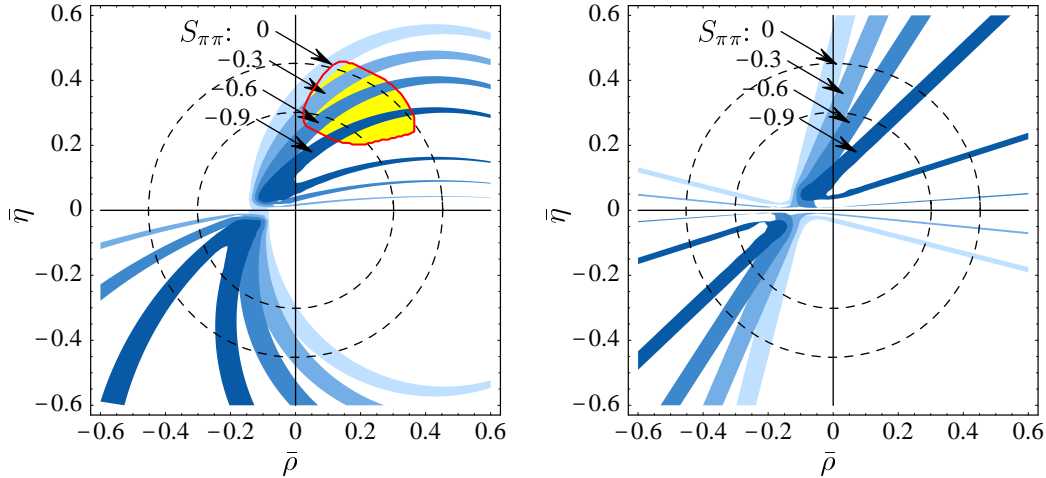}}
\centerline{\parbox{14cm}{\caption{\label{fig:fancy}
Allowed regions in the $(\bar\rho,\bar\eta)$ plane corresponding to
constant values of the mixing-induced asymmetry $S_{\pi\pi}$, assuming
the Standard Model (left), and using a fixed value $\sin2\phi_d=0.48$ 
(and $\cos2\phi_d>0$) for the $B_d$--$\bar B_d$ mixing phase (right). 
The widths of the bands reflect the total theoretical uncertainty. The 
corresponding bands for positive values of $S_{\pi\pi}$ are obtained by 
a reflection about the $\bar\rho$ axis. The light circled area in the
left-hand plot shows the allowed region obtained from the standard 
global fit of the unitarity triangle \protect\cite{Hoecker}.}}}
\end{figure}

Although it illustrates nicely the effect of ``penguin pollution'' on
the determination of $\sin2\alpha$, Figure~\ref{fig:sin2a} is not the
most appropriate way to display the constraint on the unitarity triangle 
implied by a measurement of $S_{\pi\pi}$. In general, there is a 
four-fold discrete ambiguity in the determination of $\sin2\alpha$, 
which we have reduced to a two-fold ambiguity by assuming that the 
triangle lies in the upper half-plane. Next, and more importantly, we 
have assumed that $|V_{ub}/V_{cb}|$ and $\beta$ are known with precision,
whereas $\alpha$ is undetermined. However, in the Standard Model 
$|V_{ub}/V_{cb}|$ and the angles $\alpha,\beta,\gamma$ are all functions 
of the Wolfenstein parameters $\bar\rho$ and $\bar\eta$. It is thus more 
appropriate to represent the constraint implied by a measurement of 
$S_{\pi\pi}$ as a band in the $(\bar\rho,\bar\eta)$ plane. To this end, 
we write
\begin{equation}
   e^{\mp i\gamma}
   = \frac{\bar\rho\mp i\bar\eta}{\sqrt{\bar\rho^2+\bar\eta^2}}
    \,, \quad
   e^{-2i\beta}
   = \frac{(1-\bar\rho)^2-\bar\eta^2-2i\bar\eta(1-\bar\rho)}
          {(1-\bar\rho)^2+\bar\eta^2} \,, \quad
   \frac{P_{\pi\pi}}{T_{\pi\pi}}
   = \frac{r_{\pi\pi}\,e^{i\phi_{\pi\pi}}}
          {\sqrt{\bar\rho^2+\bar\eta^2}} \,,
\end{equation}
where $r_{\pi\pi}\,e^{i\phi_{\pi\pi}}$ parameterizes the large fraction
in (\ref{TPpipi}) without the prefactor $1/R_b$ and is thus independent 
of $\bar\rho$ and $\bar\eta$. We now insert these relations into 
(\ref{Spipi}) and draw contours of constant $S_{\pi\pi}$ in the
$(\bar\rho,\bar\eta)$ plane. The result is shown by the bands in the 
left-hand plot in Figure~\ref{fig:fancy}. The widths of the bands 
reflect the total theoretical uncertainty (including power corrections).
For clarity we show only bands for negative values of $S_{\pi\pi}$; those
corresponding to positive $S_{\pi\pi}$ values can be obtained by a 
reflection about the $\bar\rho$ axis (i.e., $\bar\eta\to-\bar\eta$). Note
that even a rough measurement of $S_{\pi\pi}$ would translate into a
rather narrow band in the $(\bar\rho,\bar\eta)$ plane (this was also
noted in \cite{Gerhard}), which intersects the ring representing the 
$|V_{ub}/V_{cb}|$ constraint at almost right angle. This would therefore
provide a very powerful constraint on the Wolfenstein parameters. We 
also stress that the sign of $S_{\pi\pi}$, when combined with a 
measurement of $|\gamma|$ (using, e.g., the method described in 
Section~\ref{sec:NRdet}), can potentially determine whether the 
unitarity triangle lies in the upper or lower half-plane, and thus 
provide a nontrivial test of the CKM model of CP violation.

The strategy outlined above remains useful even in the hypothetical
case where $B_d$--$\bar B_d$ mixing is affected by new physics beyond
the Standard Model (this scenario has recently be discussed in 
\cite{KaNe00,Joao,Yossi,Xing,Buras2}). Then the factor $e^{-2i\beta}$ 
in the expression for $\lambda_{\pi\pi}$ in (\ref{Spipi}) must be 
replaced with the mixing phase $e^{-2i\phi_d}$, where $\phi_d\ne\beta$ 
due to the presence of new physics. It is the value of $\sin2\phi_d$ 
that is measured in the time-dependent CP asymmetry in the decays 
$B^0,\bar B^0\to J/\psi\,K_S$. From (\ref{Spipi}), we then obtain
\begin{equation}
   \lambda_{\pi\pi}^{\rm NP}
   = \Big[ \pm \sqrt{1-\sin^2\!2\phi_d} - i\sin2\phi_d \Big]\,
   \frac{\bar\rho-i\bar\eta + r_{\pi\pi}\,e^{i\phi_{\pi\pi}}}
        {\bar\rho+i\bar\eta + r_{\pi\pi}\,e^{i\phi_{\pi\pi}}} \,,
\end{equation}
which still implies a constraint in the $(\bar\rho,\bar\eta)$ plane for
each pair of experimental values $S_{\pi\pi}$ and $\sin2\phi_d$. The
$\pm$ sign refers to the sign of $\cos2\phi_d$. The result obtained for 
$\sin2\phi_d=0.48$ is shown in the right-hand plot in 
Figure~\ref{fig:fancy}. In the figure we assume that $\cos2\phi_d>0$; 
the resulting bands for $\cos2\phi_d<0$ are, once again, obtained by a 
reflection about the $\bar\rho$ axis. We observe that the bands in the 
first quadrant of the $(\bar\rho,\bar\eta)$ plane intersect the rings 
from the $|V_{ub}/V_{cb}|$ constraint at almost the same places as in 
the Standard Model case, implying that for $S_{\pi\pi}<1$ (and 
$\cos2\phi_d>0$) potential new physics effects in $B_d$--$\bar B_d$ 
mixing have a minor impact on the results. On the other hand, the impact 
would be significant if $S_{\pi\pi}$ turned out to be positive.

\subsection{Predictions for CP-averaged branching ratios}
\label{sec:BRs}

Whereas so far in this section we have focused on methods that require 
minimal input from the QCD factorization approach, we now discuss in 
detail our theoretical predictions for the $B\to\pi K,\pi\pi$ branching 
ratios. These predictions follow from the theory described in this work
without relying on further phenomenological input. 
All branching fractions discussed in this section are averaged over 
CP-conjugate modes (even though for neutral $K$ and $B$ mesons this is
not indicated by the notation). We use $\tau_{B^+}=1.65$\,ps and 
$\tau_{B^0}=1.56$\,ps for the $B$-meson lifetimes.

\subsubsection*{Absolute predictions for branching fractions}

Two out of the seven decay modes, $B^\pm\to\pi^\pm\pi^0$ and 
$B^\pm\to\pi^\pm K^0$, are (almost) independent of the CKM phase 
$\gamma$, since the corresponding decay amplitude have to a very good 
approximation only a single weak phase. The predicted branching 
fractions for these modes are (setting $\gamma=55^\circ$ for 
concreteness)
\begin{eqnarray}\label{BRs}
   10^6\,\mbox{Br}(B^\mp\to\pi^\mp\pi^0)
   &=& \Big[ 5.3_{\,-0.4}^{\,+0.8}\,(\lambda_B,\alpha_2^\pi)
    \pm 0.3\,(X_H) \Big]\times 
    \left[ \frac{|V_{ub}|}{0.0035}\,\frac{F_0^{B\to\pi}(0)}{0.28}
    \right]^2 \,, \nonumber\\
   10^6\,\mbox{Br}(B^\mp\to\pi^\mp\bar K^0)
   &=& \Big[ 14.1_{\,-4.0}^{\,+6.4}\,(m_s)\,
    \phantom{}_{\,-3.6}^{\,+8.1}\,(X_A) \Big]\times
    \left[ \frac{F_0^{B\to\pi}(0)}{0.28} \right]^2 \,,
\end{eqnarray}
where the first error is due to parameter variations as shown in
Table~\ref{tab:inputs}, whereas the second one accounts for the 
uncertainty due to power corrections from weak annihilation and twist-3 
hard spectator contributions. The dominant contributions to the
uncertainty are shown in parentheses. Note that the decays 
$B^\pm\to\pi^\pm\pi^0$ do not receive weak annihilation contributions.
The $B^\pm\to\pi^\pm K^0$ branching ratio is to a good approximation 
proportional to $(m_s/110\,\mbox{MeV})^{-1.35}$. The largest 
uncertainty for the $\pi^\pm\pi^0$ final state is a $50\%$ 
normalization uncertainty from the current errors on $|V_{ub}|$ and the 
$B\to\pi$ form factor. The systematics of theoretical errors is 
different for the $\pi K$ final states. Their absolute branching 
fractions are (approximately) proportional to the square of the penguin 
amplitude $a_4+r_\chi^K a_6+r_A b_3$, which is sensitive to the penguin 
annihilation coefficient $b_3$ (see the left-hand plot in 
Figure~\ref{fig:bi2}). The dominant theoretical errors therefore come 
from the annihilation parameter $X_A$ and from the strange-quark mass 
(through $r_\chi^K$). The theoretical errors detailed here are 
characteristic for the absolute branching fractions of all $\pi\pi$ and 
$\pi K$ final states, respectively.

The central values in (\ref{BRs}) agree well with the current data 
summarized in Table~\ref{tab:data}. In particular, we find that the 
QCD factorization approach prefers large branching fractions for 
the $\pi K$ final states. This arises due to an enhancement of the 
QCD penguin amplitude relative to naive factorization. Weak annihilation 
contributions play only a minor role in this enhancement. Without 
annihilation the $B^\pm\to\pi^\pm K^0$ branching fraction would be 
reduced to $12.1\times 10^{-6}$. Hence the effect is not negligible, 
but there is no need for a largely enhanced annihilation contribution 
(this is in contrast to the findings of \cite{KLS00}). In fact, the good
agreement of our prediction with the data provides circumstantial 
evidence against the idea that annihilation effects could be much 
enhanced with respect to our estimates. For instance, increasing the 
parameter $\varrho_A$ in (\ref{XAparam}) from 1 to 2 would increase the
corresponding error on the branching ratio from 
$\phantom{}_{\,-3.6}^{\,+8.1}$ to 
$\phantom{}_{\,-\phantom{2}9.2}^{\,+26.4}$, in which case it would 
require considerable fine-tuning of the strong-interaction phase of 
$X_A$ to reproduce the experimental value of the branching ratio.

It has also been suggested in the literature that one needs a large 
enhancement from penguin diagrams with a charm loop (so-called 
``charming penguins'') to understand the overall $\pi K$ branching 
fractions \cite{Ciuc97}. The leading perturbative contribution from 
the penguin diagrams in Figure~\ref{fig:penguin}, however, turns out to 
be very small. There exists also a power-suppressed contribution from 
these diagrams, when one of the quarks the gluon decays into becomes 
soft. In this case, the gluon is ``semi-hard'' and probes the 
charm-quark loop at a scale of order $\mu_h\sim
(\Lambda_{\rm QCD}\,m_b)^{1/2}$. The penguin function $G(s,x)$ in 
(\ref{penfunction2}) tends to a constant for small $x$, which implies 
that the contribution from the semi-hard region to the function $G_K(s)$ 
in (\ref{penfunction1}) is suppressed by two (not one!) powers of the 
heavy-quark mass relative to the leading perturbative contribution. 
Hence, although the strong coupling constant in the semi-hard region is 
larger than $\alpha_s(m_b)$, a large nonperturbative enhancement of the 
charm-penguin contribution appears implausible. (It is possible to 
obtain a first-order power correction by invoking a higher Fock 
component in the wave function of the emission meson, as discussed in 
\cite{BBNS2}. However, in this case the penguin loop continues to be a 
hard subprocess, and so contributes a factor of $\alpha_s(m_b)$.) 
In a recent article \cite{Cetal01} a non-perturbative charming penguin
contribution has been fitted to experimental data under the 
assumption that this effect is responsible for any deviation of 
the measured branching fractions from those expected within the
Standard Model with $\bar{\rho}$ and $\bar{\eta}$ determined by the 
standard unitarity-triangle fit. It it worth noting that such a fit 
is technically equivalent to fitting the annihilation contribution 
to the QCD penguin amplitude, indirectly related to $X_A$ 
[eq.~(\ref{XAparam})] in our notation. If a modification of this 
amplitude were required (for which the present data does not provide 
strong motivation, as will become more evident below), we would 
attribute its physical origin to weak annihilation rather than 
charm penguins, given the power-counting detailed above.

\begin{figure}[p]
\epsfxsize=15.6cm
\centerline{\epsffile{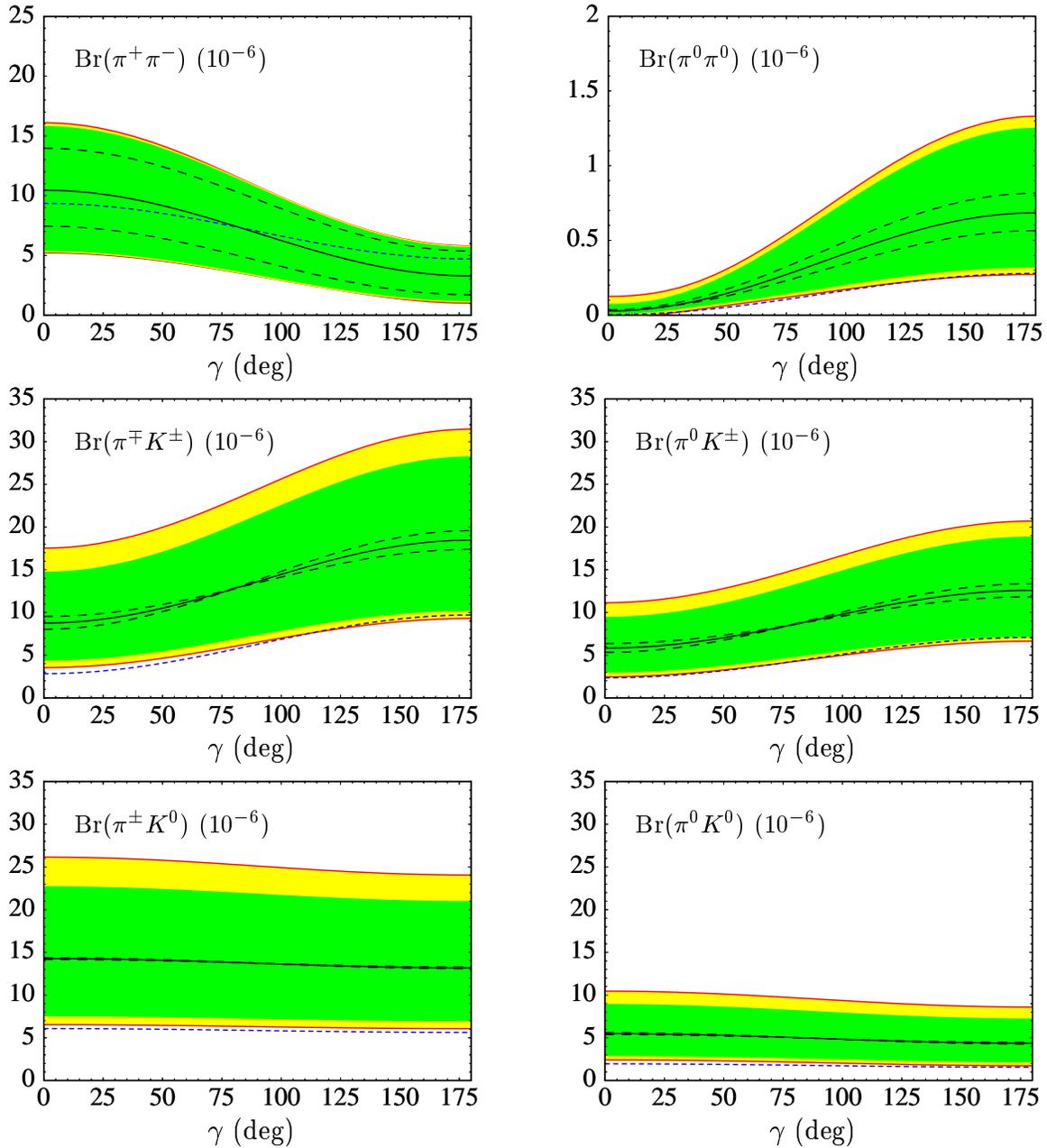}}
\centerline{\parbox{14cm}{\caption{\label{fig:brav1}
Absolute CP-averaged branching fractions as functions of $\gamma$. 
Central values are shown by the solid line. The inner (dark) band 
corresponds to the variation of the theory input parameters (including 
$|V_{ub}/V_{cb}|$), whereas the outer (light) band also includes the 
uncertainty from weak annihilation and twist-3 hard spectator 
contributions (see text for details). The long-dashed lines indicate the
sensitivity to $|V_{ub}/V_{cb}|$. The short-dashed curve shows the 
result obtained using naive factorization.}}}
\end{figure}

In Figure~\ref{fig:brav1}, we show the dependence of the absolute 
branching fractions of the various $B\to\pi K,\pi\pi$ decay modes 
(except for $B^\pm\to\pi^\pm\pi^0$) on the weak phase $\gamma$. In each 
plot, the solid line gives the central prediction of the QCD 
factorization approach at next-to-leading order in $\alpha_s$. For 
comparison, the short-dashed line shows the result at leading order, 
corresponding to naive factorization. The dark-shaded band is obtained 
by varying all input parameters as specified in Table~\ref{tab:inputs}, 
and by varying the renormalization scale between $m_b/2$ and $2 m_b$. 
It also includes the uncertainties due to the errors on the CKM 
parameters $|V_{ub}/V_{cb}|=0.085\pm 0.017$ and 
$|V_{cb}|=0.041\pm 0.003$. The variation due to $|V_{ub}/V_{cb}|$ alone 
is indicated by the long-dashed lines. The light-shaded band adds to 
this the uncertainties inherent to our modelling of power corrections 
due to twist-3 hard spectator and weak annihilation corrections, as 
discussed in Sections~\ref{subsec:hardspec} and 
\ref{subsec:annihilation_numerics}. The annihilation contributions 
dominate the uncertainty in all cases (see Figure~\ref{fig:pars}). They 
imply a considerable uncertainty in the overall normalization of the 
$\pi K$ modes. For the purposes of our discussion here the different 
sources of theoretical uncertainty are added in quadrature. Later, when 
our focus is on constraining CKM parameters, we will be more 
conservative and scan over the entire theory parameter space 
(corresponding to linear addition of errors).

The next-to-leading order effects included in the QCD factorization 
approach significantly enhance the branching fractions for the 
$B\to\pi K$ modes with respect to their values obtained in the naive 
factorization model. No such enhancement occurs for the decays 
$B^0\to\pi^+\pi^-$ and $B^\pm\to\pi^\pm\pi^0$. As a result, the 
empirical finding that $B\to\pi K$ branching ratios are larger than the 
$B\to\pi\pi$ branching ratios is reproduced in our approach without any 
tuning of parameters.  

\subsubsection*{Predictions for ratios of branching fractions}

\begin{table}[t]
\centerline{\parbox{14cm}{\caption{\label{tab:rcps}
Predicted ratios of CP-averaged branching fractions for selected values 
of $\gamma$. The last column shows the experimental values deduced from 
Table~\protect\ref{tab:data}. Our averages ignore correlations between 
the individual measurements.}}}
\begin{center}
\begin{tabular}{|c|cccc|c|}
\hline\hline
Ratio & $40^\circ$ & $70^\circ$ & $100^\circ$ & $130^\circ$ 
& Experiment \\
\hline
&&&&&\\[-0.5cm]
$\frac{2\mbox{\footnotesize Br}(\pi^0 K^\pm)}
      {\mbox{\footnotesize Br}(\pi^\pm K^0)}$
 & $0.94\pm 0.07$ & $1.16\pm 0.07$ & $1.44\pm 0.16$ & $1.70\pm 0.25$ 
 & $1.41\pm 0.29$ \\[0.1cm]
\hline
&&&&&\\[-0.5cm]
$\frac{\mbox{\footnotesize Br}(\pi^\mp K^\pm)}
      {2\mbox{\footnotesize Br}(\pi^0 K^0)}$
 & $0.92\pm 0.08$ & $1.17\pm 0.08$ & $1.50\pm 0.19$ & $1.83\pm 0.34$ 
 & $0.83\pm 0.22$ \\[0.1cm]
\hline
&&&&&\\[-0.5cm]
$\frac{\tau_{B^+}}{\tau_{B^0}}\,
 \frac{\mbox{\footnotesize Br}(\pi^\mp K^\pm)}
      {\mbox{\footnotesize Br}(\pi^\pm K^0)}$
 & $0.74\pm 0.07$ & $0.91\pm 0.04$ & $1.12\pm 0.07$ & $1.32\pm 0.12$ 
 & $1.06\pm 0.18$ \\[0.1cm] 
\hline
&&&&&\\[-0.5cm]
$\frac{\mbox{\footnotesize Br}(\pi^+\pi^-)}
      {\mbox{\footnotesize Br}(\pi^\mp K^\pm)}$
 & $0.96\pm 0.60$ & $0.67\pm 0.38$ & $0.43\pm 0.25$ & $0.27\pm 0.18$ 
 & $0.26\pm 0.06$ \\[0.1cm] 
\hline
&&&&&\\[-0.5cm]
$\frac{\tau_{B^+}}{\tau_{B^0}}\,
 \frac{\mbox{\footnotesize Br}(\pi^+\pi^-)}
      {2\mbox{\footnotesize Br}(\pi^\pm\pi^0)}$
 & $0.96\pm 0.25$ & $0.83\pm 0.20$ & $0.66\pm 0.15$ & $0.50\pm 0.13$ 
 & $0.42\pm 0.14$ \\[0.1cm] 
\hline\hline
\end{tabular}
\end{center}
\end{table}

\begin{figure}[p]
\epsfxsize=15.6cm
\centerline{\epsffile{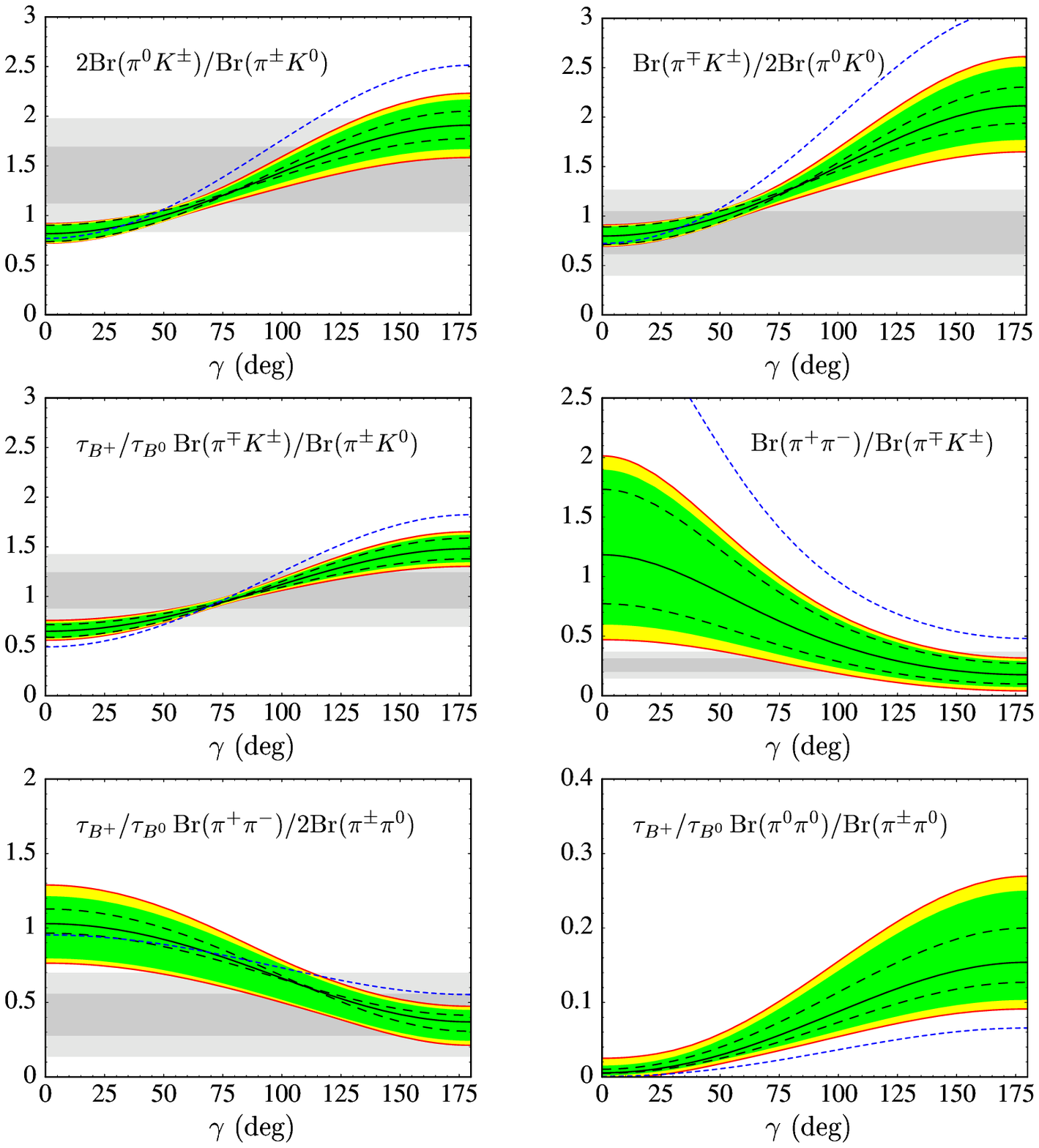}}
\centerline{\parbox{14cm}{\caption{\label{fig:rcp1}
Ratios of CP-averaged branching fractions as functions of $\gamma$. The
meaning of the curves and bands is the same as in 
Figure~\protect\ref{fig:brav1}. The horizontal bands show the $1\sigma$ 
and $2\sigma$ error bands for the current experimental results.}}}
\end{figure}

The uncertainty in the overall magnitude of the penguin amplitude due 
to weak annihilation effects, and the overall scale uncertainty in the 
predictions for the branching fractions due to hadronic form factors 
and CKM elements, are largely eliminated by taking ratios of branching 
fractions. The dependence of the six relevant ratios on the weak phase 
$\gamma$ is displayed in Figure~\ref{fig:rcp1}, in which the curves and 
bands have the same interpretation as in the previous figure. (The 
results presented here are consistent with the preliminary results 
published in \cite{BBNS3}, which did not include annihilation 
contributions. However, the treatment of theoretical uncertainties in 
\cite{BBNS3} differs from the present work.) Table~\ref{tab:rcps} gives 
numerical predictions and theoretical uncertainties for some selected 
values of the angle $\gamma$. It is evident that annihilation effects 
are less important for the ratios. The small error on the ratios 
involving $\pi K$ final states shows the potential of these ratios to 
constrain $\gamma$. Comparing our results with the preliminary 
experimental data collected in Tables~\ref{tab:data} and \ref{tab:rcps} 
(and shown by the horizontal bands), we note a preference for large 
values of $\gamma$; however, the experimental errors are still too large 
to assign much significance to this observation. 

An important remark in this context concerns the ratio involving the 
neutral mode $B^0\to\pi^0 K^0$, which seems to prefer a small value of 
$\gamma$. Our theoretical results indicate that to a good approximation 
(compare the two plots in the first row in Figure~\ref{fig:rcp1})
\begin{equation}\label{isospin}
 \frac{2\mbox{Br}(B^\pm\to \pi^0 K^\pm)}
      {\mbox{Br}(B^\pm\to\pi^\pm K^0)} \approx
 \frac{\mbox{Br}(B^0\to\pi^\mp K^\pm)}
      {2\mbox{Br}(B^0\to\pi^0 K^0)} \,.
\end{equation}
This relation is in slight disagreement (albeit by no more than
$2\sigma$) with present data. Some authors have interpreted this fact 
as an indication of large rescattering phases (see, e.g., \cite{George}).
However, the validity of (\ref{isospin}) is a model-independent 
consequence of isospin symmetry, which is valid to linear order in the 
small tree-to-penguin ratios \cite{Mat98}. Therefore, we expect that 
with more precise data the values of the two ratios in (\ref{isospin})
will come closer to each other.

Finally, we wish to stress that the difference between the QCD 
factorization results and naive factorization are largest for the 
ratio of the $\pi^+\pi^-$ and $\pi^\mp K^\pm$ final states. Whereas 
we can accommodate the low experimental value of this ratio for 
$\gamma>50^\circ$ (at the $2\sigma$ level), this would not be possible
for any value of $\gamma$ if one were to use the naive factorization 
model.

It is evident from Figure~\ref{fig:rcp1} that in many cases the error 
on $|V_{ub}/V_{cb}|$ constitutes one of the largest uncertainties in the
predictions for the ratios of branching fractions. This is true, in 
particular, for the interesting ratio of the $\pi^+\pi^-$ and 
$\pi^\pm K^\mp$ final states. For instance, if $|V_{ub}/V_{cb}|$ is 
20\% lower than its current central value, then our prediction for the 
$B^0\to\pi^+\pi^-$ branching ratio is reduced by 40\%, whereas the 
result for the $B^0\to\pi^\mp K^\pm$ branching fraction remains almost
unaffected. This shows that for future analyses of rare $B$ decays it 
will be important to reduce the theoretical uncertainty in the value
of $|V_{ub}|$.

\begin{figure}[p]
   \vspace*{3.5cm}
   \centerline{\epsffile{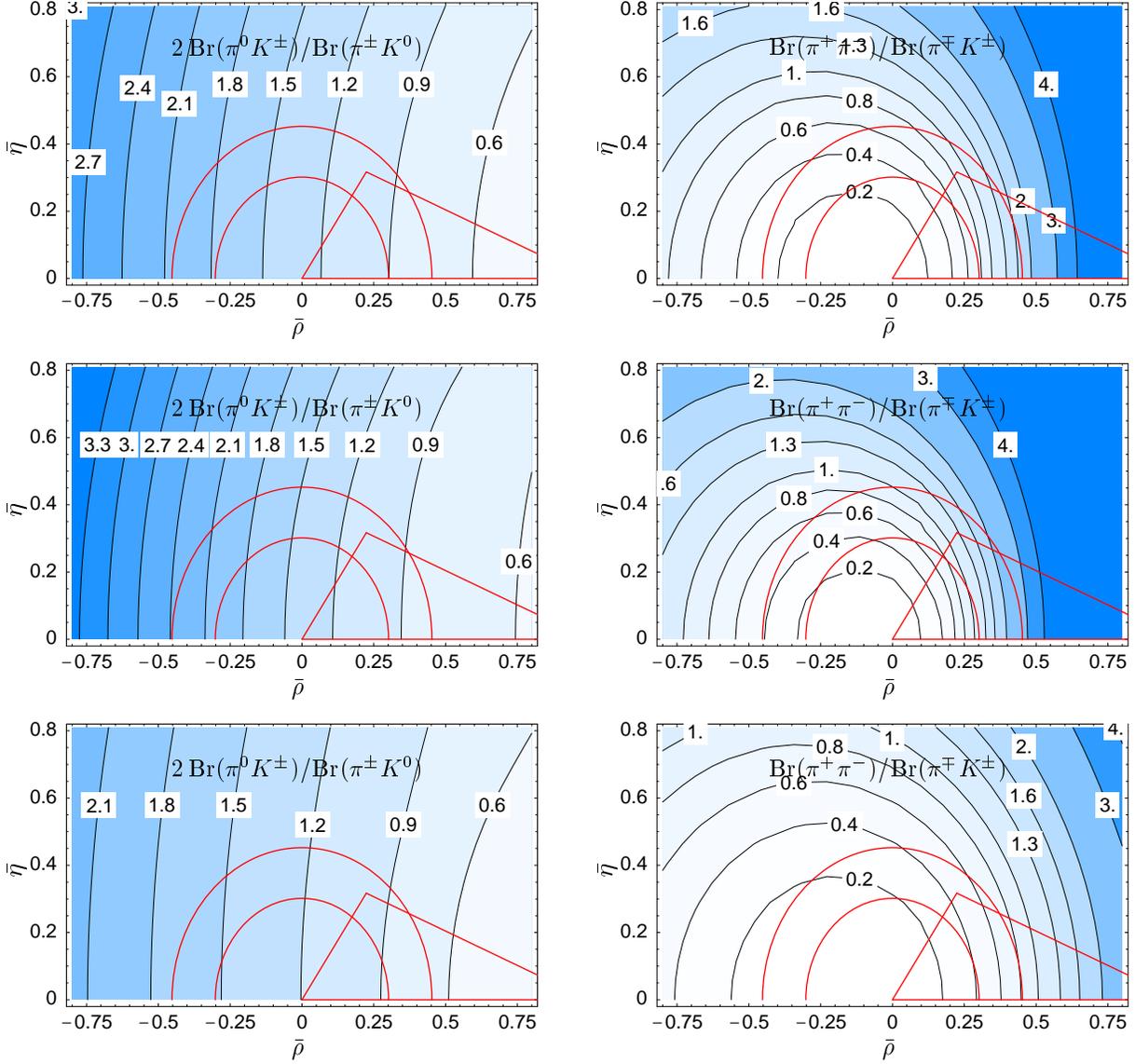}}
   \vspace*{0.3cm}
\centerline{\parbox{14cm}{\caption{\label{fig:rercp}
Two ratios of CP-averaged branching fractions as functions of the 
Wolfenstein parameters $\bar\rho$ and $\bar\eta$. The upper row 
corresponds to our central values; the middle (lower) row corresponds 
to the upper (lower) theoretical value including all uncertainties. We 
only show the upper half-plane ($\bar\eta>0$). The current standard 
best-fit unitarity triangle and the annulus given by 
$|V_{ub}/V_{cb}|$ are overlaid to guide the eye.}}}
\end{figure}

We can unfold the strong dependence of the ratios on $|V_{ub}/V_{cb}|$ 
by presenting the predictions for the branching fractions as functions 
of the Wolfenstein parameters $\bar\rho$ and $\bar\eta$.
This is illustrated in Figure~\ref{fig:rercp} for the two cases of the
ratios $2\mbox{Br}(\pi^0 K^\mp)/\mbox{Br}(\pi^\mp\bar K^0)$ and 
$\mbox{Br}(\pi^+\pi^-)/\mbox{Br}(\pi^\pm K^\mp)$. The other two ratios 
involving $\pi K$ final states exhibit a dependence similar to the
first of these two examples. For not too large 
values of $\bar\eta$, any of the three ratios involving only $\pi K$ 
final states is a direct measure of $\bar\rho$, since the dominant 
dependence on the CKM parameters arises from 
$\epsilon_{\rm KM}\cos\gamma=\tan^2\!\theta_C\,\bar\rho$. 
For positive $\bar\rho$ the theoretical uncertainty is 
relatively small, as can be deduced by comparing the three panels of 
Figure~\ref{fig:rercp} that refer to 
$2\mbox{Br}(\pi^0 K^\mp)/\mbox{Br}(\pi^\mp\bar K^0)$. (Note that the 
contour lines in this plot do not coincide with those of constant 
$R_*^{-1}$ inferred from Figure~\ref{fig:CKMfit}, since there the 
parameter $\varepsilon_{3/2}$ is assumed to be fixed and extracted from 
data, whereas here it is determined from theory and hence proportional 
to $|V_{ub}/V_{cb}|$.) The ratio 
$\mbox{Br}(\pi^+\pi^-)/\mbox{Br}(\pi^\mp K^\pm)$ exhibits a rather
different behaviour. The circular contours reflect the increase of the 
$\pi^+\pi^-$ branching fraction with $|V_{ub}|$. The offset of the 
center to negative values of $\bar\rho$ results from the interference of 
the two contributions with different weak phases to the decay 
amplitudes, which for $\bar\rho>0$ is constructive (destructive) for the 
$\pi\pi$ ($\pi K$) final states (and vice versa for $\bar\rho<0$).

\begin{figure}[p]
\epsfxsize=15.6cm
\centerline{\epsffile{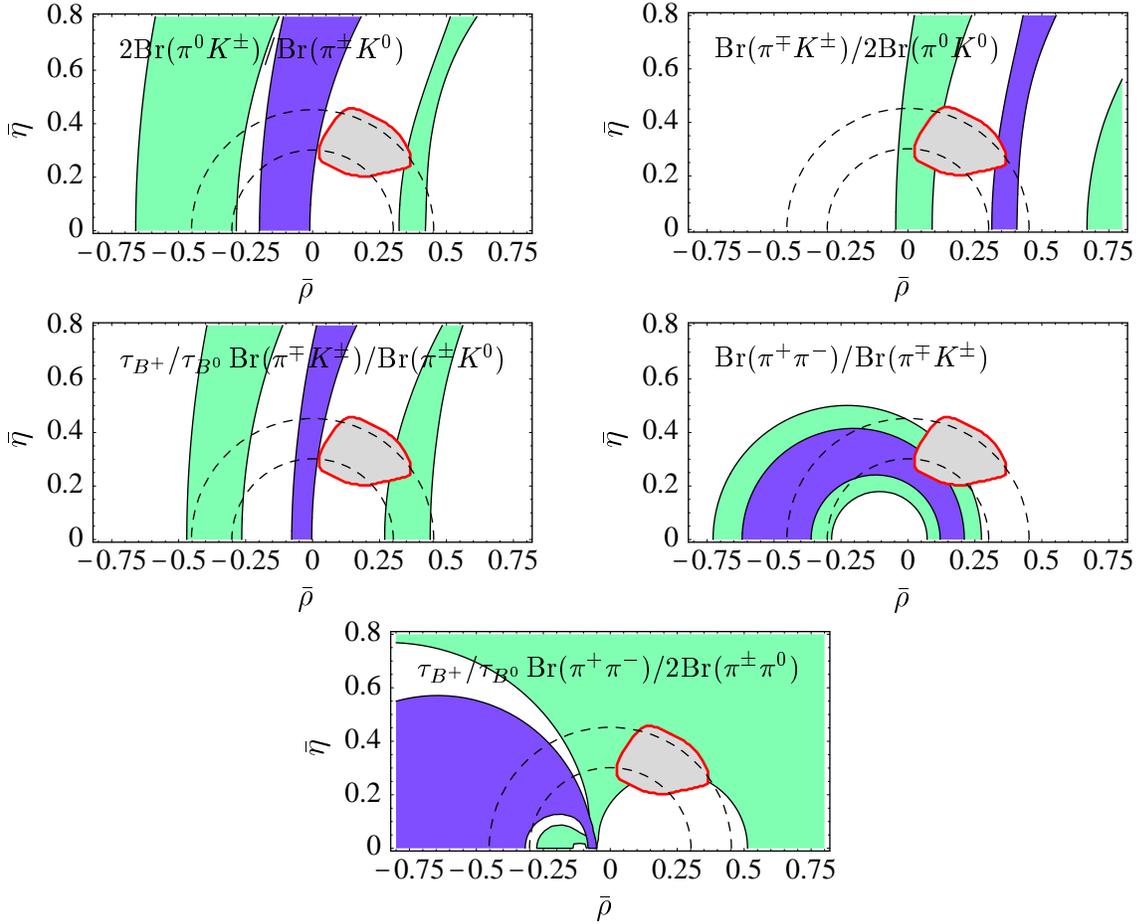}}
\centerline{\parbox{14cm}{\caption{\label{fig:reband}
Regions in the $(\bar\rho,\bar\eta)$ plane allowed by theory given 
the current central experimental values of the ratios of branching 
fractions (dark bands), and the experimental values $\pm 2\sigma$ 
standard deviations (light bands). Only the upper half-plane 
($\bar\eta>0$) is shown. The dashed circles indicate the allowed region 
implied by the measurement of $|V_{ub}/V_{cb}|$ in semileptonic $B$ 
decays. The light circled area shows the allowed region obtained from 
the standard global fit of the unitarity triangle 
\protect\cite{Hoecker}.}}}
\end{figure}

The current 
experimental values for the CP-averaged branching fractions are still 
afflicted by large errors, so a fit of $(\bar\rho,\bar\eta)$ to these 
ratios may appear premature. Nonetheless, it is useful to carry out 
such an analysis at the present stage, not only to exhibit its potential 
once the errors decrease, but also to gauge the limitations eventually 
set by the theoretical uncertainties. Figure~\ref{fig:reband} shows the
results separately for each of the five ratios of CP-averaged branching 
fractions that have been measured so far. Consider first the 
dark-shaded band in each plot, which covers the region allowed by 
theory given the current central experimental value. 
The light-shaded bands in Figure~\ref{fig:reband} represent the 
theoretically allowed regions obtained by using experimental values for
the ratios shifted up or downwards by two standard deviations. Hence, 
the width of each band thus reflects the (largely irreducible) theory 
uncertainty given a certain value of the observable, while the current 
constraint on $\bar\rho$ and $\bar\eta$ provided by each observable 
separately can roughly be taken to be the entire region bounded by the 
two light-shaded bands. 

\subsubsection*{\boldmath Global fit in the $(\bar\rho,\bar\eta)$ 
plane\unboldmath}

The information from all CP-averaged $B\to\pi K,\pi\pi$ branching 
fractions (and, in the future, from the corresponding CP asymmetries) 
can be combined into a single global fit, giving regions for the
Wolfenstein parameters $\bar\rho$ and $\bar\eta$ that are allowed by 
theory. We will now determine these regions, using two slightly 
different treatments of theoretical uncertainties. Because our focus 
is to determine fundamental parameters of the Standard Model, we adopt 
a more conservative error analysis than in the previous paragraphs of 
this section and scan all theory input parameters over their respective 
ranges of uncertainty (rather than adding theoretical uncertainties in 
quadrature). The same conservative treatment was used in the analysis 
of the weak phase $\gamma$ in Section~\ref{sec:NR}.

It is evident from our previous discussion that ratios of CP-averaged
branching fractions provide the most powerful constraints on the 
Wolfenstein parameters, since they are afflicted with relatively small 
theoretical uncertainties. However, there are two reasons for not
performing a global fit to the ratios directly (see also the discussion 
in Appendix~C of \cite{Hoecker}): first, with the present large 
experimental errors there are significant differences in the results of 
the fit depending on whether the ratios or inverse ratios are used, 
leading to an element of arbitrariness; secondly, even if two 
quantities have gaussian errors, their ratio does not. It is therefore 
preferable to perform the global fit for the individual branching 
ratios, despite the fact that the predictions for the branching 
fractions have larger (and correlated) theoretical uncertainties. 

\begin{figure}[t]
\vspace*{4.4cm}
\centerline{\epsffile{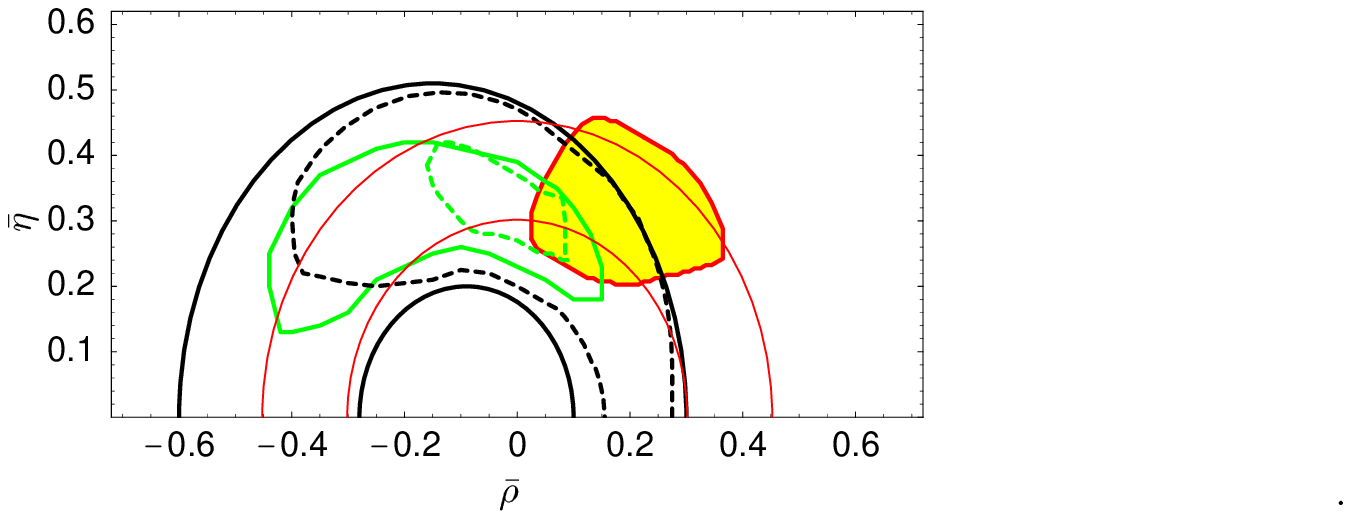}}
\centerline{\parbox{14cm}{\caption{\label{fig:chi2fit}
95\% confidence region in the $(\bar\rho,\bar\eta)$ plane obtained from 
a fit to the CP-averaged $B\to\pi K,\pi\pi$ branching fractions. The 
dark curve gives the envelope of the 95\% confidence level contours, 
the lighter curve the envelope of the set of $\chi_{\rm min}^2$ points. 
Here the solid lines refer to a cut on the $\chi_{\rm min}^2$
value, which corresponds to a 5\% confidence level of the fit for a
given model. The dashed lines refer to all models which have a 
$\chi_{\rm min}^2$ per degree of freedom of less than 1. See text for further 
explanations.}}}
\end{figure}

We now present two versions of such a fit. In the first method, we find 
the allowed region from $\bar\rho$ and $\bar\eta$ by determining first 
the 95\% confidence level contour for a given theoretical input, and 
then scanning over all models in the theory parameter space that survive 
a 5\% confidence cut. The theory parameter space is given by the set of 
all input parameters confined to the ranges specified in 
Table~\ref{tab:inputs}, variation of the renormalization scale $\mu$ 
between $m_b/2$ and $2m_b$, and scanning of the parameters $X_H$ and 
$X_A$ in the ranges described in (\ref{XHparam}) and (\ref{XAparam}). 
The resulting $95\%$ confidence region is bounded by the dark solid 
lines in Figure~\ref{fig:chi2fit}. The grey solid line inside this
region envelopes the set of $\chi_{\rm min}^2$ points of all models 
that pass the 5\% confidence cut. The dashed dark and grey lines 
have the same interpretation except that only theory models 
with a $\chi_{\rm min}^2$ per degree of freedom of less than 1 are 
selected. The $(\bar\rho,\bar\eta)$ range so determined is compared 
with the standard CKM fit (also at 95\% confidence level) \cite{Hoecker}, 
which uses information from semileptonic $B$ decays ($|V_{ub}|$ and 
$|V_{cb}|$), $K$--$\bar K$ mixing ($\epsilon_K$), and $B$--$\bar B$ 
mixing ($\Delta m_{B_d}$, $\Delta m_{B_s}$, and $\sin2\beta$). The 
constraint corresponding to $|V_{ub}/V_{cb}|=0.085\pm 0.017$ is 
indicated by the solid annulus.  

A fit to CP-averaged branching fractions cannot distinguish between 
$\bar\eta$ and $-\bar\eta$, and so for simplicity only the upper half 
of the $(\bar\rho,\bar\eta)$ plane is shown in the figure. At present 
our contours are symmetric about the $\bar\rho$ axis. In the future, 
data on direct CP asymmetries in the various $B\to\pi K,\pi\pi$ modes 
should be included in the fit, in which case it will become possible to 
obtain information about the sign of $\bar\eta$.

\begin{figure}[t]
\epsfxsize=10cm
\centerline{\epsffile{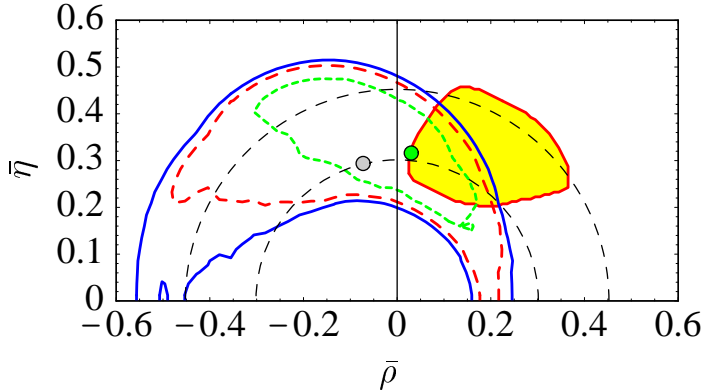}}
\centerline{\parbox{14cm}{\caption{\label{fig:chi2BaBar}
95\% (solid), 90\% (dashed) and 68\% (short-dashed) confidence level 
contours in the $(\bar\rho,\bar\eta)$ plane obtained from a global 
fit to the CP-averaged $B\to\pi K,\pi\pi$ branching fractions, using 
the scanning method of \protect\cite{Hoecker}. The dark dot shows the
overall best fit, whereas the light dot indicates the best fit for the
default parameter set. See text for further explanations.}}}
\end{figure}

Before interpreting the results of this analysis, we discuss a second 
method for performing the global fit, which follows the strategy 
outlined in \cite{Hoecker}. It differs from the method explained above
in two ways. First, the large overall scale uncertainty (of almost 40\%)
due to the error on the value of the form factor $F_0^{B\to\pi}(0)$, 
which is common to the theoretical predictions for all branching 
fractions (see, e.g., (\ref{BRs})), is eliminated by fitting the form 
factor to the data. In that way, the 
sensitivity to the Wolfenstein parameters in increased in the same way 
as it would be by using ratios; however, the statistical problems of 
using ratios (see above) are avoided. Specifically, we start from the 
$\chi^2$ function
\begin{equation}
   \chi^2(S) = \sum_i \left( \frac{E_i-S\,T_i}{\sigma_i} \right)^2 ,
\end{equation}
where $E_i$ are the experimental values for the branching ratios, 
$\sigma_i$ are their errors, and $T_i$ are the corresponding theoretical
predictions obtained with the fixed form factor $F_0^{B\to\pi}(0)=0.28$. 
The scale factor $S$ parameterizes the deviations of the product 
$(F_0^{B\to\pi}(0)\,|V_{cb}|)^2$ from its central value. For a given set 
of theory parameters, we determine 
\begin{equation}\label{Sdet}
   S = \frac{\sum_i E_i\,T_i/\sigma_i^2}{\sum_i\,(T_i/\sigma_i)^2}
\end{equation}
such that $\chi^2(S)$ is minimized, provided that the value of $S$ is
allowed by the theory ranges for the form factor (see 
Table~\ref{tab:inputs}) and $|V_{cb}|$. For most choices of theory 
parameters this condition is satisfied. If it is not, we choose the
value of $S$ inside the allowed range that is closest to the optimal 
value in (\ref{Sdet}). Next, we scan over all theory parameters and
determine the global minimum of $\chi^2$ in the $(\bar\rho,\bar\eta)$ 
plane. We then plot contours of constant 
$\Delta\chi^2=\chi^2-\chi_{\rm min}^2$ corresponding to the 68\%, 90\%
and 95\% confidence levels. The second difference with respect to our 
first scanning method is that these contours refer to a constant 
$\chi^2$, whereas in the first method the $\chi^2$ is different for
each of the theory models considered. The results of this analysis are 
shown in Figure~\ref{fig:chi2BaBar}. The best fit has an excellent 
$\chi^2=2.1$ (for 3 degrees of freedom), corresponding to a confidence 
level of 54\%. The parameters of the particular theory model 
corresponding to this fit are all very reasonable. In particular, the 
quantity $X_A$ parameterizing the weak annihilation terms is even 
smaller in magnitude than our default value. Hence, a good global fit 
to the data can be obtained without using extreme choices of input 
parameters, or invoking phenomenological recipes such as largely 
enhanced annihilation or charm-penguin contributions. The default 
parameter set in Table~\ref{tab:inputs} also yields a good fit, which 
has $\chi^2=4.57$ and 21\% confidence level. Note that at 90\% 
confidence level the allowed range obtained by combining our results 
with the measurement of $|V_{ub}/V_{cb}|$ in semileptonic $B$ decays 
excludes $\bar\eta=0$, thus establishing the existence of a CP-violating 
phase in $b\to u$ transitions.

The results of the two scanning methods shown in 
Figures~\ref{fig:chi2fit} and \ref{fig:chi2BaBar} are similar (though
the range obtained in the first method is somewhat more conservative). 
The allowed region for $\bar\rho$ and $\bar\eta$ obtained from our 
analysis of rare hadronic $B$ decays is compatible with the standard 
global fit using information from semileptonic $B$ decays, $K$--$\bar K$ 
mixing and $B$--$\bar B$ mixing. However, the best fits to the 
rare-decay branching fractions prefer a larger value of $\gamma$ (of 
about $90^\circ$). This has been noted by many other authors in the 
past, usually based on the naive factorization approximation and without 
a theoretical error estimate. Here we have put this analysis on a firmer 
theoretical footing. On the other hand, we see that small values of 
$\gamma$ are also allowed, provided the value of $|V_{ub}|$ is lower 
than its standard central value. Combining our results with the standard 
fit would reduce the allowed region in the $(\bar\rho,\bar\eta)$ plane 
by about a factor 2, indicating that even with present experimental (and 
theoretical) errors the information obtained from rare hadronic decays 
provides important constraints on the Wolfenstein parameters.

It will be interesting to follow how the comparison of the standard fit
and the fit to rare hadronic $B$ decays will develop as the data become
more precise. If the two fits remain consistent with each other even as
the allowed regions shrink in size, this would constitute a highly 
nontrivial test of the CKM model, in which for the first time the 
phase of the $V_{ub}$ matrix element (which enters through tree--penguin
interference) is restricted by both direct and indirect measurements.
If, on the other hand, the two regions would not overlap at a reasonable
confidence level, this may suggest that the standard theory of weak 
interactions does not account completely for either the $B$--$\bar B$ 
mixing amplitude (since this is what determines the left-most side of
the allowed region in the standard fit) or the loop-induced 
flavour-changing amplitudes in $B\to\pi K,\pi\pi$ decays. (Another 
option would be to abandon the theoretical framework advocated here, but 
we understandably leave it to others to pursue this avenue.)

\subsection{Predictions for CP asymmetries}
\label{sec:CP}

\begin{figure}[p]
\epsfxsize=15.6cm
\centerline{\epsffile{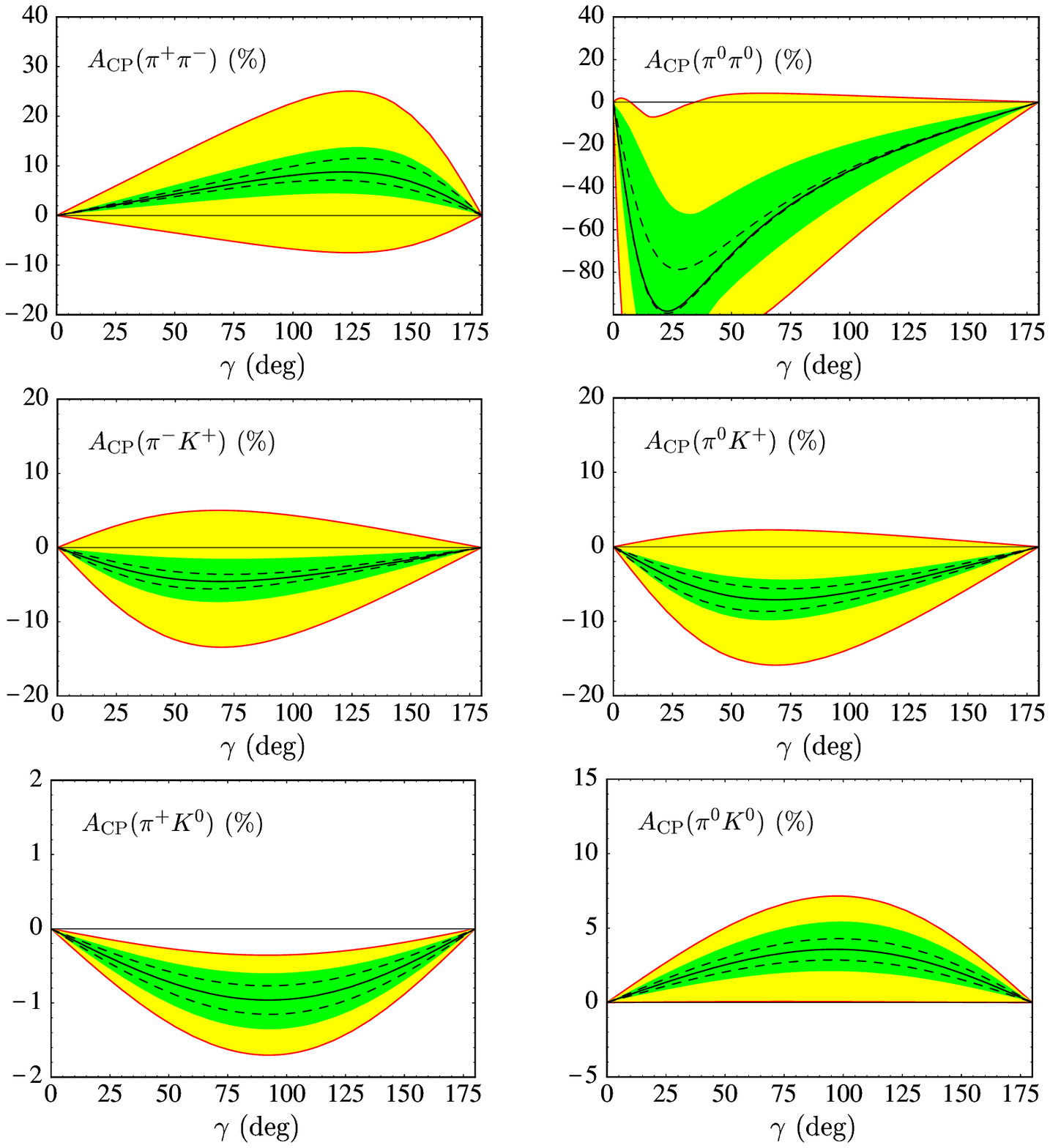}}
\centerline{\parbox{14cm}{\caption{\label{fig:acp1}
Direct CP asymmetries as functions of $\gamma$, assuming $\gamma>0$. 
For negative $\gamma$, the signs of the asymmetries are reversed. The 
meaning of the curves and bands is the same as in 
Figure~\protect\ref{fig:brav1}. The asymmetries vanish in naive 
factorization, so no corresponding line is drawn.}}}
\end{figure}

The QCD factorization approach predicts the strong-interaction phases 
and hence the direct CP asymmetries in the heavy-quark limit. In the
following, we define the asymmetries as
\begin{equation}
   A_{\rm CP}(f) =
   \frac{\mbox{Br}(B\to f)-\mbox{Br}(\bar B\to \bar f)}
        {\mbox{Br}(B\to f)+\mbox{Br}(\bar B\to \bar f)} \,,
\end{equation}
where $\bar B=\bar B^0,B^-$ contains a $b$ quark (rather than 
antiquark). The dependence of the CP asymmetries on the weak phase 
$\gamma$ is displayed in Figure~\ref{fig:acp1}. The asymmetries are 
typically predicted to be small, concurrent with the fact that 
strong-interaction phases are suppressed in the heavy-quark limit 
\cite{BBNS1}. Even the largest asymmetries for the $\pi K$ final states
are predicted to be of order 10\% or less. An exception to this rule is 
the final state $f=\pi^0\pi^0$, for which the asymmetry is sensitive to 
QCD penguins and to the coefficient $a_2$, both of which have large 
uncertainties and a potentially large relative phase. 

While our approach predicts the generic magnitude of the CP asymmetries, 
their precise values remain uncertain. This is not surprising, since in 
contrast to the branching ratios the asymmetries are more sensitive to 
-- and in fact generated by -- corrections to naive factorization. Hence,
they are subject to larger relative uncertainties. In particular, the 
CP asymmetries are proportional to the sines of strong-interaction
phases, which are of order $\alpha_s(m_b)$ or $\Lambda_{\rm QCD}/m_b$. 
Only the leading perturbative contributions to these phases are 
calculable. On the contrary, the CP-averaged branching fractions depend 
on the cosines of strong-interaction phases, which are equal to 1 in the
heavy-quark limit. 

First measurements of CP asymmetries have been published by the CLEO 
\cite{CLEOAcp} and Belle \cite{Belle01} Collaborations. All results are 
compatible with no asymmetry, with typical $1\sigma$ errors of about 
$20\%$. These errors are not yet small enough to draw meaningful 
conclusions, except that very large asymmetries appear already 
improbable. In the future, accurate measurements of CP asymmetries will 
test the generic prediction of the QCD factorization approach that 
strong-interaction phases are small. If this prediction is confirmed, 
further analysis of the CP asymmetries should help to determine the sign 
of $\gamma$, which cannot be probed using CP-averaged branching 
fractions. In addition, data on CP asymmetries may help to further 
constrain annihilation contributions and other theoretical input 
parameters in the QCD factorization approach.

\section{Comparison with other work}
\label{sec:comp}

A general overview of the various approaches to the evaluation of the 
matrix elements in exclusive hadronic $B$ decays has been given in 
\cite{BBNS2}. In that reference we have commented on the strategies and 
methods that had been employed previously and explained how they are 
related to QCD factorization. In this section we discuss recent papers
on the subject and clarify the differences with our work. We divide 
these papers into four categories: those using our QCD factorization 
formalism, those using a perturbative PQCD approach, those in which 
nonperturbative effects are estimated using light-cone sum rules or 
model calculations, and finally those containing phenomenological 
analyses. We consider each of these in turn.

Before discussing other approaches to nonleptonic $B$-decays it may be
helpful to summarize again the conceptual basis of the QCD factorization
approach. The key ingredient is the systematic analysis of Feynman 
graphs in the heavy-quark limit, from which we deduce the factorization
of infrared singularities into hadronic light-cone distribution 
amplitudes and form factors. This enables us to establish the 
factorization formula (\ref{fact}). The separation of long- and 
short-distance contributions to the decay amplitudes, necessary to 
establish factorization, holds only to leading power in 
$\Lambda_{\rm QCD}/m_b$ and is based on considerations analogous to 
those used to demonstrate factorization in other applications of QCD to
hard processes (such as deep inelastic scattering, Drell--Yan 
production, and electromagnetic form factors of hadrons at large 
momentum transfer). The factorization formula leads to a 
model-independent treatment of exclusive hadronic $B$ decays in the 
heavy-quark limit. A consistent counting scheme for powers of 
$\Lambda_{\rm QCD}/m_b$ and a systematic identification of all the 
leading contributions are crucial for establishing this result. The 
framework proposed in \cite{BBNS1,BBNS2} is general and provides a
starting point for further theoretical developments, such as the
improved understanding of the nonperturbative input (e.g., the 
$B\to\pi$ form factor and the light-cone distribution amplitudes) and 
estimates of power corrections.

\subsection{Analyses within QCD factorization}

Several authors have used the framework of QCD factorization for 
applications to two-body hadronic $B$ decays 
\cite{MSYY,YY1,YY2,HMW,DYZ,CY2}. We now compare the results reported in 
these papers with ours and comment on apparent differences and 
discrepancies.

Muta et al.\ have generalized the results of our previous work 
\cite{BBNS1} by including electroweak penguin contributions, and have 
applied the QCD factorization approach to the decays $B\to\pi K$, 
$\pi\pi$. We would like to point out the following differences with 
respect to the present work:
\begin{itemize}
\item
The hard-scattering kernels are derived only for symmetric light-cone
distribution amplitudes.
\item
In evaluating the $\alpha_s$ corrections to the penguin coefficients 
$a_4$ and $a_6$, the existence of the two distinct types of penguin 
contractions shown in Figure~\ref{fig:penguin} is not taken into 
account. As discussed in Section~\ref{subsec:comments}, this leads to 
incorrect terms proportional to $\alpha_s(C_4+C_6)$ in these 
coefficients.
\item
An incomplete projector for the twist-3, two-particle distribution
amplitude of the pion is employed. This gives an incorrect contribution 
proportional to $C_{8g}^{\rm eff}$ in $a_6$.
\item
We also observe a disagreement with the remaining terms in the 
correction of order $\alpha_s$ to $a_6$, which concerns the function
$G'_{M_2}(s_q)$ in \cite{MSYY}. All three components, the 
$s_q$-dependent part, the constant, and the coefficient of the logarithm, 
differ from our findings. We note that the $\mu$-dependence of $G'_{M_2}$
is in conflict with the requirement of renormalization-group invariance 
of the product $r_\chi^M(\mu)\,a_6(\mu)$. In addition, the contributions
from the radiative vertex corrections to $a_6$ and $a_8$ are missing.
\item
A minor difference comes from the fact that we neglect electroweak 
penguin contributions to $a_4$ and $a_6$, while these are included in 
\cite{MSYY}. On the other hand, Muta et al.\ omit the contributions 
proportional to $C_1$, $C_2$ and $C_{7\gamma}^{\rm eff}$ to the 
electromagnetic penguin coefficients $a_8$ and $a_{10}$.
\item
The twist-3 contributions to the spectator hard-scattering amplitudes
and annihilation effects are not discussed. As we have seen, these 
corrections are the potentially most important source of theoretical 
uncertainty.
\end{itemize}
The formulae given in \cite{MSYY} have also been applied to discuss
$B$ decays into vector--pseudoscalar final states \cite{YY1} and final
states containing $\eta$ and $\eta'$ \cite{YY2}. The previous comments 
apply also to these papers, as they do to \cite{HMW}, which relies on 
the same expressions except for dropping the formally power-suppressed 
terms proportional to $a_6$ and $a_8$.

In a series of papers, Du et al.\ have discussed $B$ decays into two 
light pseudoscalar mesons \cite{DYZ}. We focus on the two last papers 
in \cite{DYZ}, which contain the most complete and updated results. 
Similarly to the present work, electroweak penguin contributions and 
chirally-enhanced twist-3 components of the pion distribution amplitude 
are included in these papers, but no weak annihilation effects are 
considered. The hard-scattering kernels are given only for symmetric 
distribution amplitudes, and explicit results are presented for the 
case of $B\to\pi\pi$ decays. We will therefore not distinguish between
$\pi\pi$ and $\pi K$ final states in the present discussion. The
formulae for the coefficients $a_i$ can be directly compared with our 
results. A minor difference concerns the tiny electroweak penguin 
contributions to the coefficients $a_1,\dots,a_6$, which we decided to 
neglect in our approximation scheme, but which are retained in 
\cite{DYZ}. Next we recall from Section~\ref{subsec:comments} that in 
the approximation of including twist-3 contributions only when they are 
chirally enhanced, the equations of motion require the use of 
asymptotic distribution amplitudes $\Phi_p$, $\Phi_\sigma$. Du et al.\
have considered a more general situation, which would affect the
cancellation of endpoint singularities and the renormalization-scale
dependence of some results. Use of the appropriate asymptotic
distribution amplitudes eliminates these spurious ambiguities. Finally, 
we note two minor discrepancies:
\begin{itemize}
\item 
Our expression for the twist-3, hard spectator interaction in 
(\ref{Hiexpr}) contains a factor
\begin{equation}
   \frac{2\mu_P}{m_b}\,\frac{\Phi_M(x)}{x}\,\frac{\Phi_p(y)}{\bar y}
\end{equation}
in the integrand, whereas the corresponding result given in eq.~(37) of
 of the third paper in \cite{DYZ} has 
\begin{equation}
   \frac{2\mu_P}{M_B}\,\frac{\Phi_M(x)}{x}\,
   \frac{\Phi_\sigma(y)}{6\bar y^2} \,.
\end{equation}
Using the asymptotic forms of the twist-3 distribution amplitudes 
$\Phi_p$ and $\Phi_\sigma$, we find that the two results differ by a 
factor of $y$.
\item 
In our result for the twist-3 penguin contribution $P_3^p$ in 
(\ref{hatPK}), the coefficient of $C_1$, for example, is
\begin{equation}
   \frac43 \ln\bigg( \frac{m_b}{\mu} \bigg) + \frac23 - \hat G(s_p) \,.
\end{equation}
The equivalent expression given in eq.~(23) of the third paper in 
\cite{DYZ} reads
\begin{equation}
   \left( 1 + \frac23\,A_\sigma \right)
   \ln\bigg( \frac{m_b}{\mu} \bigg) + \frac{7}{12}
   + \frac12\,A_\sigma - G'(s_p) - G^\sigma(s_p) \,.
\end{equation}
Using asymptotic twist-3 distribution amplitudes, one may check that
$G^\prime(s_p)+G^\sigma(s_p)=\hat G(s_p)$ and $A_\sigma=1/2$. The 
two results then agree up to a constant $1/6$. This difference can be 
traced back to an inconsistent use of twist-3 projectors in four 
space-time dimensions within a $d$-dimensional loop calculation, before 
the subtraction of ultraviolet poles is performed (see the last 
reference in \cite{DYZ}). Analogous differences occur in the other 
terms in $a_6$ and $a_8$.
\end{itemize}
Apart from these discrepancies, the expressions agree with our 
results for symmetric distribution amplitudes.

Cheng and Yang have applied the QCD factorization approach to a study 
of the decays $B\to\phi K$ \cite{CY2}. Annihilation topologies are 
discussed and argued to be important numerically. A few minor 
discrepancies with our results occur in the expressions for the 
coefficients $a_{3,\ldots,10}$. The penguin contraction of the operators 
$Q_{4,6}$ is treated incorrectly, as discussed in 
Section~\ref{subsec:comments}. Photonic penguin contractions of 
the operators $Q_{1,2}$ contributing to $a_{8,10}$ are omitted. Also, 
QCD corrections to $a_{6,8}$ are not considered, presumably because the
corresponding amplitudes are formally power suppressed. Similarly to the 
results of \cite{DYZ}, the twist-3 kernel for the hard spectator 
interaction contains an additional factor of $y$. Finally, we briefly 
comment on the analysis of weak annihilation contributions. In 
\cite{CY2} the final state consists of a vector meson and a kaon rather 
than two pseudoscalar mesons. The decay amplitudes are estimated in the
approximation of using leading-twist distribution amplitudes for the 
$\phi$ meson, but including chirally-enhanced twist-3 contributions for 
the kaon. Taking this systematic difference into account, we can still
compare the results of \cite{CY2} with our expressions. The two agree 
in the case of $(V-A)\otimes(V-A)$ operators, but the results for 
$(S-P)\otimes(S+P)$ operators are different. The latter are the ones 
that depend on the twist-3 contributions, but we do not have sufficient 
information to trace back the origin of the discrepancy.

\subsection{The PQCD approach}

Recently, Keum, Li and Sanda have presented an extensive study of 
$B\to\pi K$ decays within a perturbative hard-scattering (or ``PQCD'') 
approach \cite{KLS00}. While the underlying goal of a separation of soft 
and hard physics in the $B$-decay matrix elements is similar in spirit 
to QCD factorization, there are fundamental differences in the 
implementation of this idea.

In the PQCD approach, the $B\to\pi$ and $B\to K$ form factors are 
assumed to be perturbatively calculable. This assumption is justified by
invoking Sudakov effects to regulate the infrared-sensitive contribution
from the endpoint region in the integral over the light-cone momentum of 
the outgoing spectator quark. In other words, the contribution from the 
region where the spectator quark is soft is supposed to be strongly 
suppressed, and therefore the exchange of a hard gluon is always 
required. Instead of being of leading order (in the QCD coupling 
constant), as in the generic case of a $B\to\pi$ form factor dominated 
by soft physics, the form factor is now counted as being of order 
$\alpha_s$ in perturbation theory. Thus, the hierarchy of the various
contributions to the decay amplitudes in PQCD is very different from 
that in our approach:
\begin{itemize}
\item 
In the PQCD approach, the $B\to\pi K$ matrix element is a quantity of
order $\alpha_s$. To this order, therefore, all the Wilson coefficients
can then be taken in the leading logarithmic approximation.
\item
The ``nonfactorizable'' hard gluon-exchange contributions to the 
kernels $T_i^{\rm I}$ in (\ref{fact}) enter at order $\alpha_s^2$ and 
are therefore omitted. Note, however, that it is essential to consider 
such effects in order to establish factorization. This has never been 
done explicitly in the PQCD framework.
\item
The form-factor terms containing the kernels $T_i^I$ and the 
hard-scattering terms containing the kernels $T_i^{\rm II}$ in the 
factorization formula are treated as being of the same order in the PQCD
approach. The former are called ``factorizable'' and the latter 
``nonfactorizable''. Because both terms vanish simultaneously in the 
limit $\alpha_s\to 0$, naive factorization is not recovered in any 
limit of QCD. This is in contrast to our interpretation of the QCD 
factorization formula, for which naive factorization is recovered in the 
heavy-quark limit. 
\end{itemize}
After this brief synopsis of the key features of the PQCD scheme, we now 
examine critically the main assumptions of this approach in the context
of $B\to\pi K$ decays.

\subsubsection*{\boldmath Importance of Sudakov effects and 
calculability of the $B\to\pi$ form factor\unboldmath}

In the PQCD approach the transverse momenta of the quarks are kept 
explicitly when evaluating contributions that are potentially infrared 
sensitive. For instance, the gluon-exchange kernel for the $B\to\pi$ 
form factor is written as
\begin{equation}
   H \sim \frac{1}{(x_3 m_b^2 + \vec k_{3\perp}^2)
                [x_1 x_3 m_b^2 + (\vec k_{1\perp} - \vec k_{3\perp})^2]}
   + \ldots \,,
\end{equation}
where the ellipses represent less singular terms. Here $x_1$ ($x_3$) is 
the light-cone momentum fraction of the spectator quark in the $B$ 
meson (pion). In the conventional collinear expansion adopted in our
work, the transverse momenta $\vec k_{i\perp}^2\sim\Lambda_{\rm QCD}^2$ 
would be treated as higher-twist effects and dropped (more precisely, 
the amplitude would be expanded in powers of transverse momenta). It 
then follows that $H\sim 1/x_3^2$, and so the convolution with the 
leading-twist pion distribution amplitude $\Phi_\pi(x_3)\sim x_3(1-x_3)$ 
would lead to an integral $\int d x_3/x_3$, which is infrared divergent. 
This well-known result can be interpreted as a sign of the dominance of 
soft endpoint contributions in the $B\to\pi$ form factor \cite{SHB,BD}.

In \cite{KLS00} (see also \cite{LY}) this divergence is regulated by
keeping the transverse momenta. Then a Fourier transform from transverse
momentum space into impact parameter space ($b$ space) is performed, and 
a Sudakov factor is included for each of the meson distribution 
amplitudes. This factor strongly suppresses the region of large $b$, and 
the $B\to\pi$ form factor is therefore assumed to be perturbatively
calculable. 

It should be noted that retaining the transverse momentum dependence in 
$H$, which is only part of the complete higher-twist contribution, is a
model-dependent procedure. It is precisely in the critical endpoint 
region $x_3\to 0$ that the leading two-particle Fock-state description 
that underlies this analysis breaks down. A further puzzle to us is the 
assumption made in \cite{LY} that the spectator quark in the $B$ meson 
has a large momentum component in the ``$-$'' direction, 
$k^-_1=x_1 m_b/\sqrt{2}$, when the pion momentum is in the ``$+$''
direction. This is used to derive the Sudakov factor for the $B$ meson
in analogy to the case of a fast light meson. Since the spectator quark 
in the $B$ meson is intrinsically soft, we see no justification for such 
an assumption.

In conclusion, we believe that the perturbative evaluation of the 
$B\to\pi$ form factor, which is one of the central ingredients of the 
PQCD analysis in \cite{KLS00}, is not justified. The relevance of 
Sudakov form factors in hadronic $B$ decays should be investigated in a 
more systematic way. However, it seems unlikely to us that Sudakov 
logarithms are sufficiently important at the scale $m_b$ to eliminate
the soft contributions to heavy-to-light form factors and thus render 
them calculable in a model-independent way. A complete description 
of the $B\to\pi$ transitions at large recoil in the heavy-quark limit 
and the derivation of a (hypothetical) factorization theorem for these 
processes has, to our knowledge, never been presented. For recent 
progress in this direction and further critical discussions of the 
problem see \cite{BF00,SR}.

\subsubsection*{Dynamical enhancement of penguin contributions (``fat 
penguins'')}

Keum et al.\ claim a dynamical enhancement of (factorizable) 
contributions to the matrix elements of penguin operators (such as $Q_4$ 
and $Q_6$) in the effective weak Hamiltonian. The first reason given in
support of this assertion is a dynamical enhancement of the $B\to\pi$ 
form factor of the scalar density $\bar u b$ contained in (the 
Fierz-transformed form of) $Q_6$, relative to the form factor of the 
vector current $\bar u\gamma^\mu b$ relevant for other operators. A
dynamically different structure for the form factor of the scalar 
density in PQCD is claimed, and a strong sensitivity of the effect to 
the model employed for the $B$-meson light-cone distribution amplitude 
is noted. However, an enhancement of this nature is not possible, 
because the form factor of the scalar density is related to that of the 
vector current by the equations of motion, leading to
\begin{equation}
   \langle\pi^+|\bar u b|\bar B^0\rangle
   = F_0^{B\to\pi}(q^2)\,\frac{m^2_B}{m_b} \,.
\end{equation}
The erroneous conclusion in \cite{KLS00} is the consequence of an
incorrect treatment of the twist-3 contribution to the pion distribution
amplitude (see below).

The second reason given in favor of a penguin enhancement is the choice 
of a low scale $\mu$ in the (leading-order) penguin coefficients 
$a_4(\mu)$ and $a_6(\mu)$. This is motivated by arguing that $\mu$ 
should be a typical scale intrinsic to the dynamics of the $B\to\pi$ 
form factor in the PQCD evaluation. However, the scale $\mu$ is 
unphysical and must be canceled by vertex corrections to the operator 
matrix elements, and in the case of $a_6(\mu)$ also by the running quark 
masses. Such corrections have been properly included in the present 
work, but they were neglected in \cite{KLS00}. The scale 
$\mu$ in $a_{4,6}$ may depend on the scale of non-factorizable hard 
spectator interactions, but it is clearly independent 
of the internal dynamics of the form factor. The arguments made in 
favour of a low scale $\mu$ and correspondingly increased penguin 
coefficients are therefore not well justified.

\subsubsection*{Relevance of annihilation topologies} 

In their analysis, Keum et al.\ find that penguin annihilation 
contributions are numerically very important and give a dominant
contribution to some of the $B\to\pi K$ decay amplitudes. We have 
investigated these effects in the present work and found them to give a
significant correction (compatible with being a power correction of 
canonical size), but not a dominant contribution. Nevertheless, it must  
be emphasized that weak-annihilation diagrams contribute at subleading 
power in the heavy-quark expansion and, in general, are not calculable 
in a QCD-based factorization approach. Therefore, the question about
the relevance of these effects warrants further investigation.

\subsubsection*{Generation of large, calculable strong-interaction 
phases}

It is claimed that large strong-interaction phases are generated by
hard gluon exchange with the spectator quark in the $B$ meson, as well
as by weak annihilation diagrams. Let us consider the hard spectator 
interactions in detail. In that case the effect is calculable within the 
QCD factorization approach and found to be real to leading order. The
source of the imaginary part found in \cite{KLS00} is as follows. The 
quark propagator entering the hard-scattering diagram is written as
\begin{equation}
   H_q \sim
   \frac{1}{x_2 x_3 m^2_b
            - \big(\vec k_{2\perp}+\vec k_{3\perp}\big)^2 +i\epsilon}
   \,.
\end{equation}
Working to leading power we would drop the transverse components and 
find a real contribution to the kernel. Keum et al., on the other hand, 
keep the transverse momenta and hence generate an imaginary part 
proportional to $\delta(\vec k_\perp^2-x_2 x_3 m^2_b)$, where 
$k_\perp=k_{2\perp}+k_{3\perp}$. Again, this treatment is model 
dependent. The Fourier transform of the kernel into impact parameter 
space results in a Bessel function with the imaginary part proportional 
to $m_b b\,J_0(\sqrt{x_2 x_3}m_b b)$. In the evaluation of the matrix 
element this function is convoluted with the Sudakov factor. Since the 
Bessel function is oscillating rapidly, with the amplitude of 
$m_b b\,J_0$ growing like $\sqrt{m_b b}$, the result of this convolution 
is very sensitive to the details of the $b$-space cut-off. This implies 
that the estimate of the strong-interaction phase is both model 
dependent and numerically sensitive to effects that are poorly under 
control. Similar comments hold for the estimate of the strong-interaction 
phases from the annihilation contributions, which are generated in an 
analogous way.

Finally, we remark that the simple $\gamma_5$ structure of the twist-3
projection for the pion employed by Keum et al.\ is incomplete. The
proper treatment is discussed in Section~\ref{subsec:comments}. The 
wrong twist-3 projection is, in particular, inconsistent with gauge 
invariance. Moreover, the correct asymptotic behaviour of the twist-3 
pion distribution amplitude $\Phi_p(x)$ is proportional to a constant, 
whereas the functional form $\sim x(1-x)$ is assumed in \cite{KLS00}. 
This problem affects all decay amplitudes, including the $B\to\pi$ form 
factor, the corresponding spurious penguin enhancement, and the 
nonfactorizable spectator interactions and annihilation contributions.

\subsubsection*{Other works using the PQCD approach}

Other analyses in the spirit of the PQCD approach were presented in 
\cite{LUY,CLI,BFW,LYA}, to which similar comments apply. In \cite{BFW}, 
the presence of a ``recoil-phase'' effect was advocated, which was 
claimed to originate within the PQCD framework. This phase should 
affect, e.g., the $B\to\pi$ form factor, which was assumed to be 
dominated by hard gluon exchange. It was argued that, when the gluon is 
exchanged between the spectator quark and the $b$ quark, the $b$-quark 
propagator could go on the mass shell because $m_B>m_b+m_{\rm spect}$.
This in turn would lead to a complex phase in the form factor. In our
opinion this conclusion is unwarranted, since bound-state effects have
to be factorized before a perturbative treatment can be justified and 
thus cannot influence the hard-scattering process. Another fundamental 
objection to the ``recoil phase'' is that it would contradict the fact 
that the $B\to\pi$ form factor is a real quantity.

\subsection{Estimates of nonperturbative effects}

Recently, Khodjamirian has suggested to study hadronic matrix elements 
for two-body $B$ decays in the framework of light-cone QCD sum rules
\cite{AKH}. As an example, he discusses the matrix elements of the 
current--current operators $Q_1^u$ and $Q_2^u$ for $B\to\pi^+\pi^-$ 
decays. The idea is to generalize the light-cone QCD sum-rule analysis
of the $B\to\pi$ form factor directly to the case of hadronic two-body 
modes. The starting point is the correlation function
\begin{equation}
   F_\alpha(p,q,k) = -\int d^4x\,e^{i(p-q)x} \int d^4y\,e^{i(p-k)y}\,
   \langle 0|\,T\,[j_\alpha^{(\pi)}(y) Q_i^u(0) j^{(B)}(x)]|\pi^-(q)
   \rangle \,,
\end{equation}
where $j_\alpha^{(\pi)}=\bar u\gamma_\alpha\gamma_5 d$ and 
$j^{(B)}=m_b\,\bar b i\gamma_5 d$ are interpolating currents for the 
emission pion and the $B$ meson, respectively. The explicit pion state 
$\pi^-(q)$ represents the recoil pion that absorbs the spectator quark. 
According to the QCD sum rule philosophy, the correlator is evaluated 
in two ways: by a direct calculation in QCD, and by inserting complete 
sets of hadronic states between the operator $Q_i^u$ and the 
interpolating currents, extracting the desired ground-state contribution 
with the help of quark--hadron duality. The two sides of the equation 
are expressed in the form of dispersion relations, and the usual Borel 
transformation is applied. To leading order in $\alpha_s$, the 
factorized result for the matrix element is recovered. A particular type 
of power correction is then estimated as a further illustration. This 
contribution comes from higher twist, three-particle 
(quark--antiquark--gluon) Fock components of the recoil pion, where the 
(nonfactorizable) gluon couples to the quark lines of the emission 
current. In the heavy-quark limit the resulting expression is 
demonstrated to scale as $\Lambda_{\rm QCD}/m_b$ relative to the leading 
contribution. It is estimated to be of relative size $\lambda_E/m_B$ 
with $\lambda_E\approx 0.05$--0.15\,GeV. Note that this correction has 
no rescattering phase. Its numerical effect is small, but comparable to 
the perturbative corrections at leading power. Additional 
nonfactorizable contributions exist but have not yet been investigated. 
Examples are the gluon-exchange effects that correspond to the vertex 
corrections and the spectator scattering diagrams in the QCD 
factorization formula. In the sum-rule approach these effects are in 
principle contained in similar diagrams, together with certain hadronic
corrections of subleading power. A potential difficulty will be to
disentangle power corrections to the asymptotic result given by the 
QCD factorization formula from uncertainties intrinsic to the sum rule 
method, such as the assumption of quark--hadron duality and the 
approximation of the emission pion by an interpolating current. It is 
not fully clear how this can be achieved in a controlled and systematic 
fashion. 

Some recent papers \cite{ZZX,PZE,KTE} have tried to model soft 
final-state interactions via the rescattering of certain hadronic 
channels such as $B\to D\bar D\to\pi\pi$. We consider such an approach 
to be problematical for a variety of reasons. There are many more 
intermediate channels beyond those taken into account. Systematic 
cancellations among these channels, which are predicted to occur in the 
heavy-quark limit, are missed when only a few  intermediate states are 
retained (see the discussion in Sections~3.4 and 7.2 of \cite{BBNS2}).
Moreover, the hadronic dynamics of multi-body decays is very complicated 
and in general not under theoretical control. The main problem of such 
models is the lack of a systematic approximation scheme based on 
parametric expansions. In our opinion, the use of a purely hadronic 
language, suitable for kaon decays, is not very helpful in the case of 
$B$ decays, where the number of channels and the energy release are 
large. 

In \cite{ZZX}, the coefficients $a_i$ obtained from the QCD 
factorization approach have been used in conjunction with a hadronic 
description of final-state interactions. The general caveats concerning 
hadronic rescattering models apply also here. Moreover, such an 
approach faces a manifest double-counting problem, since rescattering
effects in the heavy-quark limit are already contained in the QCD
coefficients $a_i$.

A nonperturbative treatment of ``charming penguins'', matrix elements 
with charm loops of the type shown in Figure~\ref{fig:penguin}, was 
proposed in \cite{ILNPS}. In this calculation operators with flavour 
structure $(\bar cb)(\bar sc)$ were split into nonlocal products of 
currents $(\bar cb)$ and $(\bar sc)$. These were connected by 
intermediate virtual $D$-meson states to yield transitions such as 
$B\to D\to K\pi$ under the separate action of the currents. However, 
the operators $(\bar cb)(\bar sc)$ are local operators in the effective 
theory at the $b$-mass scale. They would become nonlocal only if probed 
at the far higher scale of $M_W$, but certainly not at a scale of order
$m_D$, as implied in \cite{ILNPS}. We therefore see no basis for the 
method adopted in this paper.

\subsection{Phenomenological analyses}

There are several recent analyses in the literature that approach 
hadronic $B$ decays in a more phenomenological way. Some of the most
extensive studies of this type have been presented by Ali et al.\ 
\cite{Alietal}, who elaborate on the concept of ``generalized
factorization'' \cite{BSW}. In these analyses, part of the 
order-$\alpha_s$ vertex and penguin contributions calculated in the 
present work are included. As a consequence, the decay amplitudes
receive strong-interaction phases due to the Bander--Silverman--Soni
mechanism \cite{BSS}, and nonzero CP asymmetries are obtained. However, 
no systematic attempt to include all such effects (or to prove 
factorization) is made. Also, these authors rely on ad hoc model 
assumptions such as an ``effective number of colours'' $N_c\ne 3$, 
which is introduced to parameterize nonfactorizable corrections. 
 
A recent study of $B\to\pi K$ decays presented in \cite{BRF} emphasizes 
general, model-independent parameterizations of the decay amplitudes 
such as our parameterization in (\ref{para}), avoiding as much as 
possible the use of dynamical input. For instance, strong-interaction 
phases are a priori allowed to take any value in the construction of 
bounds on CKM parameters. Information on these phases may then be 
inferred indirectly. Such an approach is complementary to the more 
ambitious goal of exploiting theoretical insight into the hadronic 
matrix elements. In fact, the smallness of the rescattering effects 
parameterized by $\varepsilon_a\, e^{i\phi_a}$ in (\ref{para}), which 
is a prediction of the QCD factorization approach, represents a useful 
input to such a phenomenological analysis.

Many other phenomenological studies of two-body hadronic $B$ decays 
have been performed in the literature. Recent examples are 
\cite{CY3,WZ,ZWNG,GLT,IP,CMW}, where simplifying assumptions are made 
and no systematic estimates of theoretical uncertainties are undertaken. 

\section{Conclusions}
\label{sec:conclusion}

In this paper we have presented a detailed study of
$B\to\pi K$ and $B\to\pi\pi$ decays based on the QCD factorization
formula. This approach allows us to perform a systematic and
model-independent calculation of two-body hadronic $B$ decays in the
heavy-quark limit. We have evaluated the hard-scattering kernels entering
the matrix elements for $B\to\pi K$, $\pi\pi$ decays at next-to-leading 
order in $\alpha_s$, thus obtaining predictions for the decay amplitudes 
including the leading, ``nonfactorizable'' corrections. 
We have included the contributions from electroweak
penguins, taking into account the corresponding order-$\alpha_s$
corrections to the dominant terms (enhanced by $m^2_t/M^2_W$ or
$1/\sin^2\!\theta_W$) in a consistent approximation scheme.
All hard-scattering kernels have been derived for general, asymmetric
distribution amplitudes, as appropriate for $K$ mesons.

In addition to computing the model-independent leading-twist results, we
have identified and estimated those power corrections which are expected
to be the largest. 
We have analyzed the complete set of contributions from light-cone
wave functions of twist 3 with a chiral enhancement factor
$m^2_\pi/(m_u+m_d)$ or $m^2_K/(m_q+m_s)$, 
as well as the contributions from weak annihilation
topologies, including both twist-2 and twist-3 components in the
light-meson wave functions. We distinguish three classes
of power corrections: 
the contributions from operators with (pseudo-) scalar currents 
($\sim a_{6,8}$), twist-3 effects in the hard spectator interactions, 
and weak annihilation corrections. 
While the first two effects are proportional to the chiral enhancement
factor, the annihilation terms are quadratic polynomials in this factor.
Since these are power-suppressed effects, 
we do not expect the factorization formula to work in these cases. In
fact, naively evaluating these terms in a ``hard-scattering'' framework
one encounters logarithmic endpoint singularities.
(Incidentally, the terms of order $\alpha_s$ related to $a_{6,8}$ are
free of such infrared singularities, but divergent terms are present for
the other
two contributions.) Therefore, while our results at leading twist are
model-independent predictions of QCD in the heavy-quark limit, some
model dependence is currently unavoidable in the description of power 
corrections.
We choose to regulate these divergences by introducing a simple
cutoff, leading to the complex phenomenological parameters $X_H$
and $X_A$. 
The motivation for this model-dependent approach to power
corrections is twofold: we obtain systematic order-of-magnitude
estimates, while automatically keeping track of relative suppressions
or enhancements 
from Wilson coefficients and CKM factors. In addition, we can investigate
the sensitivity of the observables on the model parameters. It turns out
that annihilation graphs may in principle give sizable corrections. On
the other hand, twist-3 effects in the hard spectator contributions
appear generally to be less prominent.
 
Using this framework, we have performed a detailed comparison with the
available experimental data on $B\to\pi K$ and $B\to\pi\pi$ branching ratios.
The data are beginning to
be sufficiently precise to allow for detailed phenomenological analyses.
An important result is that in the QCD factorization approach the
$B\to\pi K$ branching fractions can be larger than the $B\to\pi\pi$ ones 
without
any tuning of the theoretical input parameters, and without invoking large
phenomenological power corrections. An acceptable fit to the branching  
fractions is obtained even if we impose that $\gamma<90^\circ$ as implied  
by the standard constraints on the unitarity triangle. This is in 
contrast to
the naive factorization model, in which the observed branching fractions 
can either not be reproduced at all or require a large CKM angle $\gamma$,
in conflict with other indirect determinations. Encouraged by this
success, we have derived constraints in the $(\bar\rho,\bar\eta)$ plane
from a global fit to the $B\to\pi K$, $\pi\pi$ branching ratios, which 
already provide
useful additional information on the unitarity triangle.
Clearly, the experimental situation will continue to improve 
and allow us to further test our framework and exploit it in
phenomenological studies of hadronic $B$ decays. 
An important role in these analyses 
will be played by the tree-to-penguin ratio $\varepsilon_{3/2}$, which
can be reliably determined from $B^\pm\to\pi^\pm\pi^0$ and 
$B^\pm\to\pi^\pm K^0$
decays and can be unambiguously confronted with the theoretical
predictions.
This will be a valuable check that the power corrections
are not surprisingly large. 
The absence of penguin and annihilation contributions in the
$B^\pm\to\pi^\pm\pi^0$ mode is a crucial feature in this analysis.
A more precise measurement of this decay will thus be particularly
useful.

We would like to conclude by emphasizing again the strategy underlying
our treatment of weak annihilation contributions and non-leading, but  
chirally-enhanced spectator interactions. While the present data do not
require these effects to be larger than expected, the data also do not
definitively exclude this possibility. Our error analysis
is based on allowing the parameters $\varrho_A$ and $\varrho_H$ to be smaller
than 1 (reflecting the range of our expectations), but the error
estimates sometimes depend sensitively on the precise
choice of this upper limit, in particular in the case of $\varrho_A$.
With more precise data one may contemplate fitting $\varrho_A$ to data     
(in practice this implies fitting the QCD penguin amplitude) in order to
decide upon the plausibility of the adopted range of values. The discovery
of large annihilation contributions would by itself constitute valuable
insight into the strong-interaction dynamics of nonleptonic decays, but
would evidently limit the utility of the QCD factorization approach. The 
present data make this scenario appear unlikely, but we look forward to
further
information to validate our error estimation. In this case, important and
reliable information on flavour physics will become available from
$B\to\pi K$ and $B\to\pi\pi$ decays, as well as from a large class of
other two-body hadronic decays into light pseudoscalar and vector mesons.

\vspace{0.15cm}\noindent 
{\it Acknowledgments:\/}
We are grateful to A.~H\"ocker and H.~Lacker for helpful discussions 
and for providing us with a data file of their CKM fit results. We
also thank G.~Martinelli and L.~Silvestrini for useful discussions.
The research of M.N.\ is supported in part by the National Science 
Foundation. C.T.S.\ acknowledges partial support from PPARC through 
Grant No.\ GR/K55738.



\begin{thebibliography}{99}

\bibitem{NR2}
M. Neubert and J.R. Rosner, \prl{81}{1998}{5076} [\hepph{9809311}].

\bibitem{Fl96}
R. Fleischer, \plb{365}{1996}{399} [\hepph{9509204}].

\bibitem{NR1}
M. Neubert and J.R. Rosner, \plb{441}{1998}{403} [\hepph{9808493}].

\bibitem{Mat98}
M. Neubert, \jhep{02}{1999}{014} [\hepph{9812396}]; 
\npps{86}{2000}{477} [\hepph{9909564}].

\bibitem{Fl98}
R. Fleischer, \epjc{6}{1999}{451} [\hepph{9802433}];\\
A.J. Buras and R. Fleischer, \epjc{11}{1999}{93} [\hepph{9810260}].

\bibitem{FM97}
R. Fleischer and T. Mannel, \prd{57}{1998}{2752} [\hepph{9704423}].

\bibitem{Fl99}
R. Fleischer, \plb{459}{1999}{306} [\hepph{9903456}].

\bibitem{GL90}
M. Gronau and D. London, \prl{65}{1990}{3381}.

\bibitem{SiWo94}
J.P. Silva and L. Wolfenstein, \prd{49}{1994}{1151} [\hepph{9309283}].

\bibitem{GQ97}
Y. Grossman and H.R. Quinn, \prd{58}{1998}{017504} [\hepph{9712306}].

\bibitem{Ch98}
J. Charles, \prd{59}{1999}{054007} [\hepph{9806468}].

\bibitem{QS93}
A.E. Snyder and H.R. Quinn, \prd{48}{1993}{2139}.

\bibitem{Zoltan}
Z. Ligeti, M. Luke and M.B. Wise, Preprint LBNL-47551 [\hepph{0103020}].

\bibitem{BBNS1}
M. Beneke, G. Buchalla, M. Neubert and C.T. Sachrajda, 
\prl{83}{1999}{1914} [\hepph{9905312}].

\bibitem{BBNS2}
M. Beneke, G. Buchalla, M. Neubert and C.T. Sachrajda, 
\npb{591}{2000}{313} [\hepph{0006124}].

\bibitem{BBNS3}
M. Beneke, G. Buchalla, M. Neubert and C.T. Sachrajda, Preprint 
PITHA-00-13 [\hepph{0007256}], to appear in the Proceedings of the 30th 
International Conference on High-Energy Physics (ICHEP 2000), Osaka, 
Japan, 27 July--2 August 2000.

\bibitem{KLS00}
Y.Y. Keum, H.-n.\ Li and A.I. Sanda, Preprint KEK-TH-642 
[\hepph{0004004}]; 
\prd{63}{2001}{054008} [\hepph{0004173}];\\
Y.Y. Keum and H.-n.\ Li, \prd{63}{2001}{074006} [\hepph{0006001}].

\bibitem{MSYY}
T. Muta, A. Sugamoto, M.Z. Yang and Y.D. Yang, \prd{62}{2000}{094020} 
[\hepph{0006022}].

\bibitem{YY1}
M.Z. Yang and Y.D. Yang, \prd{62}{2000}{114019} [\hepph{0007038}].

\bibitem{YY2}
M.Z. Yang and Y.D. Yang, Preprint OCHA-PP-166 [\hepph{0012208}].

\bibitem{HMW}
X.G. He, J.P. Ma and C.Y. Wu, \prd{63}{2001}{094004}
[\hepph{0008159}].

\bibitem{DYZ}
D.-s.\ Du, D.-s.\ Yang and G.-h.\ Zhu, \plb{488}{2000}{46} 
[\hepph{0005006}];
Preprints \hepph{0008216}, \hepph{0102077}, \hepph{0103211}.

\bibitem{CY2}
H.-Y. Cheng and K.-C. Yang, Preprint \hepph{0012152}.

\bibitem{Bj89}
J.D. Bjorken, \npps{11}{1989}{325}.

\bibitem{DG91}
M.J. Dugan and B. Grinstein, \plb{255}{1991}{583}.

\bibitem{PW91}
H.D. Politzer and M.B. Wise, \plb{257}{1991}{399}.

\bibitem{BJLW}
A.J. Buras, M. Jamin and M.E. Lautenbacher, \npb{400}{1993}{75} 
[\hepph{9211321}]. 

\bibitem{CFMR}
M. Ciuchini, E. Franco, G. Martinelli and L. Reina, \plb{301}{1993}{263}
[\hepph{9212203}]; \npb{415}{1994}{403} [\hepph{9304257}].

\bibitem{DeHe}
N.G. Deshpande and X.-G. He, \prl{74}{1995}{26} [E: {\bf 74} (1995) 
4099] [\hepph{9408404}].

\bibitem{BBL}
G. Buchalla, A.J. Buras and M.E. Lautenbacher, \rmp{68}{1996}{1125}
[\hepph{9512380}].

\bibitem{BGH}
A.J. Buras, P. Gambino and U.A. Haisch, \npb{570}{2000}{117} 
[\hepph{9911250}].

\bibitem{GrNe}
A.G. Grozin and M. Neubert, \prd{55}{1997}{272} [\hepph{9607366}].

\bibitem{BF00}
M. Beneke and T. Feldmann, \npb{592}{2001}{3} [\hepph{0008255}].

\bibitem{BraF}
V.M. Braun and I.E. Filyanov, \zpc{44}{1989}{157}; \zpc{48}{1990}{239}.

\bibitem{Gesh}
B.V. Geshkenbein and M.V. Terentev, \plb{117}{1982}{243};
\sjnp{40}{1984}{487} [\yf{40}{1984}{758}]; \yf{39}{1984}{873}. 

\bibitem{BBN01}
T. Becher, B.D. Pecjak and M. Neubert, Preprint CLNS~01-1723 
[\hepph{0102219}].

\bibitem{BSS}
M. Bander, D. Silverman and A. Soni, \prl{43}{1979}{242}.

\bibitem{Korch}
G.P. Korchemsky, D. Pirjol and T.-M. Yan, \prd{61}{2000}{114510}
[\hepph{9911427}].

\bibitem{Khod}
A. Khodjamirian et al., \prd{62}{2000}{114002} [\hepph{0001297}];\\ 
A. Khodjamirian and R. R\"uckl, in: {\it Heavy Flavours II}, A.J.~Buras 
and M.~Lindner eds.\ (World Scientific, Singapore, 1998) pp.~345 
[\hepph{9801443}].

\bibitem{Patr}
P. Ball, \jhep{09}{1998}{005} [\hepph{9802394}].

\bibitem{Stech00}
D. Melikhov and B. Stech, \prd{62}{2000}{014006} [\hepph{0001113}].

\bibitem{Bpilat}
A. Abada et al., \npps{83}{2000}{268} [\heplat{9910021}].

\bibitem{AKH}
A. Khodjamirian, Preprint CERN-TH-2000-325 [\hepph{0012271}].

\bibitem{Burr01}
C.N. Burrell and A.R. Williamson, Preprint UTPT-01-02 [\hepph{0101190}].

\bibitem{Falk}
A.F. Falk, A.L. Kagan, Y. Nir and A.A. Petrov, \prd{57}{1998}{4290} 
[\hepph{9712225}].

\bibitem{GR}
M. Gronau and J.L. Rosner, \prd{59}{1999}{113002} [\hepph{9809384}].

\bibitem{CLEO00}
D. Cronin-Hennessy et al.\ (CLEO Collaboration), \prl{85}{2000}{515}.

\bibitem{BaBar01}
L. Cavoto (BaBar Collaboration), talk at the 36th Rencontres de Moriond 
on QCD and Hadronic Interactions, Les Arcs, France, 17--24 March 2001, 
to appear in the Proceedings.

\bibitem{Belle01}
T. Iijima (Belle Collaboration), talk at the 4th International 
Conference On $B$ Physics And CP Violation (BCP4), Ise-Shima, Japan, 
19--23 February 2001, to appear in the Proceedings.

\bibitem{UT1}
A. Ali and D. London, \epjc{9}{1999}{687} [\hepph{9903535}].

\bibitem{UT2}
S. Plaszczynski and M.-H. Schune, talk given at 8th International 
Symposium on Heavy Flavour Physics, Southampton, UK, 25--29 July 1999, 
Preprint LAL-99-67 [\hepph{9911280}].

\bibitem{UT3}
M. Ciuchini et al., Preprint LAL-00-77 [\hepph{0012308}].

\bibitem{Hoecker}
A. H\"ocker, H. Lacker, S. Laplace and F. Le Diberder, Preprint
LAL-01-06 [\hepph{0104062}].

\bibitem{GRL}   
M. Gronau, J.L. Rosner and D. London, \prl{73}{1994}{21}
[\hepph{9404282}].

\bibitem{sin2bref}
J. Beringer (BaBar Collaboration), talk at the 36th Rencontres de 
Moriond on QCD and Hadronic Interactions, Les Arcs, France, 17--24 
March 2001, to appear in the Proceedings.

\bibitem{Gerhard}
G. Buchalla and A.J. Buras, \prd{54}{1996}{6782} [\hepph{9607447}].

\bibitem{KaNe00}
A.L. Kagan and M. Neubert, \plb{492}{2000}{115} [\hepph{0007360}].

\bibitem{Joao}
J.P. Silva and L. Wolfenstein, \prd{63}{2001}{056001} [\hepph{0008004}]. 

\bibitem{Yossi}
G. Eyal, Y. Nir and G. Perez, \jhep{0008}{2000}{028} [\hepph{0008009}].

\bibitem{Xing}
Z.-z. Xing, Preprint \hepph{0008018}.
 
\bibitem{Buras2}
A.J. Buras and R. Buras, \plb{501}{2001}{223} [\hepph{0008273}].

\bibitem{Ciuc97}
M. Ciuchini, E. Franco, G. Martinelli and L. Silvestrini,
\npb{501}{1997}{271} [\hepph{9703353}].

\bibitem{Cetal01}
M. Ciuchini et al., Preprint \hepph{0104018}.

\bibitem{George}
W.-S. Hou and K.-C. Yang, \prl{84}{2000}{4806} [\hepph{9911528}].

\bibitem{CLEOAcp}
S. Chen et al.\ (CLEO Collaboration), \prl{85}{2000}{525} 
[\hepex{0001009}].

\bibitem{SHB}
A. Szczepaniak, E.M. Henley and S.J. Brodsky, \plb{243}{1990}{287}.

\bibitem{BD}
G. Burdman and J.F. Donoghue, \plb{270}{1991}{55}.

\bibitem{LY}
H.-n.\ Li and H.-L. Yu, \prd{53}{1996}{2480} [\hepph{9411308}].

\bibitem{SR}
A.P. Szczepaniak and A. Radyushkin, \prd{59}{1999}{034017}
[\hepph{9810438}].

\bibitem{LUY}
C.-D. Lu, K. Ukai and M.-Z. Yang, \prd{63}{2001}{074009}
[\hepph{0004213}].

\bibitem{CLI}
C.-H. Chen and H.-n.\ Li, \prd{63}{2001}{014003} [\hepph{0006351}].

\bibitem{BFW}
B.F.L. Ward, Preprint \hepph{0006357}.

\bibitem{LYA}
C.-D. Lu and M.-Z. Yang, Preprint HUPD-0012 [\hepph{0011238}].

\bibitem{ZZX}
Z.-z.\ Xing, \plb{493}{2000}{301} [\hepph{0007136}].

\bibitem{PZE}
P. Zenczykowski, \prd{63}{2001}{014016} [\hepph{0009054}].

\bibitem{KTE}
K. Terasaki, Preprint YITP-00-65 [\hepph{0011358}].

\bibitem{ILNPS}
C. Isola et al., Preprint BA-TH-404-00 [\hepph{0101118}].

\bibitem{Alietal}
A. Ali and C. Greub, \prd{57}{1998}{2996} [\hepph{9707251}];\\
A. Ali, G. Kramer and C.-D. Lu, \prd{58}{1998}{094009} 
[\hepph{9804363}]; \prd{59}{1999}{014005} [\hepph{9805403}].

\bibitem{BSW}
M. Bauer, B. Stech and M. Wirbel, \zpc{34}{1987}{103};\\
M. Neubert and B. Stech, in: {\it Heavy Flavours II}, ed.\ 
A.J.~Buras and M.~Lindner (World Scientific, Singapore, 1998) 
pp.~294 [hep-ph/9705292].

\bibitem{BRF}
A.J. Buras and R. Fleischer, \epjc{16}{2000}{97} [\hepph{0003323}].

\bibitem{CY3}
H.-Y. Cheng and K.-C. Yang, \prd{62}{2000}{054029} [\hepph{9910291}].

\bibitem{WZ}
Y.-L. Wu and Y.-F. Zhou, \prd{62}{2000}{036007} [\hepph{0002227}].

\bibitem{ZWNG}
Y.F. Zhou, Y.L. Wu, J.N. Ng and C.Q. Geng, \prd{63}{2001}{054011}
[\hepph{0006225}].

\bibitem{GLT}
X.H. Guo, O. Leitner and A.W. Thomas, \prd{63}{2001}{056012}
[\hepph{0009042}].

\bibitem{IP}
C. Isola and T.N. Pham, Preprint CPHT-S-085-0900 [\hepph{0009210}].

\bibitem{CMW}
W.N. Cottingham, H. Mehrban and I.B. Whittingham, Preprint 
\hepph{0102012}.

\end{thebibliography}
\end{document}